\documentclass[aps,pra,groupedaddress,superscriptaddress,twocolumn,10pt,longbibliography]{revtex4-2}

\usepackage[utf8]{inputenc}
\usepackage[T1]{fontenc}
\usepackage[english]{babel}
\usepackage{amsmath}
\usepackage{physics}
\usepackage{graphicx}
\usepackage[colorlinks=true, allcolors=black]{hyperref}

\usepackage{orcidlink}

\usepackage{xcolor}
\usepackage{graphicx}
\usepackage[left=23mm,right=13mm,top=35mm,columnsep=15pt]{geometry} 

\usepackage{placeins}
\usepackage[T1]{fontenc}
\usepackage{appendix}

\usepackage{comment}

\usepackage{color}

\usepackage{soul}

\AtBeginDocument{}
\AtBeginDocument{}
\begin{document}
    \title{Predissociation dynamics of charged long-range Rydberg molecules}
 \author{Neethu Abraham}
  \email[Contact author: ]{abraham@pks.mpg.de}
	\affiliation{Max Planck Institute for the Physics of Complex Systems,  Nöthnitzer Strasse 38, 01187 Dresden, Germany}
 \author{P. Giannakeas \orcidlink{0000-0002-1164-0201}}
	\affiliation{Max Planck Institute for the Physics of Complex Systems,  Nöthnitzer Strasse 38, 01187 Dresden, Germany}
\author{Matthew T. Eiles \orcidlink{0000-0002-0569-7551}}
 \email[Contact author: ]{meiles@purdue.edu}
	\affiliation{Max Planck Institute for the Physics of Complex Systems,  Nöthnitzer Strasse 38, 01187 Dresden, Germany}
    \affiliation{Department of Physics and Astronomy, Purdue University, West Lafayette, IN 47907, USA}
	\date{\today} 
\begin{abstract}
We investigate predissociation in homonuclear ($^{87}$Rb$^*$$^{87}$Rb$^+$) and heteronuclear ($^{87}$Rb$^*$$^{7}$Li$^+$) long-range Rydberg atom-ion molecules.
Owing to their micron-scale bond lengths, these dissociate on time scales far removed from those of more tightly bound diatomic molecules. 
We employ the eigenchannel $R$-matrix method to compute predissociation rates for a broad range of principal quantum numbers $n$.
The rates depend strongly on the mass, but more remarkably display a rapid and periodic variation as a function of $n$ as well as within a single vibrational ladder. 
A semiclassical Landau-Zener-St\"uckelberg analysis reveals that St\"uckelberg interference governs the decay process and produces the observed variation in the molecular lifetime. 
Although the heavy mass of the homonuclear Rb molecule constrains its predissociation rates to a sub-kHz level, the lighter molecule $^{87}$Rb$^*$$^{7}$Li$^+$ dissociates on time scales competitive with radiative and collisional decay. 
This can enable \textit{in situ} study of non-adiabatic decay via ion microscopy. 

\end{abstract}

\maketitle
\section{Introduction}

Electronically excited molecular levels typically exist as resonances embedded in the dissociation continua of energetically lower electronic configurations. 
Within the Born-Oppenheimer, or adiabatic approximation, different electronic states are not directly coupled and radiative processes are required for the molecule to dissociate.
However, non-adiabatic coupling permits electronic levels to relax into other configurations, particularly in the vicinity of conical intersections \cite{Domcke2004ConicalIntersections, Schuurman2011ConicalIntersections, Scheid2019NonadiabaticPhotodynamics}. 
 This leads to non-radiative molecular dissociation.
When considering the formation and dynamics of many molecular states of interest, from ion-pair molecules  \cite{Donovan2022HeavyRydberg,kirrander2010HeavyRydberg,kirrander2011IonPair,Reinhold2005HeavyRydberg} or weakly-bound Feshbach molecules \cite{kokoouline2000theoretical,mcguyer2015precise,Chin2010FeshbachResonances} to exotic ultralong-range molecules containing one or more Rydberg atoms \cite{Greene2000Trilobite,Boisseau2002Macrodimers,Deiss2021AtomIonRydberg,Duspayev2021a, Eiles2019,Dunning2024UltralongRange,Shaffer2018UltracoldRM}, their stability against predissociation is of paramount importance.

In long-range Rydberg molecules bound by multipolar forces between extended electronic charge clouds, local minima form in the potential energy curves (PECs) due to level repulsion between the repulsive and attractive interactions associated with different excited electronic states \cite{Deiglmayr2016LongRange,Sassmannshausen2016Macrodimers,Hollerith2023MacrodimerReview,Weber2017CalculationRydbergPotentials}. 
The existence of potential wells able to support vibrational levels is therefore linked to avoided crossings where non-adiabaticity is strongest \cite{Baer2006BeyondBO,Yarkony1996DiabolicalCI,BixonJortner1968}. 
Furthermore, the long radiative lifetimes of Rydberg states provide ample time for non-adiabatic processes to occur \cite{Gallagher1994,Theodosiou1984}.
For these reasons, the role of predissociation and other non-adiabatic processes in ultralong-range Rydberg molecules composed of a Rydberg atom and a ground-state atom or molecule \cite{Durst2025Nonadiabatic, Srikumar2023Nonadiabatic,Gonzalezferez2024UltralongRange}, Rydberg macrodimers composed of two Rydberg atoms \cite{Schwettmann2007LongRange, Boisseau2002Macrodimers, Deiglmayr2016LongRange}, and charged Rydberg molecules composed of a Rydberg atom and an ion \cite{Deiss2021AtomIonRydberg, Duspayev2021a,bosworth2024metastable}, has been the subject of significant recent study. 
In the latter case, a striking non-monotonic dependence of the predissociation lifetimes on both the vibrational and  principal quantum numbers was theoretically observed \cite{Duspayev2022NonadiabaticRAIM} and connected to interference between different decay pathways.

This phenomenon is familiar from earlier studies of diatomic predissociation, both from theoretical studies of IBr  \cite{Child1971CurveCrossing,Child1974Semiclassical, Child1976Predissociation, Knockel1981} and the alkali-halides \cite{Sadeghpour2000ChannelInteraction,Balakrishnan1999QuantumWavePacket,Cornett1999Interferometric,Engel1988NaIDissociation, Choi1989NaIPredissociation,Mizuno2018,Liane2024} and from  
femtosecond pump-probe spectroscopy of these molecules \cite{Cong1996NaIPredissociation,deVries1980}.
These similarities suggest that long-range Rydberg molecules offer an opportunity to re-investigate predissociation dynamics -- in particular, the effects of interference between adiabatic and diabatic pathways -- on microsecond, rather than picosecond, time scales. 
Moreover, their large sizes could enable the possibility for spatial resolution of the decay process using ion microscopy \cite{Zou2023VibrationalRAIM,Berngruber2024InSituCollisionDynamics}. 
However, the non-adiabatic decay rates computed in Ref. \cite{Duspayev2022NonadiabaticRAIM} for  $^{87}$Rb$^*$$^{87}$Rb$^+$ long-range Rydberg-atom--ion molecules (RAIMs) proved to be far too small in comparison with radiative and collisional decay for predissociation to be relevant.
While this means that rubidium RAIMs are stable with respect to predissociation and hence sufficiently long-lived for their spectroscopic observation \cite{Zou2023VibrationalRAIM,Zuber2022}, an observation of the effects of non-adiabatic coupling on their decay remain out of reach.

Hence, in this article, we re-examine the role of non-adiabatic predissociation in RAIMs by investigating the much lighter heteronuclear molecule $^{87}$Rb$^*$$^7$Li$^+$. 
The electronic energy levels are independent of the choice of ion, which is structureless at the length scales characteristic of these Rydberg molecules.
Changing the ion's mass thus allows us to independently tune the non-adiabatic coupling strength. 
We have used a combination of multichannel scattering calculations and semiclassical methods to compute the vibronic resonance parameters. 
We find that $^{87}$Rb$^*$$^7$Li$^+$ RAIMs are sufficiently light that non-adiabatic decay occurs on microsecond time scales, both imposing limits on their stability and permitting the study of predissociation dynamics on relevant time scales. 
Using the Landau-Zener-Stückelberg semiclassical treatment, we identify the role of interference between adiabatic and diabatic pathways in structuring the predissociation rates as a function of both the vibrational and electronic energy levels. 
Finally, we detail how the molecular lifetimes depend on mass and principal quantum number

The article is organized as follows. Section \ref{sec:hamiltonian} introduces the long-range Rydberg-atom--ion molecule (RAIM). 
Section \ref{sec:pecs} examines the generic properties of the potential energy curves, and Section \ref{sec:spectator} discusses the emergence and treatment of ``almost dark'' states.
Sections \ref{sec:rmatrix} and \ref{sec:wkb} present the quantum-mechanical and semiclassical methodology, respectively. 
Sections~\ref{sec:resonance} and ~\ref{sec:q_para} present closed-form semiclassical expressions. 
Readers primarily interested in the results may skip Sections \ref{sec:spectator} - ~\ref{sec:q_para} and start immediately with Section~\ref{sec:RB-Li_lftm_analy}, which analyzes the predissociation lifetimes of $^{87}$Rb$^*$$^{7}$Li$^+$. The decay rates of $^{87}$Rb$^*$$^{87}$Rb$^+$ are studied in Sec.~\ref{sec:RB-Rb_lftm_analy}. Section~\ref{sec:mass_dep_analy} investigates the dependence of the lifetimes on the mass and principal quantum number $n$. Finally, we conclude with a summary of our results and an outlook.

\section{Theoretical Background}
\label{sec:theory}
\subsection{Rydberg atom -- ion molecule}
\label{sec:hamiltonian}
The system considered here consists of a $^{87}$Rb$^*$ Rydberg 
atom interacting with a distant positively charged alkali ion, which will either be a $^{7}$Li$^+$ or $^{87}$Rb$^+$ ion the following.
As this system is discussed in detail in Refs. \cite{Deiss2021AtomIonRydberg,Duspayev2022NonadiabaticRAIM,Duspayev2021a}, we discuss here only the key details most pertinent to our study of non-adiabatic dynamics.
We refer to the long-range Rydberg-atom--ion molecule (RAIM) as either Rb$^*$Rb$^+$ or Rb$^*$Li$^+$, denoting the Rydberg species with the asterisk and suppressing the isotope labels. 
We use atomic units unless otherwise specified. 

Because the internuclear distance $R$ exceeds the LeRoy radius \cite{LeRoy1974RKR}, the interaction between the Rydberg atom and the ion is given by the multipole expansion \cite{Jackson1998}
\begin{equation}
    \hat{V}(R, \mathbf{r}) = - \sum_{L=1}^{\infty} \sqrt{\frac{4\pi}{2L+1}} \frac{r^L}{R^{L+1}} Y_{L0}(\theta, \varphi). 
\end{equation}
 The internuclear axis is parallel to the quantization axis. 
We truncate the sum at $L = 6$ to achieve converged results. 

We define, at a given $R$, an adiabatic electronic basis, $\phi_\alpha(\mathbf{r}; R)$, by solving the eigenvalue equation
\begin{equation}
    \left[\hat{H}_\mathrm{Ryd}(\mathbf{r})+\hat V(R,\mathbf{r})\right]\phi_{\alpha}(\mathbf{r};R)
    = V_{\alpha}(R)\,\phi_{\alpha}(\mathbf{r};R),
\end{equation}
where $\hat H_\mathrm{Ryd}(\mathbf{r})$ is the Hamiltonian of the Rydberg atom. The eigenvalues $V_\alpha(R)$ therefore form the Born-Oppenheimer potential energy curves (PECs) \cite{Born1927}.
We represent the adiabatic states in a basis of single-electron Rydberg states following Refs.~\cite{Duspayev2021a,Deiss2021AtomIonRydberg}. 

The total molecular wave function is expanded into the adiabatic eigenstates $\phi_\alpha(\mathbf{r};R)$, i.e.
\begin{equation}
\label{eq:fullwf}
   \Psi_j(\mathbf{r},\mathbf{R}) = \sum_\alpha Y_{00}(\hat R)\phi_\alpha(\mathbf{r};R)\frac{\chi_{\alpha j}({R})}{R},
\end{equation} 
where the sum over $\alpha$ runs over all considered states in the adiabatic basis, $\chi_{\alpha j}(R)$ is the $j$th linearly independent radial vibrational wave function in the adiabatic channel $\alpha$, and $Y_{NM_N}(\hat R)$ is a spherical harmonic.  Because of its huge size, the molecule's rotational splitting is negligible and we consider only the case of zero molecular angular momentum. 
There are as many linearly independent solutions $\Psi_j$ as there are open dissociation channels at a given energy $E$. 
Substituting Eq.~\ref{eq:fullwf} into the Schrödinger equation yields a set of coupled second-order ordinary differential equations 
\begin{equation}
\left[ -\frac{1}{2\mu}\frac{d^2}{d R^2} + V_\alpha(R) - E \right] \chi_{\alpha j}(R)
= \frac{1}{2\mu} \sum_{\alpha'} \Lambda_{\alpha\alpha'} \chi_{\alpha' j}(R),
\label{eq:nuclearSE}
\end{equation}
where $\mu$ is the molecular reduced mass. The non-adiabatic coupling operator 
\begin{equation}
    \Lambda_{\alpha\alpha'} = 2P_{\alpha\alpha'}\frac{d}{dR}+ Q_{\alpha\alpha'}
\end{equation}
consists of the derivative and scalar couplings
\begin{equation}
P_{\alpha\alpha'} = \left\langle \phi_\alpha \left| \frac{d}{dR}\right| \phi_{\alpha'} \right\rangle, \qquad
Q_{\alpha\alpha'} = \left\langle \phi_\alpha \left| \frac{d^2}{dR^2}\right| \phi_{\alpha'} \right\rangle.
\end{equation}
We compute $P_{\alpha\alpha'}$ using the Hellmann--Feynman theorem and $\underline{Q}$ from the identity $\underline{Q} = \frac{d}{dR}{\underline{P}} +\underline{P}\cdot\underline{P}$.
The molecular binding energies $E_\nu$ are obtained by dropping all terms $\Lambda_{\alpha\alpha'}$ before solving Eq.~\ref{eq:nuclearSE}. 
The Born-Huang approximation consists of keeping the diagonal terms $\Lambda_{\alpha\alpha}$ \cite{Born1954DynamicalTheory}. 
Although this accounts for some non-adiabatic effects, the potential energy curves remain decoupled, as in the Born-Oppenheimer approximation, and thus only a full solution of Eq.~\ref{eq:nuclearSE} can be used to obtain predissociation rates.

\subsection{Born-Oppenheimer potential energy curves and derivative couplings}
\label{sec:pecs}
\begin{figure}
    \includegraphics[width=0.47\textwidth]{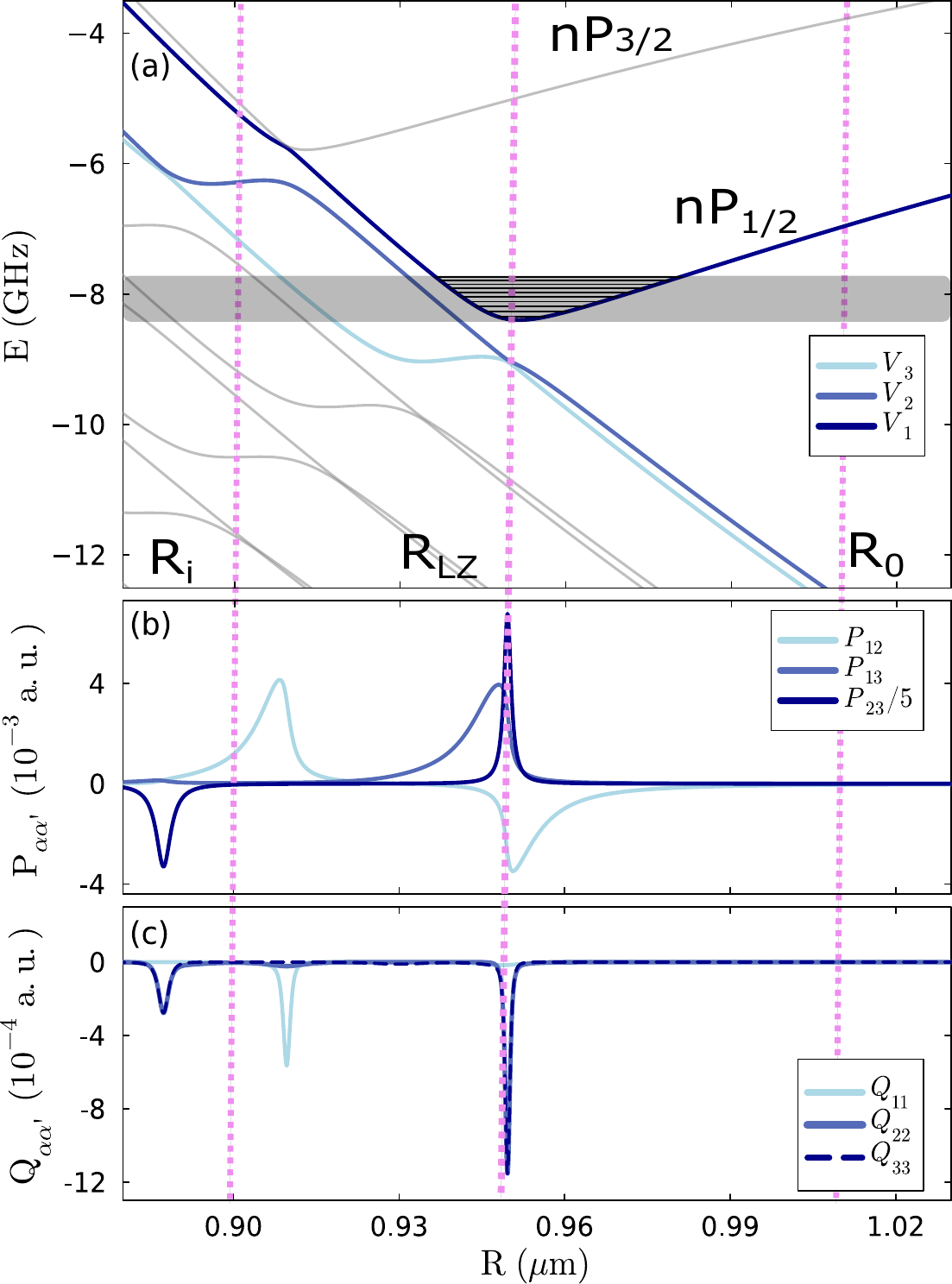}
    \caption{(a) Born--Oppenheimer potential energy curves (PECs) relevant to RAIM formation and predissociation from the $nP_{1/2}$ Rydberg state. All PECs for $m_j=1/2$ are shown. 
    The curves $V_1(R)$, $V_2(R)$, and $V_3(R)$ asymptotically approach the atomic $40P_{1/2}$ and $n=37$ levels, respectively. 
    The potential well in $V_1(R)$ supports many vibrational bound states, the first ten of which (for Rb$^*$Li$^+$) are shown in the black line spectrum. 
    We calculate the predissociation decay rate as a function of energy within the range indicated by the shaded region.  
    $R_i$ and $R_0$ denote the inner boundary and outer matching point for the $R$-matrix calculation, and $R_{LZ}$ is the position of the avoided crossing where a Landau-Zener (LZ) transition between $V_1$ and $V_2$ can occur.
    (b) The nonadiabatic derivative coupling matrix elements $P_{\alpha\alpha'}(R)$ between the labeled adiabatic states. 
    (c) The diagonal nonadiabatic scalar coupling elements $Q_{\alpha\alpha'}(R)$ for the labeled adiabatic states.}
    \label{fig:pecs}
\end{figure}

Representative Born-Oppenheimer potential energy curves $V_\alpha(R)$ are shown in Fig. \ref{fig:pecs}.
As the electronic structure is determined entirely by the Rydberg atom, the curves shown in Fig.~\ref{fig:pecs} apply to both Rb$^*$Rb$^+$ and Rb$^*$Li$^+$. 
The projection of the electronic angular momentum $j$ onto $R$, $m_j$, is a good quantum number.
For all calculations presented here we take $m_j=1/2$. 
We are interested in RAIMs which form in the potential energy curve (PEC) denoted $V_1(R)$ in Fig. \ref{fig:pecs}(a). 
This PEC approaches the energy of the $nP_{1/2}$ state asymptotically (for this specific example we have $n=40$), which serves as the zero of energy in this figure. 
At large $R$, the associated adiabatic electronic state is predominantly of $nP_{1/2}$ character and hence
the asymptotic form of this PEC is $V_1(R)\to-\alpha_P/2R^4$, where $\alpha_P$ is the polarizability of the atomic Rydberg state.

The electronic levels closest in energy to the $nP_{1/2}$ state are those of the $n-3,\ell \gg 1$ hydrogen-like states, due to the large ($\sim 2.65$) quantum defect of the $nP_{1/2}$ state.  
These high-$\ell$ states have vanishing quantum defects and therefore form polar states exhibiting a linear Stark splitting in the field of the ion, leading to PECs of the form $a/R^{2}$ at large $R$. 
At $R=R_{LZ}$, level repulsion causes the adiabatic potential $V_1(R)$ to bend away from these repulsive levels, forming a potential well but also inducing strong non-adiabatic couplings (Fig. \ref{fig:pecs}(b) and (c)). 
These are sharply localized near the avoided crossings between adjacent PECs, reflecting the rapid change of the adiabatic electronic wavefunctions in these regions.
In the energetic range (shaded region) of interest, the most significant non-adiabatic couplings to this upper state are to the adiabatic states corresponding to levels $V_2(R)$ and $V_3(R)$.

\subsection{``Almost dark'' state potential energy curves}
\label{sec:spectator}
One distinctive feature of the electronic structure of long-range Rydberg molecules is the presence of ``almost dark'' states which are almost entirely decoupled from other adiabatic states and hence exhibit very small, but finite, avoided crossings.  
These generate sharp derivative coupling elements which present several numerical and conceptual challenges in a coupled adiabatic representation. 
The electronic state evolving along the adiabatic PEC $V_3$ for $R>R_{LZ}$, before transferring at the very narrow avoided crossing to the PEC $V_2$ for $R<R_{LZ}$ is one such ``almost dark'' state in the three-level system considered here. 
It comes about because the diabatic electronic states approximately form a $\Lambda$-type three-level system. 
The dipole operator couples the $nP_{1/2}$ state to the two polar states, only couples the latter two together via their very weak fine structure mixing and coupling to states outside of this three-channel sector. 
The adiabatic PECs therefore consist of two strongly coupled levels with a large avoided crossing intersected by the nearly uncoupled dark-state PEC.  

This very weak interaction between diabatic states leads the large, but finite, $P_{23}$ derivative matrix element.
The Lorentzian $P$-matrix element expected at an avoided crossing between two levels \cite{Clark1979} is split between the two elements $P_{12}$ and $P_{13}$. 
The $P_{23}$ matrix element's sharpness requires a very dense radial grid in the region of the avoided crossing to be accurately evaluated. 
Additionally, the accuracy with which some observables are calculated can depend extremely sensitively on the magnitude of this matrix element. 
We have seen that an error of as little as one percent in the peak value of $P_{23}$ leads to changes of three orders of magnitude in the predissociation lifetimes computed for the Rb$^*$Rb$^+$ RAIM. 

To address these issues, we have numerically confirmed that the converged three-channel results including this almost singular $P_{23}$-matrix element are reproduced well by a quasi-adiabatic calculation where we interpolate between adiabatic potentials $V_{2}$ and $V_3$ at the narrow avoided crossing between them to form one adiabatic curve $U_2$ (see Fig. \ref{fig:wkbpecs}). 
The segments of $V_2$ and $V_3$ not used in constructing $U_2$ form the dark-state PEC which exhibits real crossings with $U_2$ and can therefore be excluded from the calculation. 
The $P$-matrix element coupling this adiabatic potential to the upper adiabatic potential $V_1$ is obtained from fitting a single Lorentzian profile to the wings of $P_{13}$ and $P_{12}$. 
The quantum dynamics are then solved, equivalently as above, in this effective two-channel system with one avoided crossing. 
In the semiclassical analysis described in the next section, we use this two-channel model exclusively. 
To avoid confusion, we will use the label $U_1\equiv V_1$ to refer to the upper potential curve in this effective two-level model of the full system, with $U_2$ the lower potential energy curve assembled as discussed above. 

We note that ``almost dark'' states occur in other ultralong-range Rydberg molecules and require special treatment there as well. 
These states must be treated carefully in order to compute, for example, the spectra of trilobite molecules \cite{Srikumar2025Trilobite}, butterfly molecules of singlet symmetry \cite{Exner2026SingletButterfly}, and non-polar ultralong-range Rydberg molecules \cite{Guttridge2025Tweezers}. 
In most cases, it proves accurate to follow the same approach outlined here and treat all narrow avoided crossings strictly diabatically, but a more rigorous framework would be beneficial to avoid the need for interpolation.

\subsection{Eigenchannel R-matrix solution}
\label{sec:rmatrix}
We employ the eigenchannel R-matrix method 
\cite{Aymar1996} to solve Eq.~\ref{eq:nuclearSE}. 
This method takes advantage of the natural partition of configuration space into two regions based on the rapid decay of the non-adiabatic coupling elements for $R>R_{LZ}$.
The matching distance $R_0$ is chosen such that (a) all non-adiabatic couplings are vanishingly small for $R>R_0$, and (b) the upper channel is energetically closed, i.e. we have $\chi_{1j}(R\ge R_0) = 0$. 
Criterion (a) also implies that the solutions evolve independently on the uncoupled potential energy curves $V_\alpha(R)$ for all $R>R_0$. 
The inner boundary $R_i$ is chosen to be small enough that we can impose the boundary condition $\chi_{\alpha j}(R_i) = 0$. 

Following Refs.~\cite{Aymar1996,Lane1958,Burke2011, WangND,Burke1999}, we obtain the eigenchannel $R$-matrix at a fixed energy $E$ for either one closed channel ($V_1$) and two open channels ($V_2,V_3$), for the full solution, or for one closed channel ($U_1$) and one open channel ($U_2$) for the two-channel approximation. 
The $R$-matrix fully determines the eigenchannel solutions $\chi_{\alpha j}$ at the boundary $R=R_0$, since the eigenvalues of the R-matrix, $b_\alpha$, determine the logarithmic derivative at $R_0$. 
We match the open channel solutions with $\alpha\ge 2$ solutions to a linear combination of regular $f_{\alpha,E}(R)$ and irregular $g_{\alpha, E}(R)$ long-range solutions to determine the coefficient matrices $\underline{I}$ and $\underline J$: 
\begin{align}
    \chi_{\alpha j}(R_0) &= f_{\alpha,E}(R_0)I_{\alpha j}- g_{\alpha,E}(R_0)J_{\alpha j}\\
    -b_\alpha \chi_{\alpha j}(R_0) &=  \left.\frac{d}{dR}f_{\alpha,E}(R)I_{\alpha j}- \frac{d}{dR}g_{\alpha,E}(R)J_{\alpha j}\right|_{R=R_0}.\nonumber
\end{align}

The long-range solutions for $R\ge R_0$ are determined numerically. 
As $R\to\infty$, the leading order behavior of the PECs $V_{\alpha=2,3}(R)$ is that of a dipole potential $V(R) \sim \frac{a_\alpha}{2R^{2}}$. 
The regular and irregular solutions to the radial Schrödinger equation with this potential are \cite{Watson1922}
\begin{subequations}
\begin{align}
f_{\alpha,E}(R) &=\sqrt{R}\, J_{\sqrt{1 + 4 a_\alpha \mu}}\!\left(\sqrt{2 \mu \epsilon_\alpha}\, R\right), \\
g_{\alpha,E}(R) &= \sqrt{R}\, Y_{\sqrt{1 + 4 a_\alpha \mu}}\!\left(\sqrt{2 \mu \epsilon_\alpha}\, R\right),
\end{align}
\label{eq:fandg}
\end{subequations}
where the energy in channel $\alpha$ with respect to its threshold energy is $\epsilon_\alpha = E - V_\alpha(R\to\infty)$. 
We use Numerov's algorithm to backpropagate the radial solutions in each channel using Eq.~\ref{eq:fandg} as the initial condition at a large $R$. 
Thus, we account for deviations from a pure $R^{-2}$ power law in the PECs arising from the higher-order multipole terms.  

The scattering matrix is $\underline{S} = (\underline{I}-i\underline J)(\underline I+i\underline J)^{-1}$  \cite{Aymar1996}.
The resonance positions and widths are readily obtained by considering the time-delay matrix 
\cite{Wigner1955,Smith1960}
\begin{equation}
\underline{\tau}_{PD}(E) = -i\,\underline{S}^\dagger\frac{d\underline{S}}{dE}. 
\label{time_del_qtm}
\end{equation}
For a multichannel system, the largest eigenvalue of $\underline{\tau}_{PD}$ traces a Lorentzian profile about each resonance. 
We numerically extract the resonance position and width from this Lorentzian form, after first identifying the approximate resonance position from the ''Born--Huang'' bound states obtained by solving Eq. \ref{eq:nuclearSE} for $\alpha=1$ alone \cite{Born1954DynamicalTheory}.
The actual resonance positions are shifted away from this value by a small amount $\Delta$ due to the coupling to the continuum.

\subsection{Multichannel semiclassical approach}
\label{sec:wkb}

\begin{figure}
    \centering
    \includegraphics[width=0.45\textwidth]{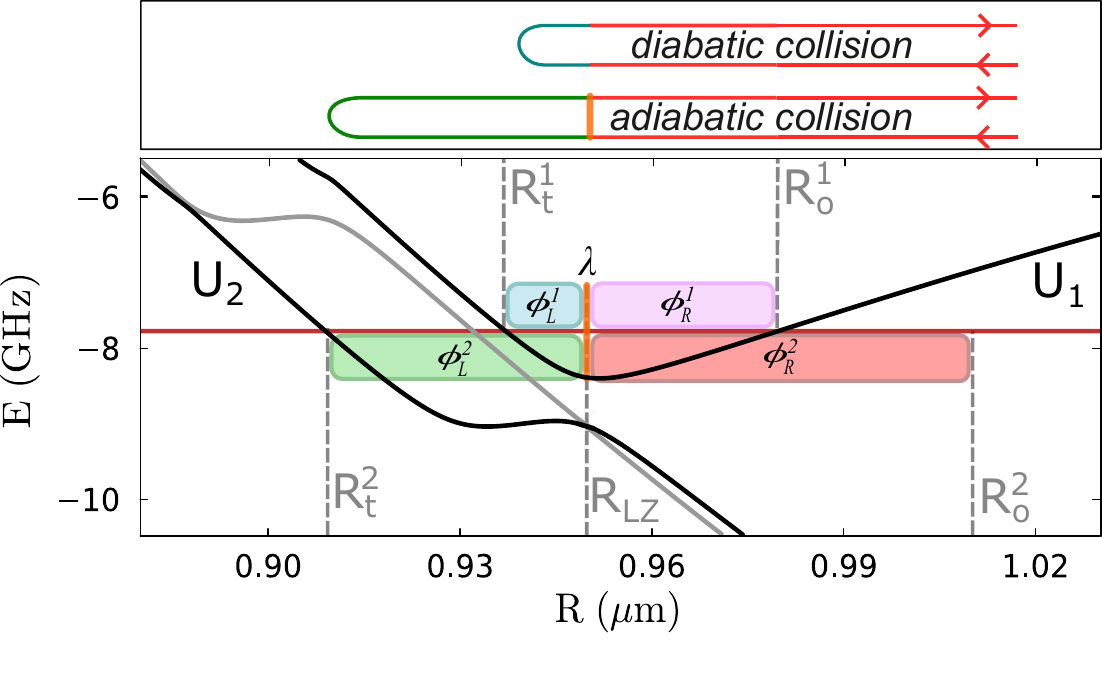}
    \caption{Effective two-channel potential energy curves used in the semiclassical LZS analysis after elimination of the dark state. The resulting model consists of one closed channel ($U_1$) supporting the quasibound molecular states and one open channel ($U_2$) describing the dissociation continuum. Superscripts denote the channel index and subscripts $L$, $R$ denote phases accumulated to the left and right of the avoided crossing, respectively. The horizontal red line indicates the energy $E$ at which the phases are evaluated: phase labels are placed above and below the line only to distinguish those associated with the upper and lower channels. The left-side phases are integrated from the respective classical turning point $R_t^\alpha$ to the avoided crossing, and right-side phases from the avoided crossing to the outer turning point or matching radius. The two dominant Stückelberg pathways 
are illustrated above the potentials, with the phases accumulated along 
each pathway color-coded according to the same convention. The phase difference between these pathways, divided by two to account for predissociation as a half-collision, determines the interference responsible for the resonance widths.}

    \label{fig:wkbpecs}
\end{figure}

To complement this quantum treatment, we utilize a semiclassical Landau-Zener-Stückelberg (LZS) \cite{Landau1932,Zener1932,Stueckelberg1932} description of the radial solutions to Eq.~\ref{eq:nuclearSE}. 
This permits a semi-analytical study of the dependence of the predissociation linewidths on principal quantum number $n$ and vibrational excitation $\nu$. 
 M. S. Child \cite{Child1971CurveCrossing,Child1974Semiclassical,Child1974MolecularCollisionTheory} and H. Nakamura \cite{ZhuNakamura1997, ZhuNakamura21997, ZhuNakamura1992_Stokes,ZhuNakamura1992_LZ, Nakamura2002Nonadiabatic} developed the LZS formalism into a general multichannel method able to treat the mixing of vibrational wave functions at each avoided crossing. 
The primary advantage of the semiclassical description is that key observables can be expressed through the phases accumulated as the system evolves along the different PECs \cite{Giannakeas2020EfimovLineshapes}. 
These depend on the semiclassical momentum in the $\alpha$th channel,
\begin{equation}
    k_\alpha(R)=\sqrt{2\mu[E - U_\alpha(R)]},
\end{equation}
where we include the Langer correction $(8\mu R^2)^{-1}$ in the effective potential $U_\alpha(R)$. 
The fundamental phases $\phi_{L,R}^{1,2}$ (see Fig. \ref{fig:wkbpecs}) corresponding to the segments to the left $(L)$ and right $(R)$ of the avoided crossing are defined as
\begin{align}
\label{eq:phasesL}
    \phi_L^\alpha &= \int_{R_t^\alpha}^{R_{LZ}}k_\alpha(R')dR'\\
    \label{eq:phasesR}
     \phi_R^\alpha &= \int_{R_{LZ}}^{R_o^\alpha}k_\alpha(R')dR'-k_\alpha(R_o^\alpha)R_o^\alpha. 
\end{align}
The inner classical turning point in each channel is obtained by solving $E = U_{\alpha}(R_t^\alpha)$, the outer classical turning point in the upper channel is obtained by solving $E = U_{1}(R_o^1)$, the outer distance at which the lower channel solution is matched to the asymptotic solution is $R_o^2$, and the location of the avoided crossing $R_{LZ}$ is determined by the peak position of the derivative coupling element $P_{12}$. 
The phase at the boundary $k_o^\alpha\equiv k_\alpha(R_o^\alpha)$ is subtracted in Eq.~\ref{eq:phasesR} to facilitate the matching to asymptotic solutions, as discussed in Appendix ~\ref{sec:wkbderivation}.

The other important phase, the Stokes phase, comes from the non-adiabatic transition between potential curves and is given by \cite{Child1991Semiclassical}
\begin{equation}
\lambda(\delta)
=
\arg\Gamma(i\delta)
-
\delta\ln\delta
+
\delta
+
\frac{\pi}{4},
\label{eq:stokes}
\end{equation}
where $\delta$ is the Landau Zener parameter. 
This can be expressed as \cite{Clark1979}
\begin{equation}
    \delta=\frac{\Delta}{8 P_{12}(R_{LZ}) v},
    \label{eq:lzparameter}
\end{equation}
where 
\begin{equation}
    \Delta = U_1(R_{LZ}) - U_{2}(R_{LZ})
\end{equation}
and
\begin{equation}
\label{eq:velocity}
v = \sqrt{\left[ 2E - \left( U_1(R_{LZ}) + U_2(R_{LZ}) \right) \right]/\mu}
\end{equation}
are, respectively, the energy gap and the semiclassical velocity at the avoided crossing. 
The Landau Zener parameter also defines the Landau Zener probability for a transition between the two adiabatic PECs at the avoided crossing, 
\begin{equation}
    P_{LZ} =  \exp(-2\pi\delta).
    \label{eq:lzprob}
\end{equation}

The total phase accumulated along each of four different physically relevant pathways are given as combinations of these five fundamental phases. 
The background phase 
\begin{align}
    X &= \frac{\pi}{4} - \lambda + \phi_{R}^2 + \phi_L^2
    \label{eq:xphase}
\end{align}
 accounts for total phase accumulated in the lower adiabatic potential $U_2$ including the $\pi/4$ phase shift associated with reflection off of the soft wall at $R_t^2$.

 Similarly, 
 \begin{equation}
    Z = \lambda + \phi_L^1 + \phi_R^1,
    \label{eq:zphase}
\end{equation}
is the ``adiabatic phase'' accumulated across one pass of the upper \textit{adiabatic} potential curve $U_1(R)$. Following the Bohr-Sommerfeld quantization rule, a vibrational bound state occurs when $Z = (\nu+1/2)\pi$ for any integer $\nu$ \cite{Child1991Semiclassical,LandauLifshitzQM}. 
The presence of the Stokes phase mirrors the role of the diagonal $Q$-matrix correction in the quantum treatment, i.e. it gives a correction to the binding energy approximately equivalent to that of the Born-Huang approximation. 

The two-level molecular system can also be described in a diabatic representation consisting of two diabatic PECs $\tilde{U}_1(R)$ and $\tilde{U}_2(R)$ which cross at $R_{LZ}$ and are coupled by an off-diagonal potential $\tilde{U}_{12}(R)$. 
The ``diabatic phase'' accumulated across one pass of the upper \textit{diabatic} potential curve $\tilde{U}_1(R)$ is
\begin{align}
    Y &= \phi_L^2 + \phi_R^1. 
    \label{eq:yphase}
\end{align}
A bound state exists in $\tilde{U}_1(R)$ at every energy satisfying $Y = (\nu+1/2)\pi$. 
The diabatic phase analogous to $X$ is 
\begin{equation}
    W = \frac{\pi}{4} + \phi_{R}^2 + \phi_L^1.
    \label{eq:wphase}
\end{equation}
accumulated across one pass of the diabatic potential $\tilde{U}_2(R)$. 
Finally, 
\begin{equation}
    \theta = X - W = -\lambda + \phi_L^2 - \phi_L^1.
    \label{eq:thetaphase}
\end{equation}
is half of the phase difference between the two collision pathways illustrated in the upper panel of Fig.~\ref{fig:wkbpecs}.

These phases, alongside the Landau Zener probability, fully determine the semiclassical $S$-matrix at a given energy $E$, 
\begin{align}
S &= e^{2 i X}\,
\frac{1 + C\,e^{-i\theta}}{1 + C\,e^{i\theta}},
\label{eq:S_analytic}
\end{align}
where \begin{align}
        C &= \frac{P_{LZ}}{1-P_{LZ}} \cos Y\,\sec Z. 
        \label{eq:Cdef}
\end{align}
A full derivation of Eq.~\ref{eq:S_analytic} is given in Appendix \ref{sec:wkbderivation}.  
In the vicinity of a resonance, the phases can be expanded to first-order about the resonance energy.
With this linearization, the Wigner-Smith time delay is
\begin{equation}
{\tau}_{PD}(E)=
2X'-2\,\frac{C'\sin\theta+C\,\theta'\cos\theta}{1 + 2C\cos\theta + C^{2}},
\label{TD_analy_wkb}
\end{equation}
where primes denote derivatives with respect to energy evaluated at \(E = E_{\mathrm{res}}\).
\subsection{LZS Resonance analysis}
\label{sec:resonance}
To obtain analytical expressions for the resonance width and position in the semiclassical framework, we exploit the adiabatic nature of our system: for Rb$^*$Li$^+$, $P_{LZ}$ is on the order of $10^{-2}$, and for Rb$^*$Rb$^+$ it is on the order of $10^{-6}$. 
The resonances therefore occur near the corresponding Born--Huang energy positions $E_{BH}$; 
expanding $Z(E)$ about this energy to first order yields 
\begin{equation}
Z(E)
\approx
\left(\nu+1/2\right)\pi
+
(E-E_{{BH}})Z'.
\end{equation}
From this, again to first order in $\varepsilon\equiv E - E_{BH}$, we obtain
\begin{align}
    \cos Z &=\nonumber
-\sin\!\left[(\nu+1/2)\pi\right]\sin\!\left(Z'\varepsilon\right)\\
&\approx (-1)^{\nu+1}Z'\varepsilon.
\end{align}
In this limit, the rapid energy dependence of $C$ is isolated to the simple pole $C\approx A/\varepsilon$ where 
\begin{equation}
    A=(-1)^{\nu+1}\frac{P_{LZ}}{(1-P_{LZ})\,Z'}\cos Y.
\end{equation}
Since $A$ contains only terms that slowly vary with energy, we have $C'\approx-A/\varepsilon^2$.
Substitution into Eq.~\eqref{TD_analy_wkb} yields
\begin{equation}
\tau_{PD}(E)\approx 2X'
-2\frac{A^2+A\varepsilon\,\theta'\cos\theta-A\sin\theta}
{A^2+\varepsilon^2+2A\varepsilon\cos\theta}.
\end{equation}
 The slowly varying term $2X'$ accounts for the background phase accumulated via evolution in $U_2(R)$. 
The denominator of the second, resonant, term of $\tau_{PD}(E)$ may be written $(A\cos\theta + \varepsilon)^2+A^2\sin^2\theta$.  
In the numerator, the term $A^2$ is second-order in $P_{LZ}$ while the second term is a product of small numbers $P_{LZ}$ and $\varepsilon$; both may be neglected to yield a Breit-Wigner profile \cite{SadeghpourGreeneCavagnero1992}
\begin{equation}
    \tau_{PD}(E)\approx 2X'+\frac{\Gamma}{(E-E_{\nu})^2 + (\Gamma/2)^2}.
    \label{eq:wstime}
\end{equation}
Here, $E_\nu = E_{BH}+\Delta$ is the resonance and with
\begin{subequations}
\begin{align}
\Gamma &= 2A\sin\theta\label{eq:width}\\
\Delta &= -A\cos\theta\label{eq:del}.
\end{align}
\end{subequations}

Exactly on resonance, the predissociation decay rate $\Gamma$ takes on an especially simple form. 
Here, the quantization condition $Z=(2\nu+1)\pi/2$ implies $\phi_R^1=(\nu+1/2)\pi - \lambda - \phi_L^1$, and hence 
\begin{equation}
    Y(E\sim E_\nu) = \theta-(\nu+1/2)\pi .
    \label{eq:ysimple}
\end{equation}
This yields a transparent expression for the resonance width,
\begin{equation}
    \Gamma = 2\left(\frac{P_{LZ}}{1-P_{LZ}}\right)
    \frac{\sin^{2}\theta}{Z'},
    \label{wid_path_diff}
\end{equation}
which admits a direct interpretation in terms of Stückelberg inteference caused by the two 
dominant pathways shown in Fig.~\ref{fig:wkbpecs} \cite{Giannakeas2018EfimovStueckelberg}. 
Consider the time-reverse of predissociation, an inelastic collision starting with the particles evolving along  $U_2(R)$ at large distances. 
The two shortest classical collision pathways are one in which the collision at energy $E$ takes place entirely along the lower adiabatic potential, accumulating a phase $2X-\pi/4$, or takes a diabatic path through the avoided crossing 
and accumulates a phase $2W-\pi/4$. 
Since predissociation consists of just a half collision, the relevant phase difference is $\theta = X - W$, and hence these two paths can destructively interfere when $\theta$ is an integer multiple of $\pi$. 
This interference decouples the bound state from the continuum, completely suppressing the decay out of the adiabatic channel $U_1$. 

\subsection{Fano line shapes}
\label{sec:q_para}
The molecular resonances in the photo-predissociation cross section will exhibit the characteristic
asymmetric Fano line shape \cite{Fano1961,Feshbach1962}.
The total cross section, assuming a single photon transition from the initial state $\Psi_0(\vec r,\vec R)$ \footnote{This could be the electronic ground-state for a direct transition to an electronic $p$-state, or a transition from an intermediate state to the weak $s$- and $d$-state admixture in the molecular configuration as was done in Ref.~\cite{Zuber2022}. The initial state of the nuclei is a low-energy scattering wave function.} is given by \cite{Aymar1996}
\begin{align}
    \sigma &= \frac{4\pi^2\alpha\omega}{3}\sum_{\alpha=2} |d_\alpha|^2,
\end{align}
where $\alpha$ appearing in the prefactor is the fine structure constant and should not be confused with the channel index.  $\omega$ is the photon frequency.
Here, $d_\alpha$ are the dipole matrix elements corresponding to dissociation into the $\alpha$th channel 
\begin{equation}
    d_\alpha = \sum_j(I + iJ)_{\alpha j}^{-1}d_j
\end{equation}
with 
\begin{equation}
    d_j = \langle \Psi_j(\vec r,\vec R)||r^{(1)}|| \Psi_0(\vec r,\vec R)\rangle.
\end{equation}
More concretely,
\begin{align}
    d_j&= \int_0^\infty \sum_{\beta}\chi_{j\beta}(R)\chi_0(R)\langle \phi_\beta(\vec r;R)||r^{(1)}||\phi_{0}(\vec r)\rangle dR,
\end{align}
where $\chi_0(R)$ is the initial atomic pair state of the distant atom and ion and $\phi_0(\vec r)$ the initial electronic state. 
Within the LZS treatment, we have a single open channel and can obtain an analytical form for the cross section using the $I$ and $J$ coefficients obtained in Appendix \ref{sec:wkbderivation}. 
Letting $d$ be the single $d_j$ matrix element and for energies $E\sim E_\nu$, we have
\begin{align}
    \sigma 
    &=\frac{\mu}{(1-P_{LZ})}\frac{2\pi d^2\omega\alpha }{3}\frac{1}{|Ce^{-i\theta}+1|^2}.
\end{align}
Using $\Gamma$ and $\Delta$ from Eqs.~\ref{eq:width} and \ref{eq:del}, we write this in terms of the characteristic Fano lineshape
\begin{align}
    \sigma = \frac{2\pi \mu d^2\omega\alpha}{3(1 - P_{LZ})}\frac{(\varepsilon - \Delta+q\Gamma/2)^2}{(\varepsilon - \Delta)^2 + (\Gamma/2)^2},
\end{align}
where $q = -\cot\theta$ the Fano lineshape parameter.

\begin{figure*}[th]
    \centering
    \includegraphics[width=0.9\textwidth]{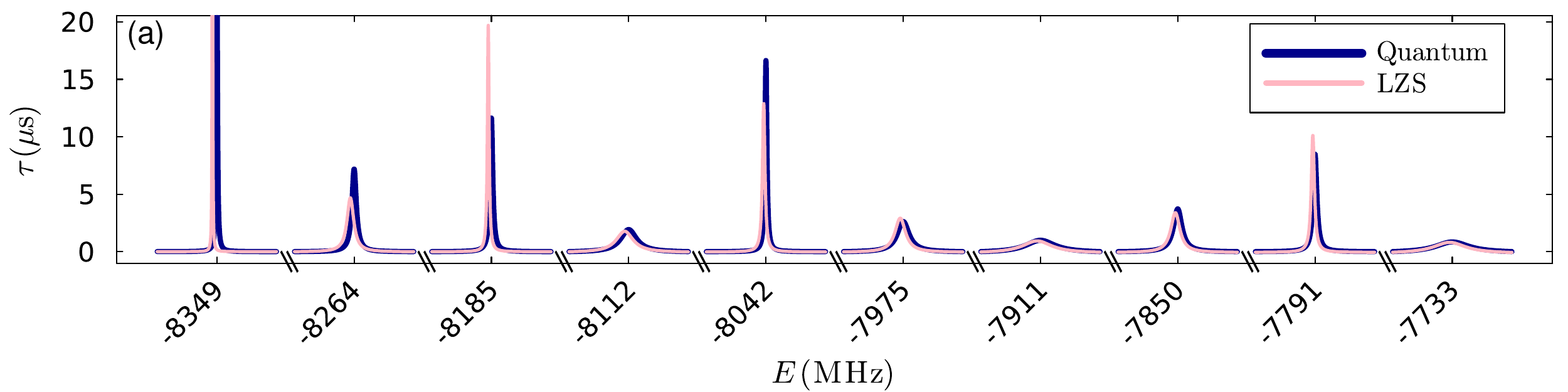}\\
    \includegraphics[width=0.9\textwidth]{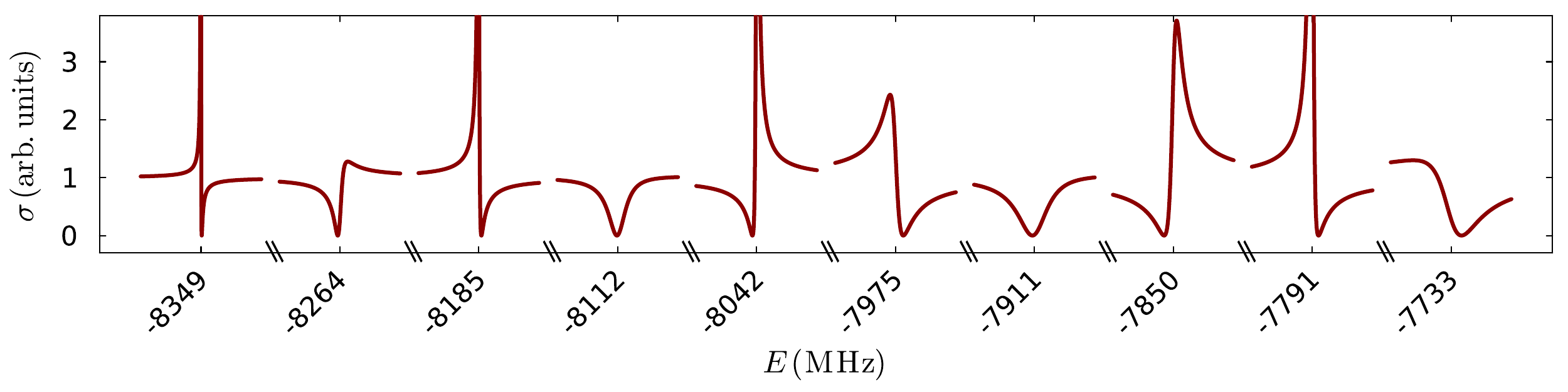}\\
    \caption{(a) Quantum (blue) and semiclassical (pink) Wigner--Smith time delays for the $40P_{1/2}$Rb$^*$Li$^+$ RAIM shown over an energy range spanning the first ten vibrational resonances. The resonance positions $E_\nu$ are labeled, and the time delay is only shown over intervale of $E_\nu\pm 1$MHz. 
    Across this range of $\nu$ the lifetimes (given by the peak value of $\tau/4$, c.f. Eq.~\ref{eq:wstime}) vary from $7\; \mu s$ for $\nu = 0$ to $0.23\; \mu s$ for $\nu=9$. 
    (b) The photo-dissociation cross section, $|(I+iJ)^{-1}|^2$, where I and J are obtained semiclassically.
    The horizontal axis is discontinuous as in (a). The Fano profiles bear clear signatures of $q$-reversal.}
    \label{fig:time_delay_lng_range_n38}
\end{figure*}

\section{Lifetime analysis for the $\mathbf{Rb^*Li^+}$ Rydberg atom-ion molecule}
\label{sec:RB-Li_lftm_analy}

We have applied both methods described in the previous section to a large-scale study of the non-adiabatic predissociation rates for the Rb$^*$Li$^+$ RAIM. The vibronic wavefunctions are represented in a B-spline basis with typically $10^4$  splines per channel, with the exact number of splines increasing slightly with $n$ to accommodate the growing spatial extent of the wavefunctions. 
The bond length scales with $n$ approximately as $n^{2.6}$; therefore we select $R_i \sim 1.3\, n^{2.6}$ and $R_0 \sim 3.0\, n^{2.6}$  as the inner and outer boundaries, respectively, of the R-matrix region. 

Fig.~\ref{fig:time_delay_lng_range_n38}(a) shows the Wigner-Smith time delay calculated in the vicinity of the first ten vibrational resonances ($\nu=0$--9) of the $n=40$ RAIM. 
 The quantum result (Eq.~\ref{time_del_qtm}) is shown in blue and the semiclassical (Eq.~\ref{TD_analy_wkb}) calculation is shown in pink. 
To facilitate comparison, the energy intervals between successive resonances have been omitted, so that the horizontal axis consists of consecutive windows of width $1$MHz to either side of the quantum resonance position (labeled for each resonance) rather than a continuous energy scale. 

The behavior shown here is characteristic of all the Rydberg levels that we studied. 
The semiclassical calculation proves very accurate, with the biggest quantitative discrepancy being that it typically underestimates the widths of very narrow resonances (c.f. $\nu=0$ and $\nu=2$).
Additionally, the predissociation lifetimes $\tau = \tau_{PD}(E_\nu)/4$ (see Eq.~\ref{eq:wstime}) vary dramatically across this energy range. 
At their shortest ($\nu=6$ and $9$), the lifetimes are less than a microsecond, substantially shorter than both the radiative lifetime of the Rydberg state and typical collisional time scales \cite{Zuber2022}. 
For other vibronic states, $\nu=0$ and $2$ for example, the lifetimes exceed several tens of microseconds.
On these time scales, non-adiabatic predissociation becomes almost irrelevant and the RAIM decays primarily through collisional loss. 
It is remarkable how the lifetimes even of adjacent molecular levels differ by an order of magnitude or more. 
This is a direct manifestation of the Stückelberg interference mechanism discussed in Sec.~\ref{sec:resonance}, in particular the fact that the resonance width (Eq.~\ref{eq:width}) will vanish if a resonance happens to occur when  the overall path difference $\theta$ is an integer multiple of $\pi$. 
This destructive interference effectively decouples the molecular bound state from the continuum states in the repulsive channels. 
The vibronic states in Fig.~\ref{fig:time_delay_lng_range_n38}(a) for which this phase condition is nearly satisfied therefore exhibit the longest predissociation lifetimes.

\begin{figure*}[t]
    \centering
    \includegraphics[width=0.95\textwidth]{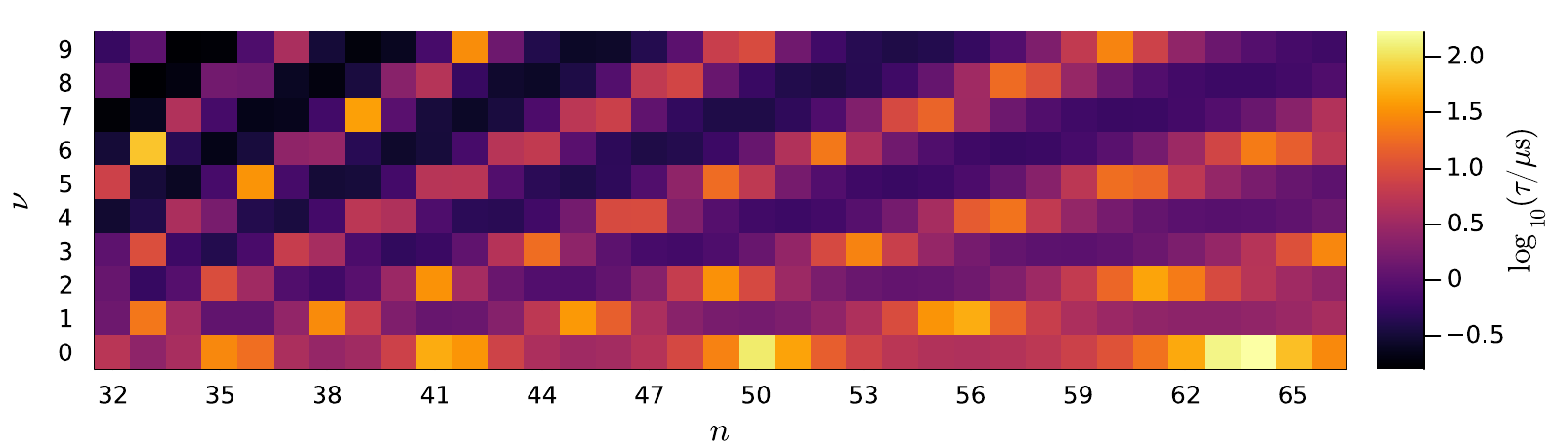}
    \caption{Predissociation lifetimes for the $nP_{1/2}$ Rb$^*$Li$^+$ molecule as a function of $\nu$ and $n$. The lifetimes span nearly 
three orders of magnitude, from $0.15\ \mu$s at $(n,\nu) = (34, 9)$ to 
$166\ \mu$s at $(n,\nu) = (64, 0)$. The diagonal streaks of enhanced 
lifetime are contours of constant path difference phase $\theta$, along which 
the destructive interference condition $\theta = m\pi$ is approximately 
satisfied, suppressing the resonance width.}
    \label{fig:density_plot_Rb_Li}
\end{figure*}

The Fano $q$-parameter also varies from one resonance to another due to Stückelberg interference.
Figure~\ref{fig:time_delay_lng_range_n38}(b) shows the semiclassical calculation of the photo-predissociation cross section $\propto|(I+iJ)^{-1}|^2$ in order to illustrate this dependence. 
As derived in Sec.~\ref{sec:q_para}, the asymmetry parameter $q=-\cot\theta$ diverges at a point of pure destructive interference. 
Consequently, states with strongly suppressed resonance widths correspond to highly asymmetric resonances, whereas the broadest levels occur for $q\approx0$ as window resonances (see $\nu=3,6$). 
The remaining vibronic resonances exhibit strongly asymmetric Fano profiles with both positive and negative values of $q$ \cite{Fano1961}.
Reversal of the Fano $q$ parameter is a familiar phenomenon from ``complex resonances'' in atomic autoionization in a three channel system, caused by an ``interloper'' resonance in one closed channel interfering with autoionizing resonances in the second closed channel \cite{Connerade1988InteractingResonances,Connerade1987QReversals}. 
Here, however, it occurs in the presence of only two channels due to the interference between the two decay pathways. 
This behavior was previously observed in the photodissociation of alkali-halides occurring via non-adiabatic transitions between ionic and covalent potential energy curves \cite{Sadeghpour2000ChannelInteraction,Balakrishnan1999QuantumWavePacket,Cornett1999Interferometric}. 
In that case, the dynamics were highly diabatic and predissociation occurs on a femtosecond timescale.

\begin{figure}[b]
    \includegraphics[width=0.47\textwidth]{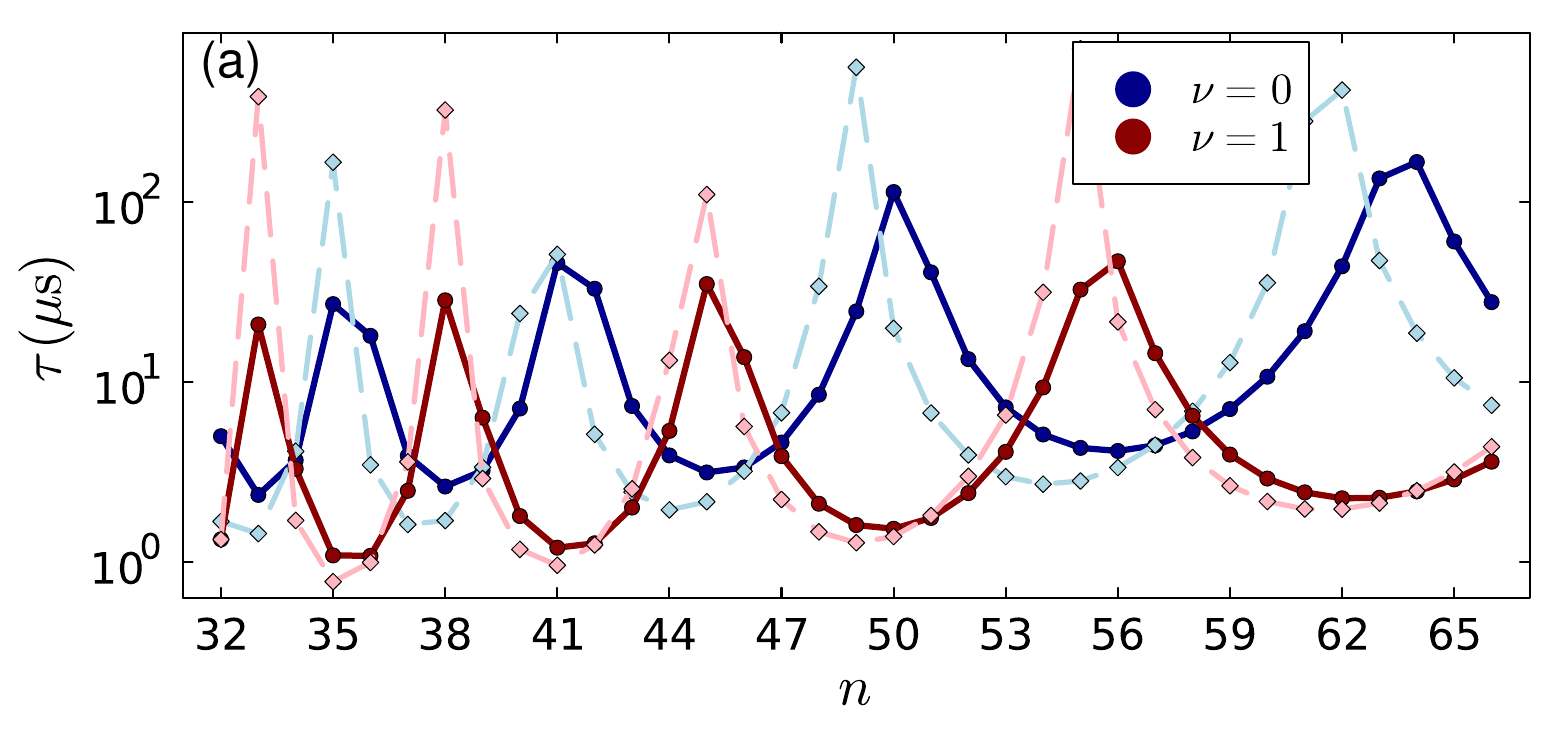}
    \includegraphics[width=0.47\textwidth]{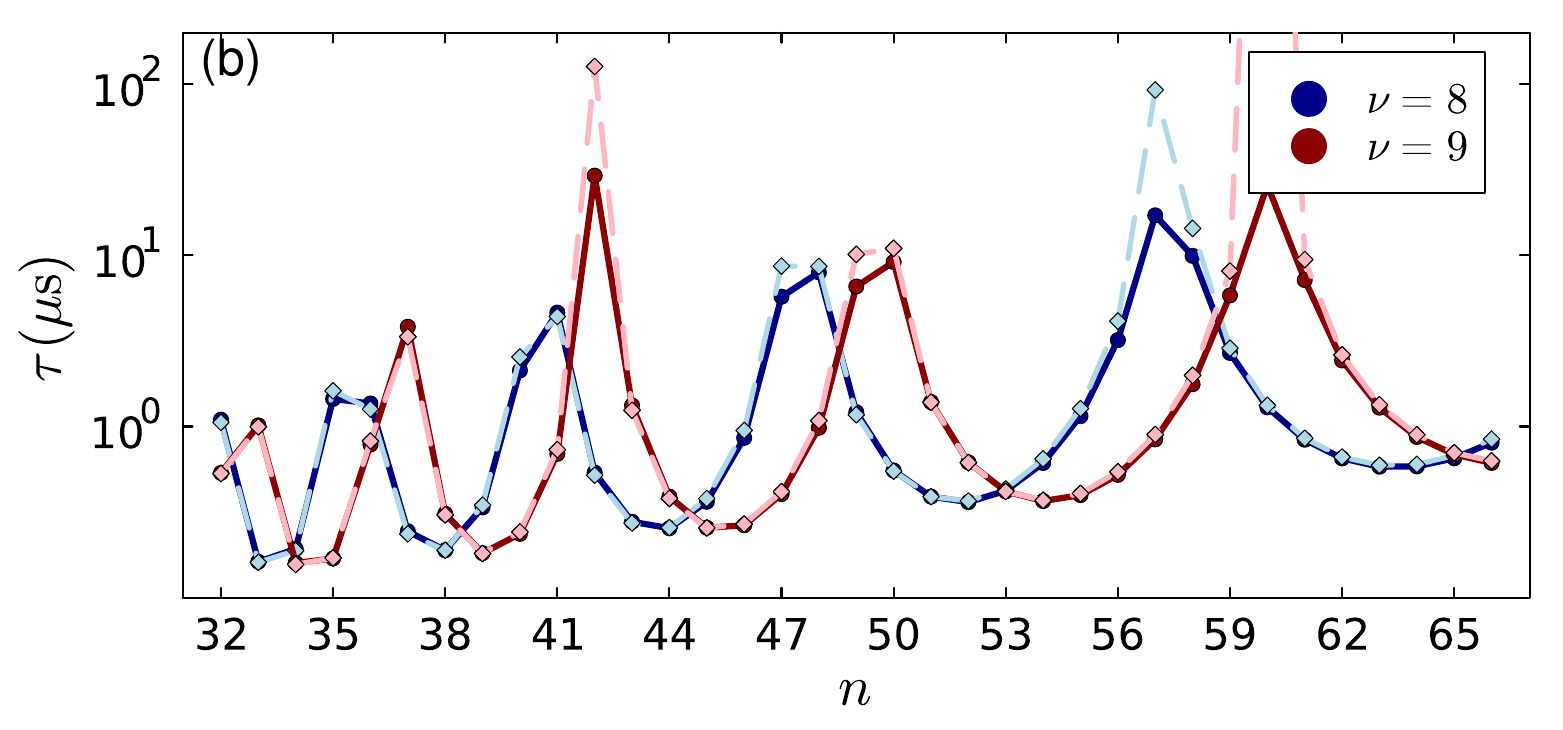}
    \caption{Predissociation lifetimes of the Rb$^*$Li$^+$ molecule as a function of principal quantum number $n$. Solid lines with circular markers (dashed lines with diamond markers) denote the quantum (semiclassical) calculations. Panel (a) shows the vibronic states $\nu=0$ (blue) and $\nu=1$ (red), while panel (b) shows $\nu=8$ (blue) and $\nu=9$ (red). The lifetimes exhibit a weak overall increase with $n$.}
    \label{fig:timedelaysvsn}
\end{figure}

\begin{figure}[t]
    \includegraphics[width=0.47\textwidth]{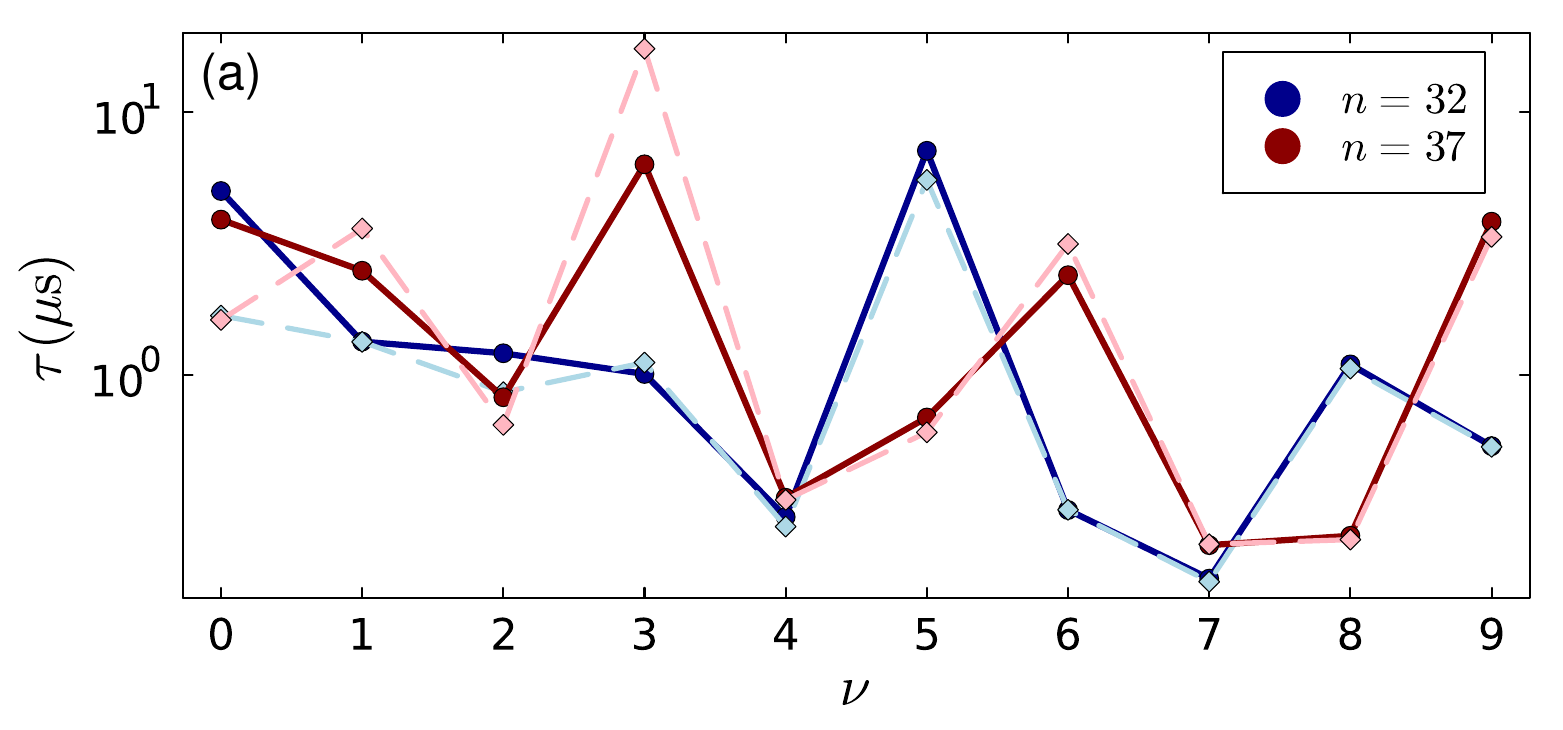}
    \includegraphics[width=0.47\textwidth]{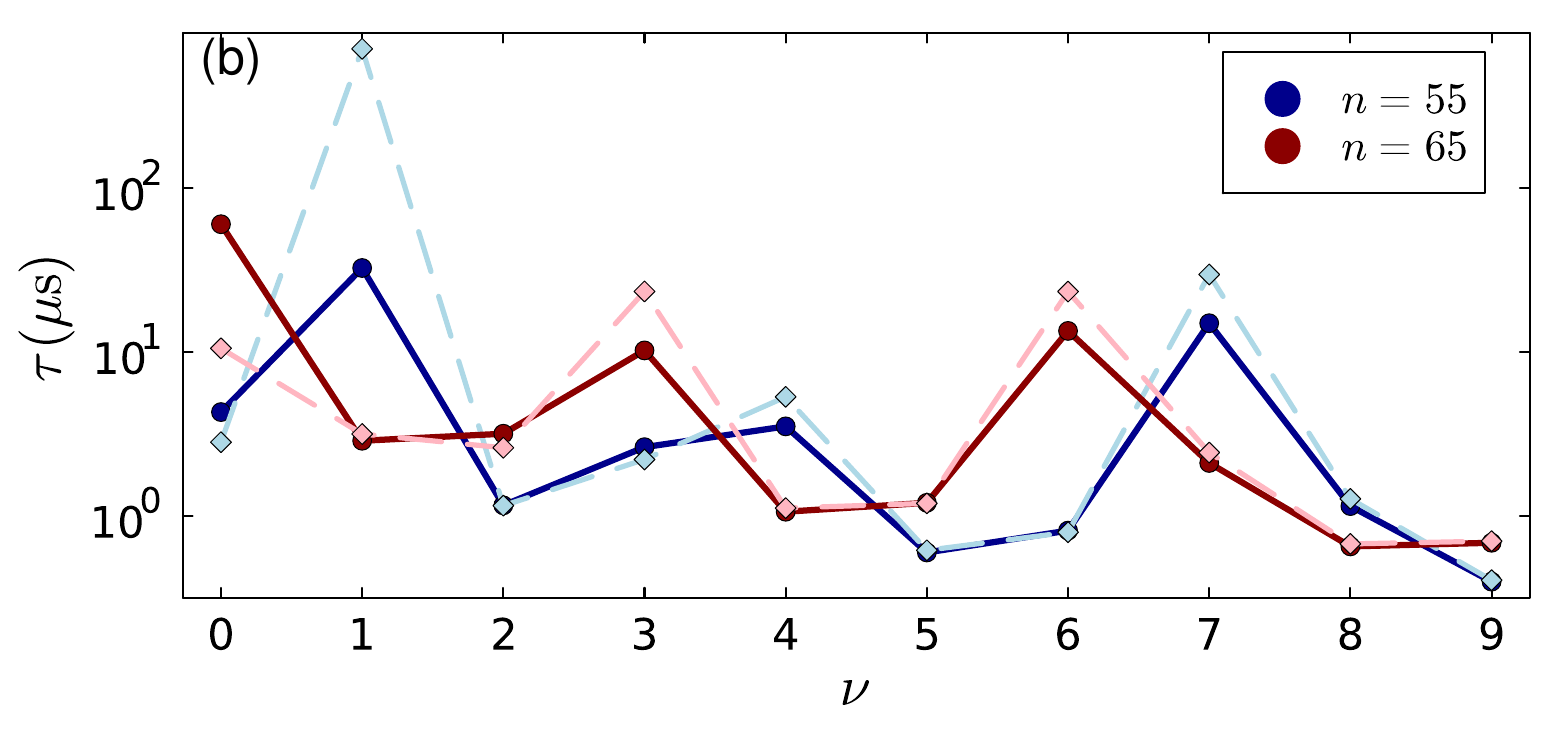}
    \caption{Predissociation lifetimes of the Rb$^*$Li$^+$ molecule as a function of the vibrational quantum number $\nu$ for various principal quantum numbers $n$. Solid lines with circular markers (dashed lines with diamond markers) denote the quantum (semiclassical) calculations. Panel (a) shows the lifetimes for $n=32$ (blue) and $n=37$ (red), while panel (b) shows the corresponding results for $n=55$ (blue) and $n=65$ (red). The lifetimes exhibit a weak overall decrease with $\nu$. }
    \label{fig:timedelaysvsnu}
\end{figure}

In alkali-halide predissociation, the effect of St\"uckelberg interference on the lineshapes and decay rates is only revealed by studying their variance across the vibrational levels of a single electronic state \cite{Cornett1999Interferometric}. In contrast, RAIMs allow us to probe this same physics along the additional axis of electronic energy as determined by the principal quantum number $n$. 
We have therefore performed time-delay calculations for principal quantum numbers $n = 32$--$65$ and across an energy range encompassing the lowest ten 
vibronic levels $\nu = 0$--$9$ for each $n$. 
The predissociation lifetimes extracted from this calculation are displayed in Fig.~\ref{fig:density_plot_Rb_Li}. 
Comparisons of these quantum results with their semiclassical counterparts along single axes are presented later in Figs.~\ref{fig:timedelaysvsn} and \ref{fig:timedelaysvsnu}. 
Several remarks are in order. 

First, the lifetimes overall tend to increase with $n$, although the large modulations due to St\"ckelberg interference make this tendency difficult to quantify.
The slight overall increase is more clearly seen in Fig.~\ref{fig:timedelaysvsn}, which shows the predissociation lifetimes for two low (panel a) and high (panel b) vibronic states as a function of $n$. 
The physical origin of the overall increase in lifetime with $n$ is discussed in Sec.~\ref{sec:mass_dep_analy}. Superimposed on this overall increase, the lifetimes exhibit a pronounced periodic modulation. This can be traced back to the slow dependence of the path difference $\theta$ on $n$. 
As a result, the interference term $\sin^2\theta$ also varies smoothly with $n$, producing the semi-periodic oscillation in the resonance widths and, consequently, the predissociation lifetimes.

Second, the predissociation lifetimes display only a weak dependence on the overall amount of vibronic excitation, decreasing slightly overall as $\nu$ increases. 
This is anticipated from the Landau-Zener probability formula (Eq.~\ref{eq:lzprob}): the velocity $v$ (Eq.~\ref{eq:velocity}) increases with energy, which in turn enhances the probability of a non-adiabatic transition into the dissociative potential curve, thus shortening the lifetimes.
Fig.~\ref{fig:timedelaysvsnu} helps to reveal this overall trend, which, as seen before, is highly non-monotonic due to the frequent appearances of exceptionally long-lived states. 

Third, Fig.~\ref{fig:density_plot_Rb_Li} exhibits pronounced diagonal stripes of alternating enhanced lifetimes viewed as a function of both $n$ and $\nu$. These arise because the adiabatic potential curves evolve only gradually with increasing principal quantum number, becoming progressively broader and shallower. Consequently, the St\"uckelberg interference condition is shifted from one vibronic level to the next as $n$ increases, producing diagonal bands of enhanced lifetimes. Equivalently, these streaks correspond to contours of approximately constant path difference phase $\theta$, along which the interference condition decoupling the bound state from the continuum is approximately satisfied. 

Finally, Figs.~\ref{fig:timedelaysvsn} and \ref{fig:timedelaysvsnu} demonstrate that the semiclassical method improves in accuracy as $\nu$ increases. 
As already seen in Fig. \ref{fig:time_delay_lng_range_n38}, the discrepancy between semiclassical and quantum methods is highest for the most long-lived states.

From the results shown here, we conclude that long-range Rb$^*$Li$^+$ Rydberg-atom--ion molecules will exhibit dramatic variation in their predissociation decay rates as a function of both $\nu$ and $n$. 
While the former variation has been observed in the predissociation of other molecules, the ability of Rydberg systems to additionally probe the systematic connection between these lifetimes and Stückelberg interference as a function of $n$ presents a new direction for study. 
Furthermore, the microsecond-scale dynamics enable \textit{in situ} study of the non-adiabatic decay mechanism, especially in conjunction with ion microscopy allowing one to study these processes on the fly.

\section{Lifetime analysis for the $\mathbf{Rb-Rb^+}$ Rydberg atom-ion molecule}
\label{sec:RB-Rb_lftm_analy}

\begin{figure}[t]
    \centering
    \includegraphics[width=0.45\textwidth]{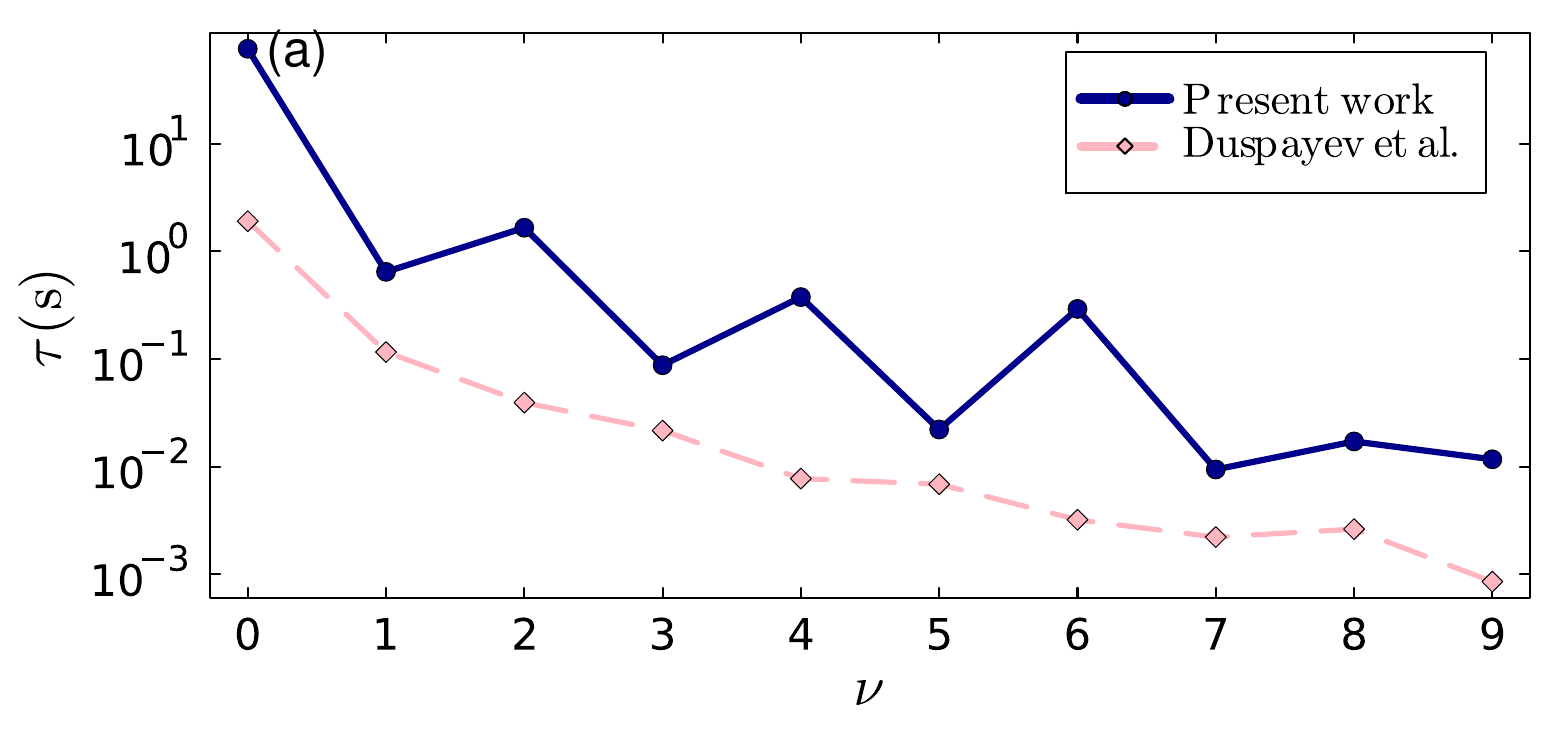}
    \includegraphics[width=0.45\textwidth]{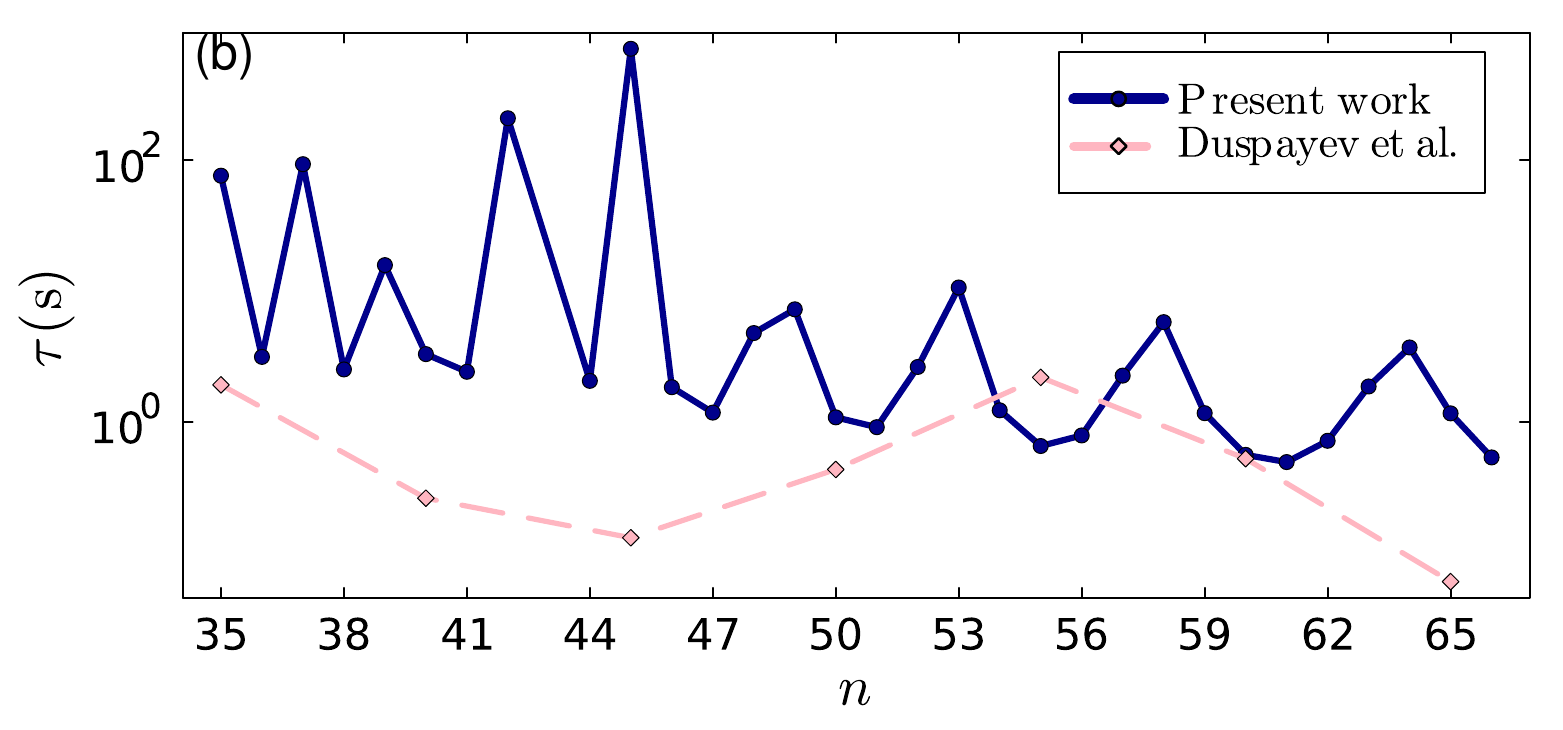}
    \includegraphics[width=0.45\textwidth]{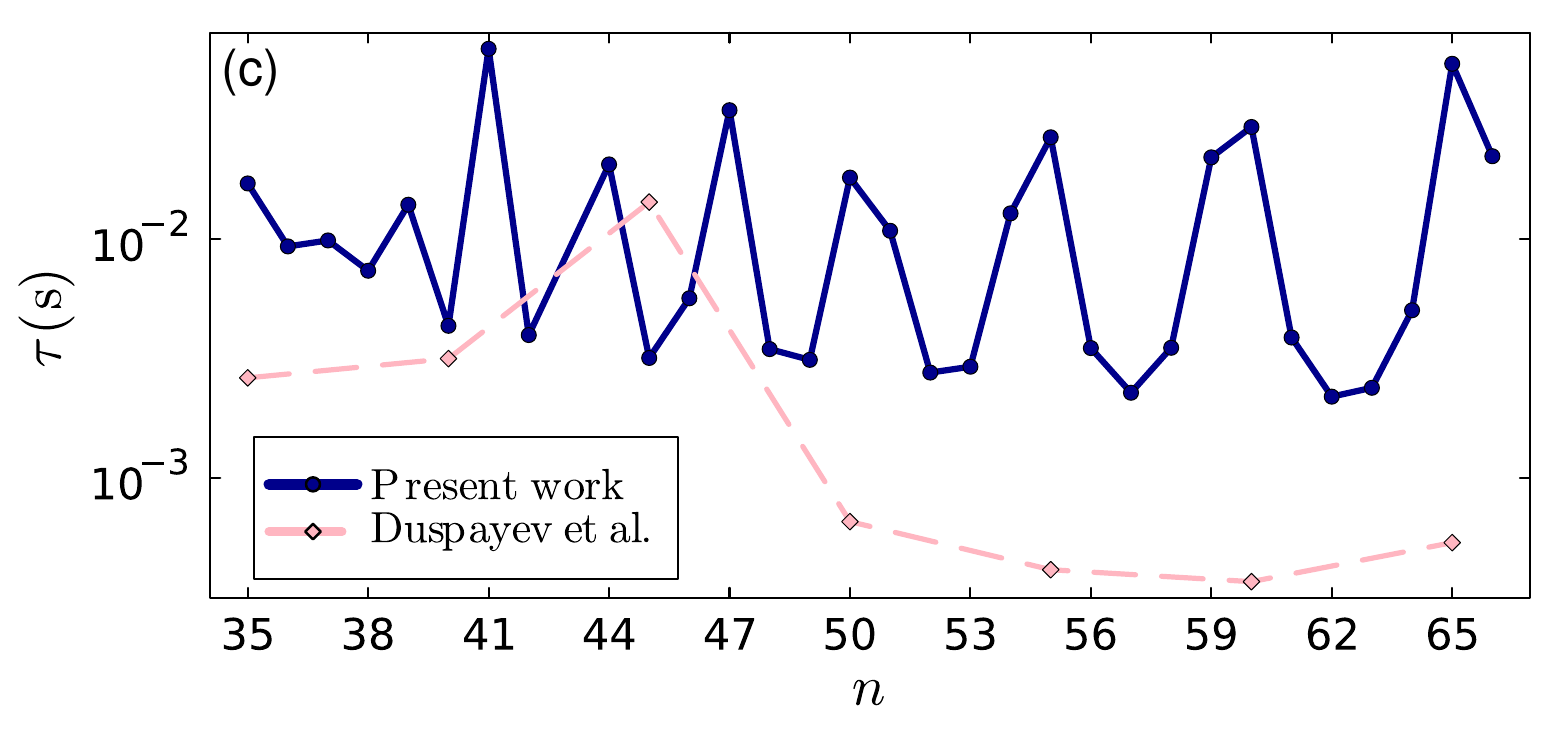}
    \caption{(a) Comparison of the predissociation lifetimes for $\nu=0-9$ obtained in the present
work with the data reported in Ref.~\cite{Duspayev2022NonadiabaticRAIM}. 
$n=35$. 
 Comparison of the vibronic state's lifetime as a function of $n$ for (b) $\nu=0$ and (c) $\nu=8$. 
In general, the lifetimes calculated in the present work are significantly larger and show a more pronounced variation due to St\"uckelberg interference. 
}
    \label{fig:rblifetimecomp}
\end{figure}

We now turn to the heavier molecular system Rb$^*$Rb$^+$ already observed experimentally \cite{Zou2023VibrationalRAIM,Zuber2022,Berngruber2024InSituCollisionDynamics} and whose predissociation lifetimes were calculated in Ref.~\cite{Duspayev2022NonadiabaticRAIM}, 
using a time-dependent study with a complex absorbing potential to remove flux in the asymptotic region. Although the resulting predissociation lifetimes were in the range of $10^{-3}-10^1$ seconds, the experimentally measured lifetimes \cite{Zou2023VibrationalRAIM} were found to be a few microseconds. 
This is much shorter than the radiative lifetime, and could be attributed to collisional processes.  
However, given the difficulties in numerically converging the predissociation lifetimes due to the ``almost dark'' states with their highly singular non-adiabatic couplings, as well as the rapid fluctuation with $\nu$ and $n$ coming from Stückelberg interference, it is worthwhile to re-investigate this molecular system with our time-independent scattering approach. 

Fig.~\ref{fig:rblifetimecomp}(a) shows a comparison between our calculations and those of Ref.~\cite{Duspayev2022NonadiabaticRAIM} for $n=35$.
The conclusions one can make from this $n$ level are generally shared by the other principal quantum numbers;  Fig.~\ref{fig:rblifetimecomp}(b) and (c) show the dependence of the lifetime of the ground vibronic state and the eighth excited state, respectively, on $n$. 
Our calculations give lifetimes that are in general one to two orders of magnitude larger than those of Ref.~\cite{Duspayev2022NonadiabaticRAIM}, and also show a more pronounced role of Stückelberg interference. 
We could not identify a clear reason for this discrepancy, but note that our investigation of the extreme sensitivity of the three coupled-channel quantum solution's to the non-adiabatic coupling $P_{23}$ suggests that a possible culprit could be a minor inaccuracy, on the order of 1\% or less, in the peak value of this coupling. 
In test calculations where we artificially decreased the $P_{23}$ coupling by this amount, we obtained molecular lifetimes that shrank as low as a few tens of microseconds.
There is also evidence suggesting that the molecular population exhibits power-law rather than exponential decay due to interference during predissociation \cite{mizuno2018power,Balakrishnan1999QuantumWavePacket}. 
As the time-dependent analysis of Ref.~\cite{Duspayev2022NonadiabaticRAIM} assumes exponential decay, this could also contribute to inaccurate extrapolation. 
We note that our calculated resonance positions are in excellent agreement with those of Refs~\cite{Duspayev2021a,Duspayev2022NonadiabaticRAIM}, consistent with the fact that the system is highly adiabatic and thus only the finite lifetimes -- which are entirely caused by non-adiabatic effects -- are sensitive to these.

\begin{figure}[b]
    \includegraphics[width=0.45\textwidth]{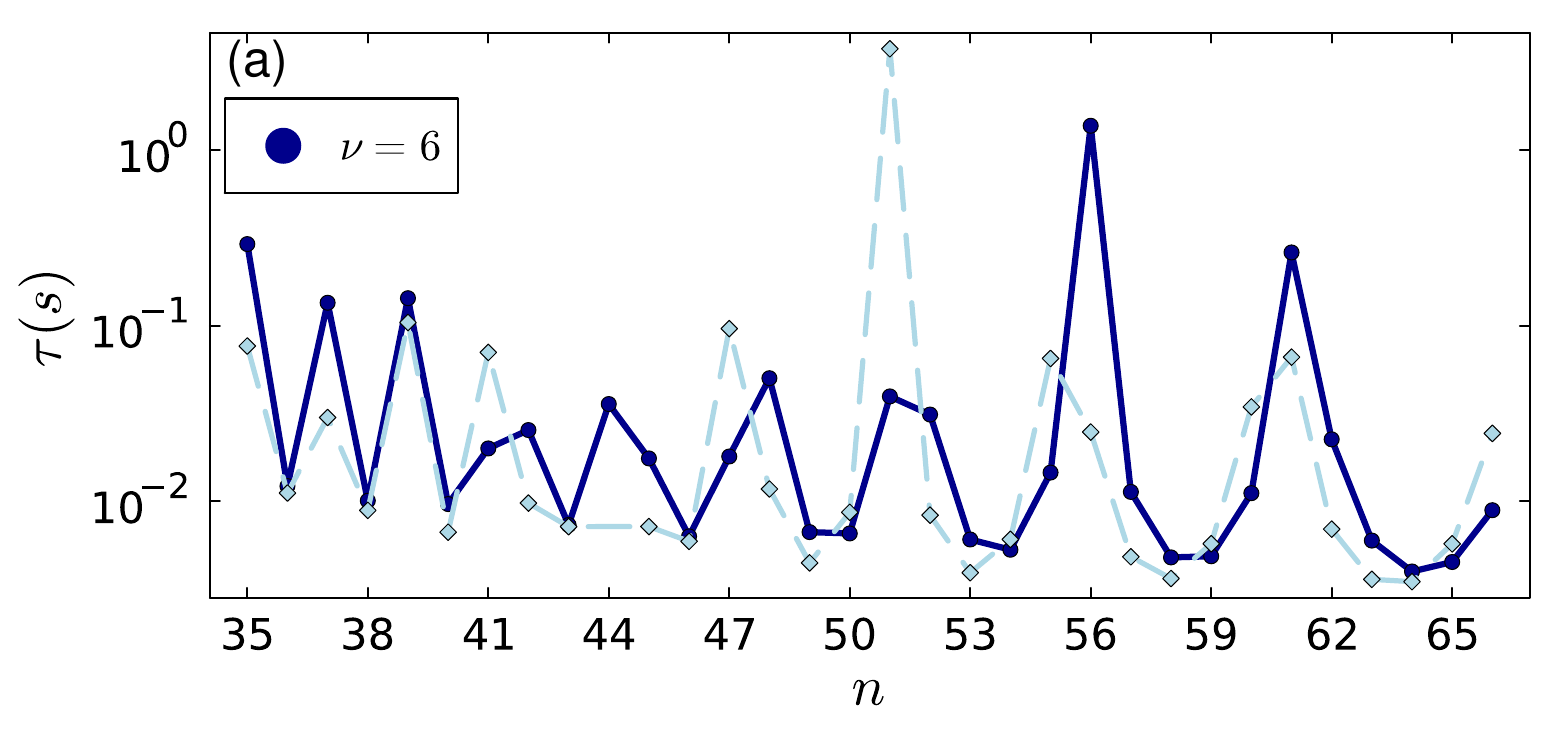}
     \includegraphics[width=0.45\textwidth]{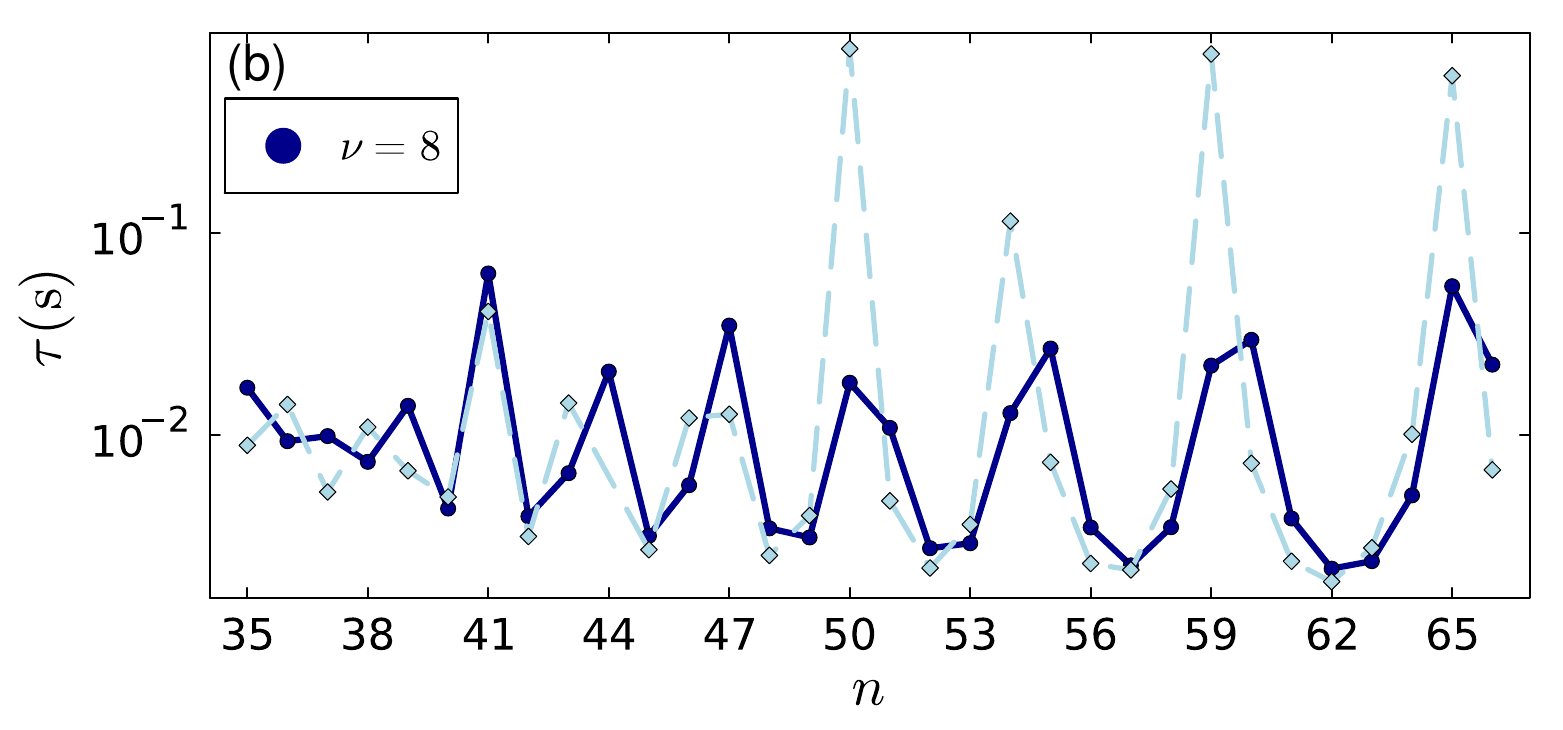}
    \caption{Comparison of quantum and LZS lifetimes for several vibronic
states $\nu$ as a function of the principal quantum number $n$ in the
$m_{J}=1/2$ manifold. While the semiclassical results capture the overall
qualitative behaviour, quantitative deviations remain.
}
    \label{fig:Rblifetimes_n}
\end{figure}

To further try to understand this discrepancy, we show in Fig.~\ref{fig:Rblifetimes_n} a comparison of the quantum and semiclassical results for the predissociation lifetimes of the $\nu=6$ and $\nu=8$ levels. 
Overall, the semiclassical calculation semi-quantitatively reproduces the quantum one, although the agreement between them is noticeably worse than that obtained for Rb$^*$Li$^+$. 
Nevertheless, this provides an order-of-magnitude check on the accuracy of our calculations.

\begin{figure}[t]
    \includegraphics[width=0.45\textwidth]{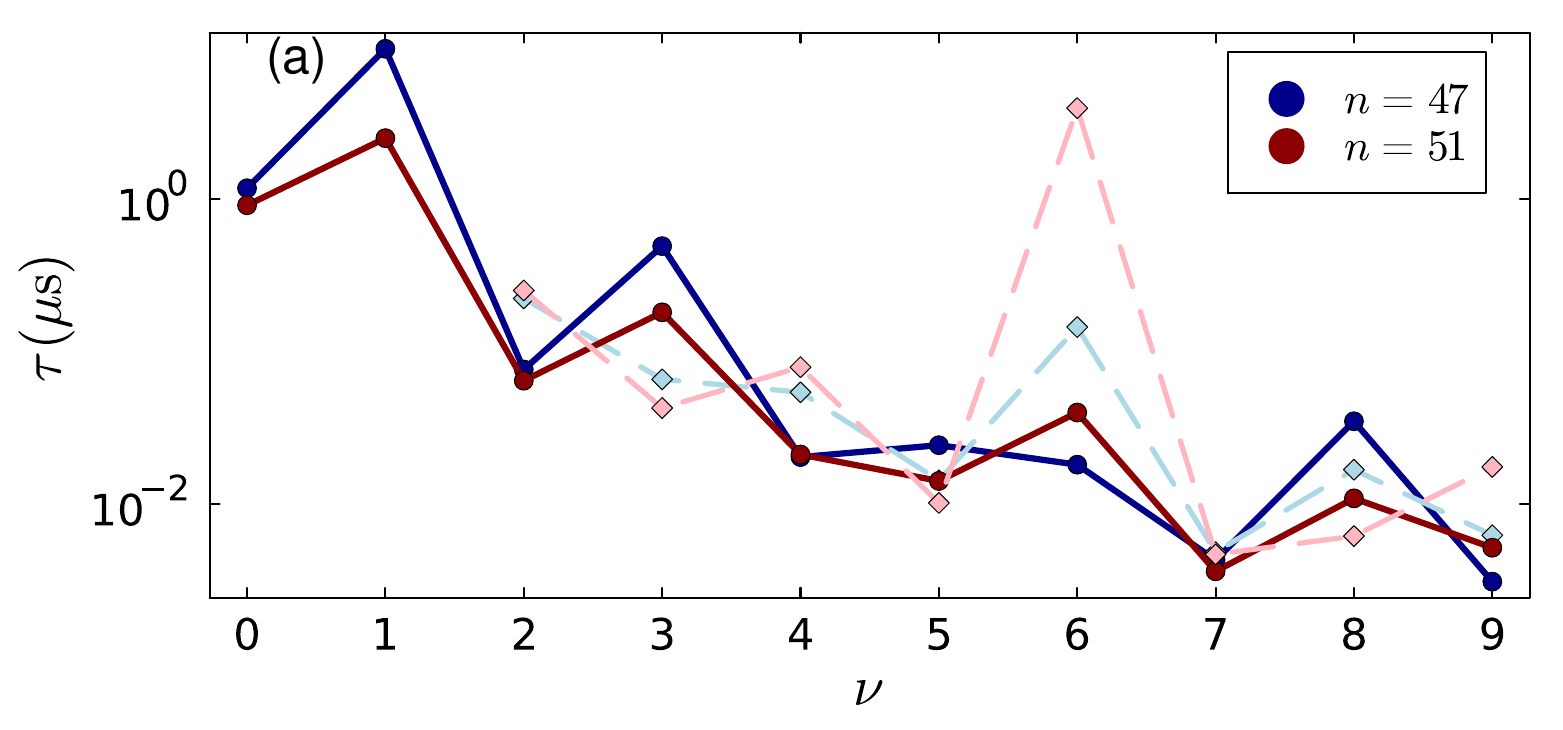}
    \includegraphics[width=0.45\textwidth]{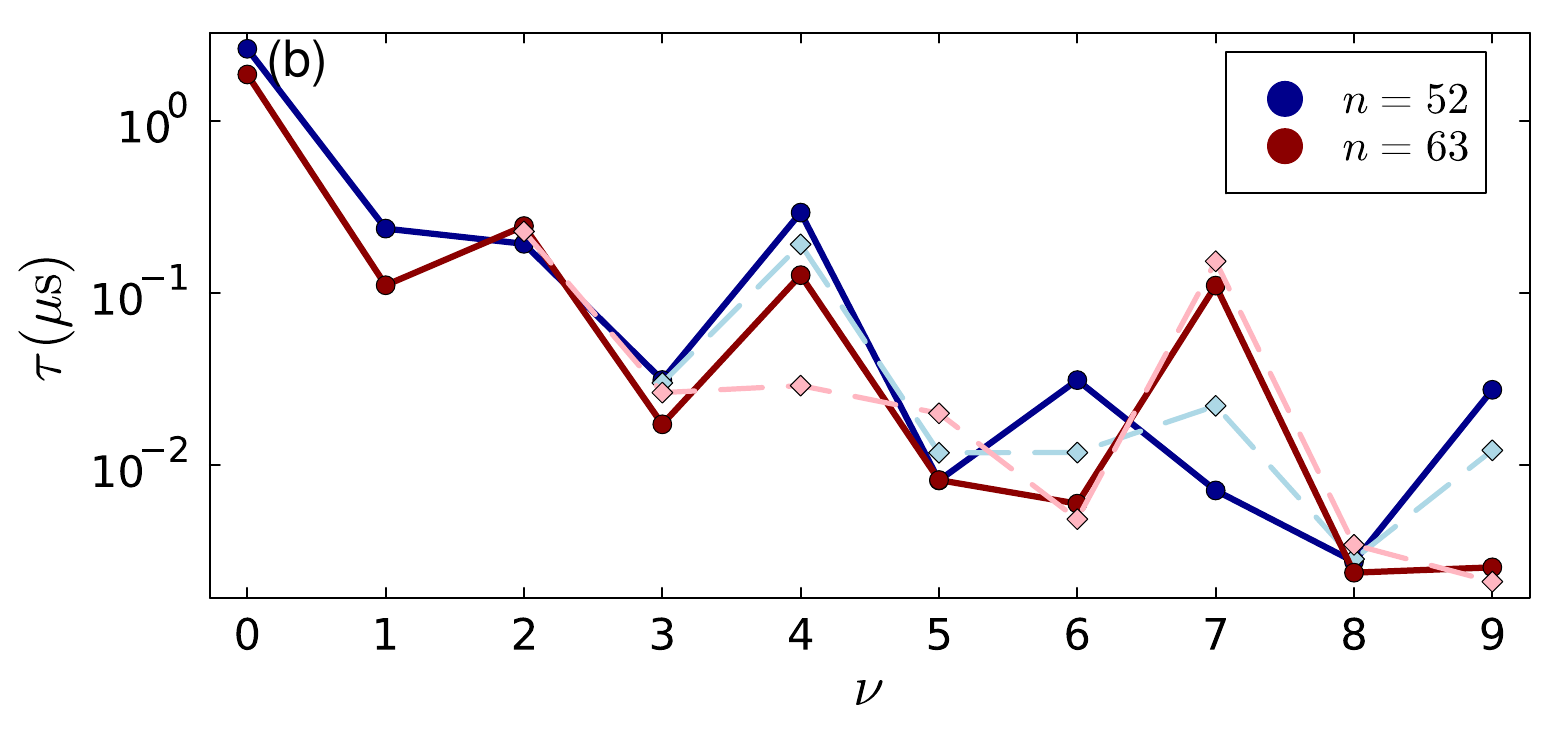}
    \caption{Molecular lifetimes as a function of vibrational quantum number $\nu$ 
for selected values of $n$, showing (a) odd--even alternation in which 
odd-$\nu$ states are longer lived than their even-$\nu$ neighbours, and 
(b) the same alternation with reversed phase, where even-$\nu$ states 
are the longer lived. In the LZS results, data are shown only from 
$\nu = 2$ onwards, since for the lowest vibronic states the inner 
classical turning point lies beyond the avoided crossing, rendering the 
Landau--Zener transition ineffective.}
    \label{fig:Rblifetimes_nu}
\end{figure}

Fig.~\ref{fig:Rblifetimes_nu} shows a similar comparison, now for several fixed $n$ values and as a function of $\nu$. 
From this and Fig.~\ref{fig:Rblifetimes_n}, we see that the rapid dependence of the predissociation lifetime on both of these quantum numbers seen for Rb$^*$Li$^+$ is also observed here.  
The periodicity is not as regular, as the path difference $\theta$ in Rb$^*$Rb$^+$ changes more rapidly with $n$. 
It is interesting to note that the lifetimes as a function of $\nu$ for $n=47$ and $n=51$ appear to depend on the parity of the vibronic state. 
Our LZS analysis however shows that this is not connected to the parity of the vibronic state at all, but is instead a manifestation of the Stückelberg interference as the phase $\theta$ accumulates at a rate of 3:2 faster than the adiabatic phase, coincidentally lining up to give the appearance of an even/odd parity dependence. 
This is confirmed by the reversal of this dependence for other $n$ levels show in Fig.~\ref{fig:Rblifetimes_nu}(b).

The discrepancy between semiclassical and quantum results seen in the homonuclear RAIM is likely caused by the strongly adiabatic
character of the relevant avoided crossings,
which leads to very small Landau--Zener transition probabilities (on the order of $10^{-6}$). 
We have tested, by artificially changing the narrowness of the avoided crossing, that the agreement improves as the Landau Zener probability increases. 
It is likely that the extreme exponential sensitivity of the Landau Zener mechanism to details of the system amplifies the errors made in the semiclassical approximation.

Despite the significant quantitative differences discussed in this section between our calculations and those presented in Ref.~\cite{Duspayev2022NonadiabaticRAIM}, our conclusion remains the same as theirs: predissociation, and by extension non-adiabatic coupling, is largely irrelevant to the stability of Rb$^*$Rb$^+$ RAIMs.
The calculated predissociation lifetimes never decrease below $10^{-3}$ seconds, making them orders of magnitude larger than radiative or collisional lifetimes
\footnote{A full dataset to compare with the results of Ref.~\cite{Duspayev2022NonadiabaticRAIM} can be found in \cite{Abraham2026Thesis}.}.

\begin{figure}
    \centering
    \includegraphics[width=0.45\textwidth]{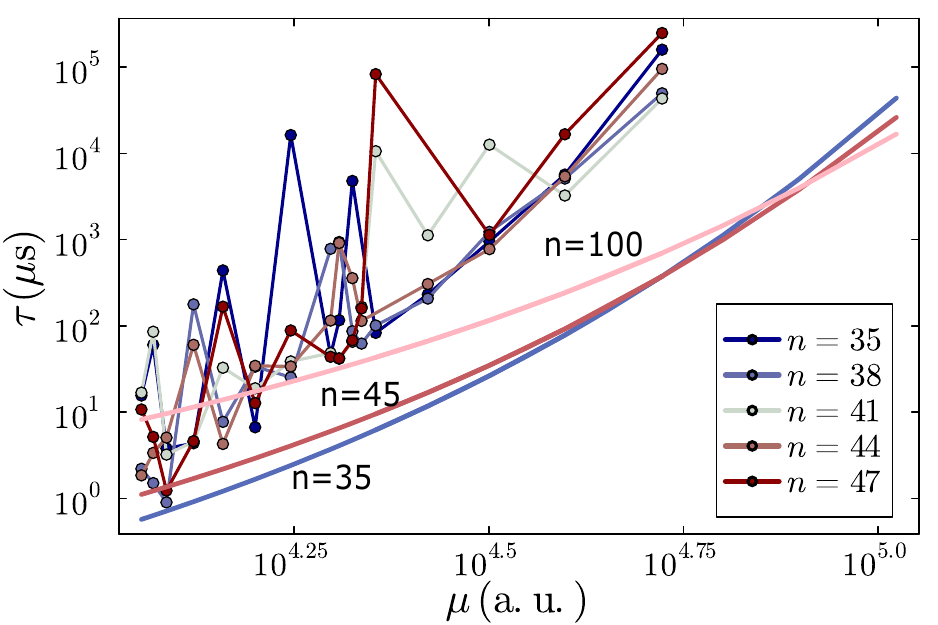}
    \caption{ Predissociation lifetime of the vibrational ground state as a function of the mass for several different principal quantum numbers. The lifetimes decrease with decreasing mass, although non-monotonically due to Stückelberg interference. The mass of Rb$^*$Li$^+$ (Rb$^*$Rb$^+$) is approximately $10^{4.1}$ ($10^{4.9}$). The solid lines show the Landau Zener estimate (Eq.~\ref{eq:LZrate}). }
\label{fig:massdependence}
\end{figure}

\section{Lifetime dependence on mass and principal quantum number}
\label{sec:mass_dep_analy}

The remarkable increase in the predissociation lifetimes with molecular mass stems from the exponential sensitivity of the Landau-Zener transition probability on the Landau Zener parameter $\delta$. 
Since the only explicit mass dependence in $\delta$ (Eq.~\ref{eq:lzparameter}) is through the semiclassical velocity, it increases as $\sqrt{\mu}$ and hence the replacement of Rb$^+$ with Li$^+$ decreases $\delta$ by almost a factor of three. 
This, in turn, increases the Landau Zener hopping probability by three to four orders of magnitude.  
To estimate the effect on the lifetime, we need to also consider the role of mass in the frequency at which the molecule has the chance to undergo a non-adiabatic transition. 
Following Refs.~\cite{Duspayev2022NonadiabaticRAIM,Deiss2021AtomIonRydberg} we estimate the predissociation rate as
\begin{equation}
     \Gamma_{LZ} \propto \Delta E  P_{LZ}
     \label{eq:LZrate}
\end{equation}
where $\Delta E$ is the vibrational frequency of the molecular level under consideration and is inversely proportional to $\sqrt{\mu}$. 
This formula gives a qualitative estimate for the lifetimes, although it overestimates the decay rate due to its neglect of destructive interference.

We have investigated the relationship between predissociation lifetimes and the molecular reduced mass in Fig.~\ref{fig:massdependence}, which shows the lifetime of the ground vibronic state as a function of $\mu$.  
To help reveal the overall trend, easily obscured by the large fluctuation in the values, the lifetimes of several $n$ levels are shown. 
These indeed indicate a rapid increase with mass, in rough qualitative agreement with the Landau Zener estimate (shown in the solid curves). 
Curiously, the overall coarse-grained -- independent of interference effects -- numerical results show a change in the $n$-dependence of the predissociation width going from the Rb$*$Li$^+$ system to the Rb$^*$Rb$^+$. 
This is shown more clearly in the Landau-Zener estimates. 
The $n=35$ and $n=45$ curves cross at {$\mu\sim 10^{4.75}$}: the lifetime increases with $n$ for $\mu<10^{4.75}$  and decreases with $n$ for $\mu>10^{4.75}$. 
This can also be seen in the $n$-dependent lifetimes of Figs. \ref{fig:timedelaysvsn} and \ref{fig:Rblifetimes_n}, where the predissociation lifetime is seen to increase with $n$ for Li$^+$ but decrease for Rb$^+$. 
This latter dependence was also seen in Ref.~\cite{Duspayev2022NonadiabaticRAIM}. 

To elucidate the complicated dependence of the lifetimes on mass and principal quantum number together, we consider again the Landau Zener estimate for the decay width (Eq. \ref{eq:LZrate}). 
This quantity depends on both $\mu$ and $n$ through the semiclassical velocity in $\delta$ and through $\Delta E$, and through $n$ alone through $\delta$'s dependence on $P_{12}(R_{LZ})$ and $\Delta$. 
We have separately fitted each of these quantities to $n$ and $\mu$ ($\Delta \sim n^{-4.1}$, $P_{12}(R_{LZ}) \sim n^{-2}$, $v\sim \mu^{-1/2}n^{-1/7}$, and $\Delta E \sim \mu^{-1/2}n^{-4}$) in order to obtain
\begin{align}
    \Gamma_{LZ} \sim A n^{-4}\mu^{-1/2}\exp\left(\frac{-B\pi n^{-b}}{4/\sqrt{\mu}}\right),
    \label{eq:lzrateapprox}
\end{align}
where $A$, $B\sim 0.25$, and $b\sim 0.4$ are positive constants.
The anticipated inverse square root dependence on the reduced mass is seen in both the velocity and energy difference contributions. 
Taking the logarithm of both sides of Eq. \ref{eq:lzrateapprox} yields
\begin{align}
    \ln \Gamma = \ln A -\frac{\ln\mu}{2} -4\ln n - \frac{B\pi}{4}n^{-b}\sqrt{\mu}.
\end{align}
From this expression, we see that while the overall decay rate is overall inversely proportional to $\sqrt\mu$, its $n$-dependence is not so straightforward as the latter two terms increase ($\ln n$) and decrease ($n^{-b}$) with $n$. 
The weight of the latter term's contribution increases with $\mu$. 
The function $\gamma(n,\mu)=-4\left(\ln n + \frac{B\pi}{16}n^{-b}\sqrt{\mu}\right)$ possesses a local maximum at 
\begin{equation}
    n^* = \left[\frac{\pi bB\sqrt\mu}{16}\right]^\frac{1}{b}.
    \label{eq:nstar}
\end{equation}
Although the asymptotic dependence of $\Gamma$ on $n$ is that it is a decreasing function of $n$ due to the logarithmic dependence of $\ln \Gamma$ on $n$, Eq.~\ref{eq:nstar} shows that the width only starts to \textit{decrease} for $n>n^*$.
The lifetimes of Rydberg states with $n<n^*$ increase until this condition is met. 
$n^*$ ranges from  $\sim 25$ for Rb$^*$Li$^+$ to $\sim 80$ for Rb$^*$Rb$^+$. 
The apparent decrease in molecular lifetimes with increasing $n$ observed in Rb is therefore a finite-$n$ effect and will reverse at $n\sim 80$.

\section{Conclusion and Outlook}
We have used a combination of quantum and semiclassical methods to solve for the positions and widths of vibronic resonances in Rb$^*$Li$^+$ and Rb$^*$Rb$^+$ long-range Rydberg atom-ion molecules. 
In both of these molecular species, the predissociation rates depend strongly and non-monotonically on the principal quantum number $n$ and vibrational level $\nu$, with lifetimes varying by orders of magnitude across the parameter space studied. 
The Landau-Zener-Stückelberg semiclassical treatment shows that these variations arise from interference between adiabatic and diabatic scattering pathways.
This interference strongly suppresses predissociation by decoupling the molecular bound state from the dissociation continuum. In Rb$^*$Li$^+$, the predicted lifetimes span from hundreds of nanoseconds to hundreds of microseconds, placing the dynamics of the predissociation process squarely within experimentally accessible timescales. Experiments should therefore observe pronounced state-to-state variations in the molecular lifetime as the principal quantum number or vibrational level is varied, providing a direct signature of Stückelberg interference. In contrast, for Rb$^*$Rb$^+$ the calculated lifetimes are on the order of seconds throughout the parameter range considered. These timescales are far longer than those associated with typical experimental limitations, indicating that nonadiabatic predissociation is not expected to be a significant decay mechanism in this system, consistent with earlier findings \cite{Duspayev2022NonadiabaticRAIM}.

The quantum and semiclassical framework developed in this work can be extended to Rydberg atom–ion molecular systems confined in hybrid atom–ion platforms. In particular, it would be interesting to investigate how the time-dependent quadrupole fields of Paul traps modify the molecular potential-energy landscape, non-adiabatic couplings, and predissociation dynamics.
The multichannel Landau-Zener-Stückelberg treatment could also be applied to atom--ion collision dynamics in the 
regime studied experimentally by the Stuttgart group 
\cite{Berngruber2024}, where non-adiabatic transitions across a series of 
parellel avoided crossings between a nonpolar Rydberg $S$-state and strongly 
polar Stark states govern the collision dynamics.
Finally, in an extreme limit of the mass scaling considered here, it would be interesting to investigate if the non-adiabatic treatment used here could be used to treat doubly excited states of atomic strontium which are in many ways analogous to the molecules studied here
\cite{Genevriez2025}, with an electron replacing the cation and a Rydberg ion instead of the neutral Rydberg atom. 
Such calculations could also be relevant in the study of impurity purification in exciton systems \cite{tiwari2026low}. 
External fields offer another avenue for control: 
while fields applied parallel to the molecular axis leave the avoided crossing 
structure largely unchanged due to the conservation of $m_j$, a more general treatment involving arbitrarily oriented fields could reveal field-induced coupling variations and offer additional control over the decay dynamics. 
Finally, investigating Rydberg–Rydberg–ion trimers, where Rydberg–Rydberg interactions compete with atom–ion polarization forces, represents an intriguing long-term direction.

\begin{acknowledgments}
We are grateful for helpful discussions with E. Naravecius, M. Berngruber, F. Meinert, A. Duspayev, M. Deiss, C. H. Greene, S. T. Rittenhouse, F. Remacle and R. \v{C}ur\'{\i}k.
\end{acknowledgments}
\newpage
\appendix
\begin{widetext}
\section{Semiclassical derivation of the scattering matrix}
\label{sec:wkbderivation}
In this appendix, we provide a pedagogical derivation of the semiclassical scattering matrix used in the main text. 
Our presentation closely follows the steps outlined in Refs.~\cite{Child1974Semiclassical,Child1991Semiclassical,Berry1966PhaseShifts}, where the final results are similarly obtained. 
Note that, since our molecule is so large that rotational splitting is negligible, we consider only the $N = 0$ non-rotating state throughout this section. 

We use the two-channel approximation described in section \ref{sec:spectator} in which the nuclear wave function evolves along the two potential energy curves $U_1(R)$ and $U_2(R)$ exhibiting an avoided crossing at $R=R_{LZ}$. 
We include in the definition of $U_\alpha(R)$ the Langer correction (for zero rotational angular momentum), i.e. $U_\alpha(R)\to  U_\alpha(R) + 
\frac{1}{8\mu R^2}$.
At energy $E$ and for $R<R_{LZ}$, there are classical turning points in both $\alpha$ channels at $R_t^\alpha$ defined by solving $E = U_\alpha(R_t^\alpha)$. At $R>R_{LZ}$, there is an additional classical turning point in the upper channel at $R_o^1$, and we define $R_o^2$ as the point where we match our solution to the asymptotic solutions in order to determine the $S$-matrix. 

Our goal is to construct the $2\times 2$ solution matrix $\chi_{\alpha j}(R)$ matrix to the radial Schrödinger equation (Eq.~\ref{eq:nuclearSE}), where $j=1,2$ labels the two linearly independent solutions (after imposing regularity of the solution at the origin to eliminate the other two solutions) and $\alpha=1,2$ labels the two channels. 
At $R<R_{LZ}$, the semiclassical wave function -- within the Wentzel-Kramers-Brillouin (WKB) description -- satisfying regularity at the origin is an exponential function
\begin{align}
    \chi_{\alpha j}(R)=\frac{\delta_{\alpha j}}{\sqrt{|k_{\alpha}(R)|}}e^{\int_0^R|k_\alpha(R')|dR'},\,\,\,\,\,\,0<R<R_t^\alpha. 
\end{align}
The channel-dependent semiclassical momenta are 
\begin{equation}
k_\alpha(R)=
\sqrt{
2\mu\left[
E-
U_\alpha(R)
\right]}.
\end{equation}
Since there is no coupling between channels in the Landau-Zener-Stueckelberg (LZS)  treatment except exactly at $R = R_{LZ}$, the two linearly independent solutions are diagonal in each channel. 
To the right of each turning point, the WKB solution becomes oscillatory. 
Because the LZS treatment of the avoided crossing involves a mixing of incoming and outgoing plane wave solutions between the different channels, we will express the solution in this region as
\begin{align}
    \chi_{\alpha j}(R) &= \frac{1}{\sqrt{k_\alpha(R)}}\left(A_{\alpha j}^+e^{i\int_{R_t^\alpha}^Rk_\alpha(R')dR'}+A_{\alpha j}^-e^{-i\int_{R_t^\alpha}^Rk_\alpha(R')dR'}\right),\,\,\,\,\,R_t^\alpha<R<R_{LZ}
    \label{eq:chiregion1}
\end{align}
where the amplitudes $A_{\alpha j}^\pm$ are determined by the WKB connection formula for a classically forbidden region to the left of a classically allowed region. Imposition of this connection formula yields
\begin{align}
   \ \begin{pmatrix}
        A_{1j}^+\\
        A_{1j}^-\\
         A_{2j}^+\\
       A_{2j}^-
    \end{pmatrix}=
  \underbrace{ \begin{pmatrix}
    e^{-i\pi/4} & 0\\e^{i\pi/4} & 0 \\0 & e^{-i\pi/4}\\0 &e^{i\pi/4}
    \end{pmatrix}}_{\underline{\mathrm{C}}}\begin{pmatrix} \delta_{1j} \\ \delta_{2j}\end{pmatrix}.
\end{align}
Next, we express the solution Eq.~\ref{eq:chiregion1} so that it explicitly depends on the phase accumulated with respect to $R_{LZ}$. 
We rewrite the phase integral as
\begin{align}
    \int_{R_t^\alpha}^Rk_\alpha(R')dR' = \phi_\alpha^L+\int_{R_{LZ}}^{R}k_\alpha(R')dR',
\end{align}
with $\phi_L^\alpha$ defined as in Eq.~\ref{eq:phasesL} in the text,
\begin{equation}
  \phi_L^\alpha=  \int_{R_t^\alpha}^{R_{LZ}}k_\alpha(R')dR'.
\end{equation}
Eq.~\ref{eq:chiregion1} is thus rewritten as
\begin{align}
    \chi_{\alpha j}(R) &= \frac{1}{\sqrt{k_\alpha(R)}}\left(B_{\alpha j}^+e^{i\int_{R_{LZ}}^Rk_\alpha(R')dR'}+B_{\alpha j}^-e^{-i\int_{R_{LZ}}^Rk_\alpha(R')dR'}\right),\,\,\,\,\,R_t^\alpha<R<R_{LZ}
    \label{eq:chiregion1_lat}
\end{align}
where
\begin{align}
   \ \begin{pmatrix}
        B_{1j}^+\\
        B_{1j}^-\\
         B_{2j}^+\\
       B_{2j}^-
    \end{pmatrix}=
  \underbrace{\begin{pmatrix}
        e^{i\phi_L^1} & 0 & 0 & 0\\0 & e^{-i\phi_L^1} & 0 & 0 \\
        0 & 0 & e^{i\phi_L^2} & 0 \\ 0 & 0 & 0 & e^{-i\phi_L^2}
    \end{pmatrix}}_{\underline{\mathrm{L}}}\begin{pmatrix}
        A_{1j}^+\\
        A_{1j}^-\\
         A_{2j}^+\\
       A_{2j}^-
    \end{pmatrix}.
\end{align}

The amplitudes are mixed at the avoided crossing according to the LZS transfer matrix, a unitary matrix that describes how the wave function splits coherently at an avoided crossing according to the Landau Zener probability to transition between potential curves $P_{LZ}$ (Eq.~\ref{eq:lzprob}) and the Stokes phase $\lambda$ (Eq.~\ref{eq:stokes})
accumulated during the adiabatic passage by an avoided crossing 
\cite{Child1971CurveCrossing,Child1974MolecularCollisionTheory,ZhuNakamura1992_LZ,Zhu1995}. 
In terms of these quantities, the LZS transfer matrix is
\begin{equation}
\underline{\mathrm{T}}=
\begin{pmatrix}
\sqrt{1-P_{LZ}}\,e^{i\lambda} & 0& -\sqrt{P_{LZ}} & 0\\
0 & \sqrt{1-P_{LZ}}\,e^{-i\lambda} & 0 & -\sqrt{P_{LZ}} \\
\sqrt{P_{LZ}} & 0 & \sqrt{1-P_{LZ}}\,e^{-i\lambda} & 0 \\
0 & \sqrt{P_{LZ}} & 0 & \sqrt{1-P_{LZ}}\,e^{i\lambda}
\end{pmatrix}
\label{curve_cross}
\end{equation}
such that the solutions to the right of the crossing are given by
    \begin{align}
    \chi_{\alpha j}(R)&= \frac{1}{\sqrt{k_\alpha(R)}}
\left(C_{\alpha j}^+e^{i\int_{R_{LZ}}^Rk_\alpha(R')d{R'}}+
C_{\alpha j}^-e^{-i\int_{R_{LZ}}^Rk_\alpha(R')d{R'}}
\right) ,\,\,\,\,\,R_{LZ}<R<R_o^\alpha
\end{align}

with
\begin{equation}
    \begin{pmatrix}
        C_{1j}^+\\
        C_{1j}^-\\
         C_{2j}^+\\
        C_{2j}^-
    \end{pmatrix}=\underline{\mathrm{T}}\cdot \underline{\mathrm{L}} \cdot \underline{\mathrm{C}} \cdot \begin{pmatrix}
        \delta_{1j}\\\delta_{2j}
    \end{pmatrix}. 
\end{equation}
Finally, to explicitly include the phase accumulated from $R_{LZ}$ to  $R_o^\alpha$, we first write the phase accumulated out to $R<R_o^\alpha$ as
\begin{equation}
    \int_{R_{LZ}}^Rk_\alpha(R')dR'=\int_{R_{LZ}}^{R_o^\alpha}k_\alpha(R')dR'+\int_{R_o^\alpha}^{R}k_\alpha(R')dR'=\phi_R^\alpha + \int_{R_o^\alpha}^{R}k_\alpha(R')dR'.
\end{equation} 
Therefore, with one more transfer matrix $\underline{\mathrm{R}}$, equivalent to $\underline{\mathrm{L}}$ but with all $L$ labels swapped to $R$ such that the accumulated phase is $\phi_R^\alpha$ (Eq. \ref{eq:phasesR}), we obtain

\begin{align}
\label{eq:ddef}
    \begin{pmatrix}
        D_{1j}^+\\
        D_{1j}^-\\
         D_{2j}^+\\
        D_{2j}^-
    \end{pmatrix}&=\underline{\mathrm{R}}\cdot\underline{\mathrm{T}}\cdot \underline{\mathrm{L}} \cdot \underline{\mathrm{C}} \cdot \begin{pmatrix}
        \delta_{1j} \\ \delta_{2j}
    \end{pmatrix}=\begin{pmatrix}
e^{i(Z-\pi/4)} \, \delta_{1j} \sqrt{1-P_{LZ}} \;-\; e^{i(Y-\pi/4)} \, \delta_{2j} \sqrt{P_{LZ}} \\
e^{-i(Z-\pi/4)} \, \delta_{1j} \sqrt{1-P_{LZ}} \;-\; e^{-i (Y-\pi/4)} \, \delta_{2j} \sqrt{P_{LZ}} \\
-i\left[e^{iX} \, \delta_{2j} \sqrt{1-P_{LZ}} \;+\; e^{iW} \, \delta_{1j} \sqrt{P_{LZ}}\right]\\
i\left[e^{-iX} \,\delta_{2j} \sqrt{1-P_{LZ}} \;+\; e^{-iW} \, \delta_{1j} \sqrt{P_{LZ}}\right]
\end{pmatrix}
\end{align}
with $W$, $X$, $Y$, $Z$ defined in Eqs. \ref{eq:wphase}, \ref{eq:xphase}, \ref{eq:yphase}, and \ref{eq:zphase}, respectively. 
With these coefficients, the full solution for $R_{LZ}<R<R_o^\alpha$ is given by
    \begin{align}
    \chi_{\alpha j}(R)&= \frac{1}{\sqrt{k_\alpha(R)}}
\left(D_{\alpha j}^+e^{i\int_{R_o^\alpha}^Rk_\alpha(R')d{R'}}+
D_{\alpha j}^-e^{-i\int_{R_o^\alpha}^Rk_\alpha(R')d{R'}}
\right) ,\,\,\,\,\,R_{LZ}<R<R_o^\alpha.
\label{eq:chi2}
\end{align}
We can condense these incoming / outgoing waves back into cosine functions now that we are ready to approach the second classical turning point in the upper channel. 
Inserting Eq.~\ref{eq:ddef} into Eq.~\ref{eq:chi2} yields for the $\alpha=1$ solution
\begin{align}
   \nonumber \chi_{1j}(R) = \frac{2}{\sqrt{k_1(R)}}\Bigg[&\sqrt{1 - P_{LZ}}\delta_{1j}\cos(\int_{R_o^1}^Rk_1(R')dR'+Z-\pi/4)\\& \nonumber- \sqrt{P_{LZ}}\delta_{2j}\cos(\int_{R_o^1}^Rk_1(R')dR'+Y-\pi/4)\Bigg]\\
    = \nonumber\frac{2}{\sqrt{k_1(R)}}\Bigg[&\sqrt{1 - P_{LZ}}\delta_{1j}\cos(\int_{R}^{R_o^1}k_1(R')dR'-(Z-\pi/2)-\pi/4)\\& - \sqrt{P_{LZ}}\delta_{2j}\cos(\int_{R}^{R_o^1}k_1(R')dR'-(Y-\pi/2)-\pi/4)\Bigg]\nonumber\\
    = \nonumber\frac{2}{\sqrt{k_1(R)}}\Bigg[&\sqrt{1 - P_{LZ}}\delta_{1j}\left\{\cos(\int_{R}^{R_o^1}k_1(R')dR'-\pi/4)\sin Z - \sin(\int_{R}^{R_o^1}k_1(R')dR'-\pi/4)\cos Z\right\}\\& - \sqrt{P_{LZ}}\delta_{2j}\left\{\cos(\int_{R}^{R_o^1}k_1(R')dR'-\pi/4)\sin Y-\sin(\int_{R}^{R_o^1}k_1(R')dR'-\pi/4)\cos Y\right\}\Bigg]\label{eq:chi1sol1}.
\end{align}
Past the classical turning point $R_o^1$, the solution in the upper channel must remain finite, and hence for $R>R_o^1$ each of these cosine terms must connect to a decaying exponential. 
The WKB connection formula appropriate to the transition from a classically forbidden region on the right to a classically allowed region on the left reads \cite{Sakurai2011ModernQM,Berry1966PhaseShifts}
\begin{equation}
\frac{1}{\sqrt{|k_1(R)|}}\exp\left(-\int_{R_o^1}^R |k_1(R')|dR'\right)\to \frac{2}{\sqrt{k_1(R)}}\cos\left(\int_R^{R_o^1}k_1(R')dR'-\pi/4\right).\label{eq:connection}
\end{equation}
Therefore there can only be terms proportional to the cosine in Eq.~\ref{eq:connection} in Eq.~\ref{eq:chi1sol1} in order to satisfy boundary conditions. 
We must therefore take a linear combination $\chi_\alpha(R) = \sum_j[\chi_{\alpha j}(R)]A_j$ of the two linearly independent solutions to impose this boundary condition. 
From Eq.~\ref{eq:chi1sol1},
\begin{align}
    \nonumber\chi_1(R)=\frac{2}{\sqrt{k_1(R)}}\Bigg[&\sqrt{1 - P_{LZ}}A_1\left\{\cos(\int_{R}^{R_o^1}k_1(R')dR'-\pi/4)\sin Z - \sin(\int_{R}^{R_o^1}k_1(R')dR'-\pi/4)\cos Z\right\}\\& - \sqrt{P_{LZ}}A_2\left\{\cos(\int_{R}^{R_o^1}k_1(R')dR'-\pi/4)\sin Y-\sin(\int_{R}^{R_o^1}k_1(R')dR'-\pi/4)\cos Y\right\}\Bigg]\label{eq:chi1sol2},
\end{align}
and hence to eliminate all terms proportional to $\sin(\int_{R}^{R_o^1}k_1(R')dR'-\pi/4)$ we require
\begin{align}
    0 &= A_1\sqrt{1 - P_{LZ}}\cos Z - \sqrt{P_{LZ}}A_2\cos Y\nonumber\\
    \implies A_1 &= \frac{A_2\sqrt{P_{LZ}}}{\sqrt{1-P_{LZ}}}\frac{\cos Y}{\cos Z}.
\end{align}
The remaining constant $A_2$ can be set to one without loss of generality, since it is just an overall normalization that will be absorbed into the matching coefficients. Having determined the linear combination satisfying boundary conditions in the upper channel, we can now obtain the $S$-matrix from the lower channel solution $\chi_2(R)$. 

For this solution, the position $R_o^2$ is determined not by a classical turning point (since there is not one in this channel) but by the outer distance where we will match $\chi_2(R)$ to the asymptotic regular and irregular free-particle solutions to obtain the $S$-matrix.  
To facilitate this matching, we want our solution to explicitly contain the phase of a free particle at the boundary, $k_o^2 R_o^2$, where for sufficiently large $R_o^2$, $k_o^2=\sqrt{2\mu[E - U_2(R_o^2)]}$ is a constant independent of the exact value of $R_o^2$. We therefore write
\begin{align}
    \nonumber\int_{R_{LZ}}^{R_o^2}k_2(R')dR' &=\underbrace{\int_{R_{LZ}}^{R_o^2}[k_2(R')-k_o^2]dR'- k_o^2 R_{LZ}}_{\phi_R^2}+k_o^2 R_o^2\\
    &= \phi_R^2 + k_o^2 R_o^2. 
\end{align}
With this definition, and again in the limit of sufficiently large $R_o^2$, $\phi_R^2$ is independent of the specific matching point $R_o^2$. 
Inserting these phase definitions into $\chi_2(R)$ then yields
\begin{align}
\chi_2(R_o^2)&=\nonumber\frac{2}{\sqrt{k_o^2}\sqrt{1-P_{LZ}}} \left(P_{LZ}\frac{\cos Y}{\cos Z}\sin(k_o^2 R_o^2+W)+(1-P_{LZ})\sin(k_o^2 R_o^2+X)\right)\\\nonumber
    &= \frac{2}{\sqrt{k_o^2}\sqrt{1-P_{LZ}}} \Bigg(P_{LZ}\frac{\cos Y}{\cos Z}\left[\sin(W)\cos k_o^2 R_o^2+\cos(W)\sin k_o^2 R_o^2\right]\\&+(1-P_{LZ})\left[\sin( X)\cos k_o^2 R_o^2 + \sin  k_o^2 R_o^2\cos(X)\right]\Bigg),
\end{align}
where $W$, $X$, $Y$, and $Z$ are all defined in the main text. 

Finally, we match this solution to the asymptotic form, $\chi_2(R_o^2)= f_2(R_o^2)I - g_2(R_o^2)J$ where $f_2(R_o^2)= \sqrt{\frac{2\mu}{\pi k_o^2}}\sin(k_o^2 R_o^2)$ and $g_2(R_o^2)= -\sqrt{\frac{2\mu}{\pi k_o^2}}\cos(k_o^2 R_o^2)$.
The coefficients $I$ and $J$ are thus determined to be
\begin{align}
I &= \sqrt{\frac{2\pi}{\mu}}\frac{1}{\sqrt{1 - P_ {LZ}}}\left[P_{LZ}\cos( W)\cos (Y)\sec (Z) + (1 - P_{LZ})\cos(X)\right]\\
    J&=\sqrt{\frac{2\pi}{\mu}}\frac{1}{\sqrt{1 - P_{LZ}}}\left[P_{LZ}\sec (Z)\cos (Y)\sin(W) +(1-P_{LZ})\sin(X)\right].
\end{align}
To simplify these expressions, we define
\begin{align}
        C &= \frac{P_{LZ}}{1-P_{LZ}} \cos Y\,\sec Z,
\end{align}
such that 
\begin{align}
    I &= \sqrt{\frac{2\pi}{\mu}}\sqrt{1 - P_ {LZ}}\left[C\cos( W) + \cos(X)\right]\\
    J&=\sqrt{\frac{2\pi}{\mu}}\sqrt{1-P_{LZ}}\left[C\sin(W) +\sin(X)\right].
\end{align}
From $I$ and $J$, the S-matrix can be constructed using $S = (I + iJ)/(I-iJ)$, giving Eq.~\ref{eq:S_analytic}.
\end{widetext}


\begin{thebibliography}{92}%
\makeatletter
\providecommand \@ifxundefined [1]{%
 \@ifx{#1\undefined}
}%
\providecommand \@ifnum [1]{%
 \ifnum #1\expandafter \@firstoftwo
 \else \expandafter \@secondoftwo
 \fi
}%
\providecommand \@ifx [1]{%
 \ifx #1\expandafter \@firstoftwo
 \else \expandafter \@secondoftwo
 \fi
}%
\providecommand \natexlab [1]{#1}%
\providecommand \enquote  [1]{``#1''}%
\providecommand \bibnamefont  [1]{#1}%
\providecommand \bibfnamefont [1]{#1}%
\providecommand \citenamefont [1]{#1}%
\providecommand \href@noop [0]{\@secondoftwo}%
\providecommand \href [0]{\begingroup \@sanitize@url \@href}%
\providecommand \@href[1]{\@@startlink{#1}\@@href}%
\providecommand \@@href[1]{\endgroup#1\@@endlink}%
\providecommand \@sanitize@url [0]{\catcode `\\12\catcode `\$12\catcode `\&12\catcode `\#12\catcode `\^12\catcode `\_12\catcode `\%12\relax}%
\providecommand \@@startlink[1]{}%
\providecommand \@@endlink[0]{}%
\providecommand \url  [0]{\begingroup\@sanitize@url \@url }%
\providecommand \@url [1]{\endgroup\@href {#1}{\urlprefix }}%
\providecommand \urlprefix  [0]{URL }%
\providecommand \Eprint [0]{\href }%
\providecommand \doibase [0]{https://doi.org/}%
\providecommand \selectlanguage [0]{\@gobble}%
\providecommand \bibinfo  [0]{\@secondoftwo}%
\providecommand \bibfield  [0]{\@secondoftwo}%
\providecommand \translation [1]{[#1]}%
\providecommand \BibitemOpen [0]{}%
\providecommand \bibitemStop [0]{}%
\providecommand \bibitemNoStop [0]{.\EOS\space}%
\providecommand \EOS [0]{\spacefactor3000\relax}%
\providecommand \BibitemShut  [1]{\csname bibitem#1\endcsname}%
\let\auto@bib@innerbib\@empty
\bibitem [{\citenamefont {Domcke}\ \emph {et~al.}(2004)\citenamefont {Domcke}, \citenamefont {Yarkony},\ and\ \citenamefont {K{\"o}ppel}}]{Domcke2004ConicalIntersections}%
  \BibitemOpen
  \bibfield  {author} {\bibinfo {author} {\bibfnamefont {W.}~\bibnamefont {Domcke}}, \bibinfo {author} {\bibfnamefont {D.~R.}\ \bibnamefont {Yarkony}},\ and\ \bibinfo {author} {\bibfnamefont {H.}~\bibnamefont {K{\"o}ppel}},\ }\href {https://doi.org/10.1142/5346} {\emph {\bibinfo {title} {Conical Intersections: Electronic Structure, Dynamics and Spectroscopy}}}\ (\bibinfo  {publisher} {World Scientific},\ \bibinfo {year} {2004})\BibitemShut {NoStop}%
\bibitem [{\citenamefont {Schuurman}\ \emph {et~al.}(2011)\citenamefont {Schuurman}, \citenamefont {Domcke}, \citenamefont {Yarkony},\ and\ \citenamefont {K{\"o}ppel}}]{Schuurman2011ConicalIntersections}%
  \BibitemOpen
  \bibfield  {author} {\bibinfo {author} {\bibfnamefont {M.~S.}\ \bibnamefont {Schuurman}}, \bibinfo {author} {\bibfnamefont {W.}~\bibnamefont {Domcke}}, \bibinfo {author} {\bibfnamefont {D.~R.}\ \bibnamefont {Yarkony}},\ and\ \bibinfo {author} {\bibfnamefont {H.}~\bibnamefont {K{\"o}ppel}},\ }\href {https://doi.org/10.1142/7803} {\emph {\bibinfo {title} {Conical Intersections: Theory, Computation and Experiment}}}\ (\bibinfo  {publisher} {World Scientific},\ \bibinfo {year} {2011})\BibitemShut {NoStop}%
\bibitem [{\citenamefont {Scheid}\ \emph {et~al.}(2019)\citenamefont {Scheid}, \citenamefont {Goswami}, \citenamefont {Meyer},\ and\ \citenamefont {K{\"o}ppel}}]{Scheid2019NonadiabaticPhotodynamics}%
  \BibitemOpen
  \bibfield  {author} {\bibinfo {author} {\bibfnamefont {S.}~\bibnamefont {Scheid}}, \bibinfo {author} {\bibfnamefont {S.}~\bibnamefont {Goswami}}, \bibinfo {author} {\bibfnamefont {H.-D.}\ \bibnamefont {Meyer}},\ and\ \bibinfo {author} {\bibfnamefont {H.}~\bibnamefont {K{\"o}ppel}},\ }\bibfield  {title} {\bibinfo {title} {Fully quantal treatment of nonadiabatic molecular photodynamics: General considerations and application to the benzene cation},\ }\href {https://doi.org/10.1016/j.comptc.2019.01.011} {\bibfield  {journal} {\bibinfo  {journal} {Comput. Theor. Chem.}\ }\textbf {\bibinfo {volume} {1150}},\ \bibinfo {pages} {38} (\bibinfo {year} {2019})}\BibitemShut {NoStop}%
\bibitem [{\citenamefont {Donovan}\ \emph {et~al.}(2022)\citenamefont {Donovan}, \citenamefont {Kirrander},\ and\ \citenamefont {Lawley}}]{Donovan2022HeavyRydberg}%
  \BibitemOpen
  \bibfield  {author} {\bibinfo {author} {\bibfnamefont {R.~J.}\ \bibnamefont {Donovan}}, \bibinfo {author} {\bibfnamefont {A.}~\bibnamefont {Kirrander}},\ and\ \bibinfo {author} {\bibfnamefont {K.~P.}\ \bibnamefont {Lawley}},\ }\bibfield  {title} {\bibinfo {title} {Heavy {Rydberg} and ion-pair states: chemistry, spectroscopy and theory},\ }\href {https://doi.org/10.1080/0144235X.2022.2077024} {\bibfield  {journal} {\bibinfo  {journal} {Int. Rev. Phys. Chem.}\ }\textbf {\bibinfo {volume} {41}},\ \bibinfo {pages} {97} (\bibinfo {year} {2022})}\BibitemShut {NoStop}%
\bibitem [{\citenamefont {Kirrander}(2010)}]{kirrander2010HeavyRydberg}%
  \BibitemOpen
  \bibfield  {author} {\bibinfo {author} {\bibfnamefont {A.}~\bibnamefont {Kirrander}},\ }\bibfield  {title} {\bibinfo {title} {Communication: Heavy {Rydberg} states: The {H$^{+}$H$^{-}$} system},\ }\href {https://doi.org/10.1063/1.3492371} {\bibfield  {journal} {\bibinfo  {journal} {J. Chem. Phys.}\ }\textbf {\bibinfo {volume} {133}},\ \bibinfo {pages} {121103} (\bibinfo {year} {2010})}\BibitemShut {NoStop}%
\bibitem [{\citenamefont {Kirrander}\ and\ \citenamefont {Jungen}(2011)}]{kirrander2011IonPair}%
  \BibitemOpen
  \bibfield  {author} {\bibinfo {author} {\bibfnamefont {A.}~\bibnamefont {Kirrander}}\ and\ \bibinfo {author} {\bibfnamefont {C.}~\bibnamefont {Jungen}},\ }\bibfield  {title} {\bibinfo {title} {Molecular ion-pair states in ungerade {H$_2$}},\ }\href {https://doi.org/10.1103/PhysRevA.84.052512} {\bibfield  {journal} {\bibinfo  {journal} {Phys. Rev. A}\ }\textbf {\bibinfo {volume} {84}},\ \bibinfo {pages} {052512} (\bibinfo {year} {2011})}\BibitemShut {NoStop}%
\bibitem [{\citenamefont {Reinhold}\ and\ \citenamefont {Ubachs}(2005)}]{Reinhold2005HeavyRydberg}%
  \BibitemOpen
  \bibfield  {author} {\bibinfo {author} {\bibfnamefont {E.}~\bibnamefont {Reinhold}}\ and\ \bibinfo {author} {\bibfnamefont {W.}~\bibnamefont {Ubachs}},\ }\bibfield  {title} {\bibinfo {title} {Heavy {Rydberg} states},\ }\href {https://doi.org/10.1080/00268970500074427} {\bibfield  {journal} {\bibinfo  {journal} {Mol. Phys.}\ }\textbf {\bibinfo {volume} {103}},\ \bibinfo {pages} {1329} (\bibinfo {year} {2005})}\BibitemShut {NoStop}%
\bibitem [{\citenamefont {Kokouline}\ \emph {et~al.}(2000)\citenamefont {Kokouline}, \citenamefont {Dulieu}, \citenamefont {Kosloff},\ and\ \citenamefont {Masnou-Seeuws}}]{kokoouline2000theoretical}%
  \BibitemOpen
  \bibfield  {author} {\bibinfo {author} {\bibfnamefont {V.}~\bibnamefont {Kokouline}}, \bibinfo {author} {\bibfnamefont {O.}~\bibnamefont {Dulieu}}, \bibinfo {author} {\bibfnamefont {R.}~\bibnamefont {Kosloff}},\ and\ \bibinfo {author} {\bibfnamefont {F.}~\bibnamefont {Masnou-Seeuws}},\ }\bibfield  {title} {\bibinfo {title} {Theoretical treatment of channel mixing in excited {Rb$_2$} and {Cs$_2$} ultracold molecules: Determination of predissociation lifetimes with coordinate mapping},\ }\href {https://doi.org/10.1103/PhysRevA.62.032716} {\bibfield  {journal} {\bibinfo  {journal} {Physical Review A}\ }\textbf {\bibinfo {volume} {62}},\ \bibinfo {pages} {032716} (\bibinfo {year} {2000})}\BibitemShut {NoStop}%
\bibitem [{\citenamefont {McGuyer}\ \emph {et~al.}(2015)\citenamefont {McGuyer}, \citenamefont {McDonald}, \citenamefont {Iwata}, \citenamefont {Tarallo}, \citenamefont {Skomorowski}, \citenamefont {Moszynski},\ and\ \citenamefont {Zelevinsky}}]{mcguyer2015precise}%
  \BibitemOpen
  \bibfield  {author} {\bibinfo {author} {\bibfnamefont {B.~H.}\ \bibnamefont {McGuyer}}, \bibinfo {author} {\bibfnamefont {M.}~\bibnamefont {McDonald}}, \bibinfo {author} {\bibfnamefont {G.~Z.}\ \bibnamefont {Iwata}}, \bibinfo {author} {\bibfnamefont {M.~G.}\ \bibnamefont {Tarallo}}, \bibinfo {author} {\bibfnamefont {W.}~\bibnamefont {Skomorowski}}, \bibinfo {author} {\bibfnamefont {R.}~\bibnamefont {Moszynski}},\ and\ \bibinfo {author} {\bibfnamefont {T.}~\bibnamefont {Zelevinsky}},\ }\bibfield  {title} {\bibinfo {title} {Precise study of asymptotic physics with subradiant ultracold molecules},\ }\href@noop {} {\bibfield  {journal} {\bibinfo  {journal} {Nature Physics}\ }\textbf {\bibinfo {volume} {11}},\ \bibinfo {pages} {32} (\bibinfo {year} {2015})}\BibitemShut {NoStop}%
\bibitem [{\citenamefont {Chin}\ \emph {et~al.}(2010)\citenamefont {Chin}, \citenamefont {Grimm}, \citenamefont {Julienne},\ and\ \citenamefont {Tiesinga}}]{Chin2010FeshbachResonances}%
  \BibitemOpen
  \bibfield  {author} {\bibinfo {author} {\bibfnamefont {C.}~\bibnamefont {Chin}}, \bibinfo {author} {\bibfnamefont {R.}~\bibnamefont {Grimm}}, \bibinfo {author} {\bibfnamefont {P.}~\bibnamefont {Julienne}},\ and\ \bibinfo {author} {\bibfnamefont {E.}~\bibnamefont {Tiesinga}},\ }\bibfield  {title} {\bibinfo {title} {Feshbach resonances in ultracold gases},\ }\href {https://doi.org/10.1103/RevModPhys.82.1225} {\bibfield  {journal} {\bibinfo  {journal} {Reviews of Modern Physics}\ }\textbf {\bibinfo {volume} {82}},\ \bibinfo {pages} {1225} (\bibinfo {year} {2010})}\BibitemShut {NoStop}%
\bibitem [{\citenamefont {Greene}\ \emph {et~al.}(2000)\citenamefont {Greene}, \citenamefont {Dickinson},\ and\ \citenamefont {Sadeghpour}}]{Greene2000Trilobite}%
  \BibitemOpen
  \bibfield  {author} {\bibinfo {author} {\bibfnamefont {C.~H.}\ \bibnamefont {Greene}}, \bibinfo {author} {\bibfnamefont {A.~S.}\ \bibnamefont {Dickinson}},\ and\ \bibinfo {author} {\bibfnamefont {H.~R.}\ \bibnamefont {Sadeghpour}},\ }\bibfield  {title} {\bibinfo {title} {Creation of polar and nonpolar ultralong-range {Rydberg} molecules},\ }\href {https://doi.org/10.1103/PhysRevLett.85.2458} {\bibfield  {journal} {\bibinfo  {journal} {Physical Review Letters}\ }\textbf {\bibinfo {volume} {85}},\ \bibinfo {pages} {2458} (\bibinfo {year} {2000})}\BibitemShut {NoStop}%
\bibitem [{\citenamefont {Boisseau}\ \emph {et~al.}(2002)\citenamefont {Boisseau}, \citenamefont {Simbotin},\ and\ \citenamefont {C{\^o}t{\'e}}}]{Boisseau2002Macrodimers}%
  \BibitemOpen
  \bibfield  {author} {\bibinfo {author} {\bibfnamefont {C.}~\bibnamefont {Boisseau}}, \bibinfo {author} {\bibfnamefont {I.}~\bibnamefont {Simbotin}},\ and\ \bibinfo {author} {\bibfnamefont {R.}~\bibnamefont {C{\^o}t{\'e}}},\ }\bibfield  {title} {\bibinfo {title} {Macrodimers: Ultralong range {Rydberg} molecules},\ }\href {https://doi.org/10.1103/PhysRevLett.88.133004} {\bibfield  {journal} {\bibinfo  {journal} {Physical Review Letters}\ }\textbf {\bibinfo {volume} {88}},\ \bibinfo {pages} {133004} (\bibinfo {year} {2002})}\BibitemShut {NoStop}%
\bibitem [{\citenamefont {Dei{\ss}}\ \emph {et~al.}(2021)\citenamefont {Dei{\ss}}, \citenamefont {Haze},\ and\ \citenamefont {Denschlag}}]{Deiss2021AtomIonRydberg}%
  \BibitemOpen
  \bibfield  {author} {\bibinfo {author} {\bibfnamefont {M.}~\bibnamefont {Dei{\ss}}}, \bibinfo {author} {\bibfnamefont {S.}~\bibnamefont {Haze}},\ and\ \bibinfo {author} {\bibfnamefont {J.~H.}\ \bibnamefont {Denschlag}},\ }\bibfield  {title} {\bibinfo {title} {Long-range atom--ion {Rydberg} molecule: A novel molecular binding mechanism},\ }\href {https://doi.org/10.3390/atoms9020034} {\bibfield  {journal} {\bibinfo  {journal} {Atoms}\ }\textbf {\bibinfo {volume} {9}},\ \bibinfo {pages} {34} (\bibinfo {year} {2021})}\BibitemShut {NoStop}%
\bibitem [{\citenamefont {Duspayev}\ \emph {et~al.}(2021)\citenamefont {Duspayev}, \citenamefont {Han}, \citenamefont {Viray}, \citenamefont {Ma}, \citenamefont {Zhao},\ and\ \citenamefont {Raithel}}]{Duspayev2021a}%
  \BibitemOpen
  \bibfield  {author} {\bibinfo {author} {\bibfnamefont {A.}~\bibnamefont {Duspayev}}, \bibinfo {author} {\bibfnamefont {X.}~\bibnamefont {Han}}, \bibinfo {author} {\bibfnamefont {M.~A.}\ \bibnamefont {Viray}}, \bibinfo {author} {\bibfnamefont {L.}~\bibnamefont {Ma}}, \bibinfo {author} {\bibfnamefont {J.}~\bibnamefont {Zhao}},\ and\ \bibinfo {author} {\bibfnamefont {G.}~\bibnamefont {Raithel}},\ }\bibfield  {title} {\bibinfo {title} {Long-range {Rydberg}-atom--ion molecules of {Rb} and {Cs}},\ }\href {https://doi.org/10.1103/PhysRevResearch.3.023114} {\bibfield  {journal} {\bibinfo  {journal} {Physical Review Research}\ }\textbf {\bibinfo {volume} {3}},\ \bibinfo {pages} {023114} (\bibinfo {year} {2021})}\BibitemShut {NoStop}%
\bibitem [{\citenamefont {Eiles}(2019)}]{Eiles2019}%
  \BibitemOpen
  \bibfield  {author} {\bibinfo {author} {\bibfnamefont {M.~T.}\ \bibnamefont {Eiles}},\ }\bibfield  {title} {\bibinfo {title} {Trilobites, butterflies, and other exotic specimens of long-range {Rydberg} molecules},\ }\href@noop {} {\bibfield  {journal} {\bibinfo  {journal} {J. Phys. B: At. Mol. Opt. Phys.}\ }\textbf {\bibinfo {volume} {52}},\ \bibinfo {pages} {113001} (\bibinfo {year} {2019})}\BibitemShut {NoStop}%
\bibitem [{\citenamefont {Dunning}\ \emph {et~al.}(2024)\citenamefont {Dunning}, \citenamefont {Kanungo},\ and\ \citenamefont {Yoshida}}]{Dunning2024UltralongRange}%
  \BibitemOpen
  \bibfield  {author} {\bibinfo {author} {\bibfnamefont {F.~B.}\ \bibnamefont {Dunning}}, \bibinfo {author} {\bibfnamefont {S.~K.}\ \bibnamefont {Kanungo}},\ and\ \bibinfo {author} {\bibfnamefont {S.}~\bibnamefont {Yoshida}},\ }\bibfield  {title} {\bibinfo {title} {Ultralong-range {Rydberg} molecules},\ }\href {https://doi.org/10.1088/1361-6455/ad7459} {\bibfield  {journal} {\bibinfo  {journal} {Journal of Physics B: Atomic, Molecular and Optical Physics}\ }\textbf {\bibinfo {volume} {57}},\ \bibinfo {pages} {212002} (\bibinfo {year} {2024})}\BibitemShut {NoStop}%
\bibitem [{\citenamefont {Shaffer}\ \emph {et~al.}(2018)\citenamefont {Shaffer}, \citenamefont {Rittenhouse},\ and\ \citenamefont {Sadeghpour}}]{Shaffer2018UltracoldRM}%
  \BibitemOpen
  \bibfield  {author} {\bibinfo {author} {\bibfnamefont {J.~P.}\ \bibnamefont {Shaffer}}, \bibinfo {author} {\bibfnamefont {S.~T.}\ \bibnamefont {Rittenhouse}},\ and\ \bibinfo {author} {\bibfnamefont {H.~R.}\ \bibnamefont {Sadeghpour}},\ }\bibfield  {title} {\bibinfo {title} {Ultracold {Rydberg} molecules},\ }\href {https://doi.org/10.1038/s41467-018-04318-5} {\bibfield  {journal} {\bibinfo  {journal} {Nature Communications}\ }\textbf {\bibinfo {volume} {9}},\ \bibinfo {pages} {1965} (\bibinfo {year} {2018})}\BibitemShut {NoStop}%
\bibitem [{\citenamefont {Deiglmayr}(2016)}]{Deiglmayr2016LongRange}%
  \BibitemOpen
  \bibfield  {author} {\bibinfo {author} {\bibfnamefont {J.}~\bibnamefont {Deiglmayr}},\ }\bibfield  {title} {\bibinfo {title} {Long-range interactions between {Rydberg} atoms},\ }\href {https://doi.org/10.1088/0031-8949/91/10/104007} {\bibfield  {journal} {\bibinfo  {journal} {Phys. Scr.}\ }\textbf {\bibinfo {volume} {91}},\ \bibinfo {pages} {104007} (\bibinfo {year} {2016})}\BibitemShut {NoStop}%
\bibitem [{\citenamefont {Sa{\ss}mannshausen}\ and\ \citenamefont {Deiglmayr}(2016)}]{Sassmannshausen2016Macrodimers}%
  \BibitemOpen
  \bibfield  {author} {\bibinfo {author} {\bibfnamefont {H.}~\bibnamefont {Sa{\ss}mannshausen}}\ and\ \bibinfo {author} {\bibfnamefont {J.}~\bibnamefont {Deiglmayr}},\ }\bibfield  {title} {\bibinfo {title} {Observation of {Rydberg}-atom macrodimers: Micrometer-sized diatomic molecules},\ }\href {https://doi.org/10.1103/PhysRevLett.117.083401} {\bibfield  {journal} {\bibinfo  {journal} {Physical Review Letters}\ }\textbf {\bibinfo {volume} {117}},\ \bibinfo {pages} {083401} (\bibinfo {year} {2016})}\BibitemShut {NoStop}%
\bibitem [{\citenamefont {Hollerith}\ and\ \citenamefont {Zeiher}(2023)}]{Hollerith2023MacrodimerReview}%
  \BibitemOpen
  \bibfield  {author} {\bibinfo {author} {\bibfnamefont {S.}~\bibnamefont {Hollerith}}\ and\ \bibinfo {author} {\bibfnamefont {J.}~\bibnamefont {Zeiher}},\ }\bibfield  {title} {\bibinfo {title} {{Rydberg} macrodimers: Diatomic molecules on the micrometer scale},\ }\href {https://doi.org/10.1021/acs.jpca.2c08727} {\bibfield  {journal} {\bibinfo  {journal} {The Journal of Physical Chemistry A}\ }\textbf {\bibinfo {volume} {127}},\ \bibinfo {pages} {3865} (\bibinfo {year} {2023})}\BibitemShut {NoStop}%
\bibitem [{\citenamefont {Weber}\ \emph {et~al.}(2017)\citenamefont {Weber}, \citenamefont {Tresp}, \citenamefont {Menke}, \citenamefont {Urvoy}, \citenamefont {Firstenberg}, \citenamefont {B{\"u}chler},\ and\ \citenamefont {Hofferberth}}]{Weber2017CalculationRydbergPotentials}%
  \BibitemOpen
  \bibfield  {author} {\bibinfo {author} {\bibfnamefont {S.}~\bibnamefont {Weber}}, \bibinfo {author} {\bibfnamefont {C.}~\bibnamefont {Tresp}}, \bibinfo {author} {\bibfnamefont {H.}~\bibnamefont {Menke}}, \bibinfo {author} {\bibfnamefont {A.}~\bibnamefont {Urvoy}}, \bibinfo {author} {\bibfnamefont {O.}~\bibnamefont {Firstenberg}}, \bibinfo {author} {\bibfnamefont {H.~P.}\ \bibnamefont {B{\"u}chler}},\ and\ \bibinfo {author} {\bibfnamefont {S.}~\bibnamefont {Hofferberth}},\ }\bibfield  {title} {\bibinfo {title} {Calculation of {Rydberg} interaction potentials},\ }\href {https://doi.org/10.1088/1361-6455/aa743a} {\bibfield  {journal} {\bibinfo  {journal} {Journal of Physics B: Atomic, Molecular and Optical Physics}\ }\textbf {\bibinfo {volume} {50}},\ \bibinfo {pages} {133001} (\bibinfo {year} {2017})}\BibitemShut {NoStop}%
\bibitem [{\citenamefont {Baer}(2006)}]{Baer2006BeyondBO}%
  \BibitemOpen
  \bibfield  {author} {\bibinfo {author} {\bibfnamefont {M.}~\bibnamefont {Baer}},\ }\href {https://doi.org/10.1002/0471780081} {\emph {\bibinfo {title} {Beyond {Born--Oppenheimer}: Conical Intersections and Electronic Nonadiabatic Coupling Terms}}}\ (\bibinfo  {publisher} {John Wiley \& Sons},\ \bibinfo {address} {Hoboken, NJ},\ \bibinfo {year} {2006})\BibitemShut {NoStop}%
\bibitem [{\citenamefont {Yarkony}(1996)}]{Yarkony1996DiabolicalCI}%
  \BibitemOpen
  \bibfield  {author} {\bibinfo {author} {\bibfnamefont {D.~R.}\ \bibnamefont {Yarkony}},\ }\bibfield  {title} {\bibinfo {title} {Diabolical conical intersections},\ }\href {https://doi.org/10.1103/RevModPhys.68.985} {\bibfield  {journal} {\bibinfo  {journal} {Reviews of Modern Physics}\ }\textbf {\bibinfo {volume} {68}},\ \bibinfo {pages} {985} (\bibinfo {year} {1996})}\BibitemShut {NoStop}%
\bibitem [{\citenamefont {Bixon}\ and\ \citenamefont {Jortner}(1968)}]{BixonJortner1968}%
  \BibitemOpen
  \bibfield  {author} {\bibinfo {author} {\bibfnamefont {M.}~\bibnamefont {Bixon}}\ and\ \bibinfo {author} {\bibfnamefont {J.}~\bibnamefont {Jortner}},\ }\bibfield  {title} {\bibinfo {title} {Intramolecular radiationless transitions},\ }\href {https://doi.org/10.1063/1.1668615} {\bibfield  {journal} {\bibinfo  {journal} {The Journal of Chemical Physics}\ }\textbf {\bibinfo {volume} {48}},\ \bibinfo {pages} {715} (\bibinfo {year} {1968})}\BibitemShut {NoStop}%
\bibitem [{\citenamefont {Gallagher}(1994)}]{Gallagher1994}%
  \BibitemOpen
  \bibfield  {author} {\bibinfo {author} {\bibfnamefont {T.~F.}\ \bibnamefont {Gallagher}},\ }\href@noop {} {\emph {\bibinfo {title} {{Rydberg} Atoms}}}\ (\bibinfo  {publisher} {Cambridge University Press},\ \bibinfo {address} {Cambridge},\ \bibinfo {year} {1994})\BibitemShut {NoStop}%
\bibitem [{\citenamefont {Theodosiou}(1984)}]{Theodosiou1984}%
  \BibitemOpen
  \bibfield  {author} {\bibinfo {author} {\bibfnamefont {C.~E.}\ \bibnamefont {Theodosiou}},\ }\bibfield  {title} {\bibinfo {title} {Lifetimes of alkali-metal-atom {Rydberg} states},\ }\href {https://doi.org/10.1103/PhysRevA.30.2881} {\bibfield  {journal} {\bibinfo  {journal} {Physical Review A}\ }\textbf {\bibinfo {volume} {30}},\ \bibinfo {pages} {2881} (\bibinfo {year} {1984})}\BibitemShut {NoStop}%
\bibitem [{\citenamefont {Durst}\ \emph {et~al.}(2025)\citenamefont {Durst}, \citenamefont {Simi{\'c}}, \citenamefont {Abraham},\ and\ \citenamefont {Eiles}}]{Durst2025Nonadiabatic}%
  \BibitemOpen
  \bibfield  {author} {\bibinfo {author} {\bibfnamefont {A.~A.~T.}\ \bibnamefont {Durst}}, \bibinfo {author} {\bibfnamefont {M.}~\bibnamefont {Simi{\'c}}}, \bibinfo {author} {\bibfnamefont {N.}~\bibnamefont {Abraham}},\ and\ \bibinfo {author} {\bibfnamefont {M.~T.}\ \bibnamefont {Eiles}},\ }\bibfield  {title} {\bibinfo {title} {Nonadiabatic couplings as a stabilization mechanism in long-range {Rydberg} molecules},\ }\href {https://doi.org/10.1103/PhysRevA.111.032811} {\bibfield  {journal} {\bibinfo  {journal} {Physical Review A}\ }\textbf {\bibinfo {volume} {111}},\ \bibinfo {pages} {032811} (\bibinfo {year} {2025})}\BibitemShut {NoStop}%
\bibitem [{\citenamefont {Srikumar}\ \emph {et~al.}(2023)\citenamefont {Srikumar}, \citenamefont {Hummel},\ and\ \citenamefont {Schmelcher}}]{Srikumar2023Nonadiabatic}%
  \BibitemOpen
  \bibfield  {author} {\bibinfo {author} {\bibfnamefont {R.}~\bibnamefont {Srikumar}}, \bibinfo {author} {\bibfnamefont {F.}~\bibnamefont {Hummel}},\ and\ \bibinfo {author} {\bibfnamefont {P.}~\bibnamefont {Schmelcher}},\ }\bibfield  {title} {\bibinfo {title} {Nonadiabatic interaction effects in the spectra of ultralong-range {Rydberg} molecules},\ }\href {https://doi.org/10.1103/PhysRevA.108.012809} {\bibfield  {journal} {\bibinfo  {journal} {Phys. Rev. A}\ }\textbf {\bibinfo {volume} {108}},\ \bibinfo {pages} {012809} (\bibinfo {year} {2023})}\BibitemShut {NoStop}%
\bibitem [{\citenamefont {Mellado-Alcedo}\ \emph {et~al.}(2024)\citenamefont {Mellado-Alcedo}, \citenamefont {Guttridge}, \citenamefont {Cornish}, \citenamefont {Sadeghpour},\ and\ \citenamefont {Gonz{\'a}lez-F{\'e}rez}}]{Gonzalezferez2024UltralongRange}%
  \BibitemOpen
  \bibfield  {author} {\bibinfo {author} {\bibfnamefont {D.}~\bibnamefont {Mellado-Alcedo}}, \bibinfo {author} {\bibfnamefont {A.}~\bibnamefont {Guttridge}}, \bibinfo {author} {\bibfnamefont {S.~L.}\ \bibnamefont {Cornish}}, \bibinfo {author} {\bibfnamefont {H.~R.}\ \bibnamefont {Sadeghpour}},\ and\ \bibinfo {author} {\bibfnamefont {R.}~\bibnamefont {Gonz{\'a}lez-F{\'e}rez}},\ }\bibfield  {title} {\bibinfo {title} {Ultralong-range {Cs-RbCs} {Rydberg} molecules: Nonadiabaticity of dipole moments},\ }\href {https://doi.org/10.1103/PhysRevA.110.013314} {\bibfield  {journal} {\bibinfo  {journal} {Phys. Rev. A}\ }\textbf {\bibinfo {volume} {110}},\ \bibinfo {pages} {013314} (\bibinfo {year} {2024})}\BibitemShut {NoStop}%
\bibitem [{\citenamefont {Schwettmann}\ \emph {et~al.}(2007)\citenamefont {Schwettmann}, \citenamefont {Overstreet}, \citenamefont {Tallant},\ and\ \citenamefont {Shaffer}}]{Schwettmann2007LongRange}%
  \BibitemOpen
  \bibfield  {author} {\bibinfo {author} {\bibfnamefont {A.}~\bibnamefont {Schwettmann}}, \bibinfo {author} {\bibfnamefont {K.~R.}\ \bibnamefont {Overstreet}}, \bibinfo {author} {\bibfnamefont {J.}~\bibnamefont {Tallant}},\ and\ \bibinfo {author} {\bibfnamefont {J.~P.}\ \bibnamefont {Shaffer}},\ }\bibfield  {title} {\bibinfo {title} {Analysis of long-range {Cs} {Rydberg} potential wells},\ }\href {https://doi.org/10.1080/09500340701584076} {\bibfield  {journal} {\bibinfo  {journal} {J. Mod. Opt.}\ }\textbf {\bibinfo {volume} {54}},\ \bibinfo {pages} {2551} (\bibinfo {year} {2007})}\BibitemShut {NoStop}%
\bibitem [{\citenamefont {Bosworth}\ \emph {et~al.}(2024)\citenamefont {Bosworth}, \citenamefont {Eiles},\ and\ \citenamefont {Schmelcher}}]{bosworth2024metastable}%
  \BibitemOpen
  \bibfield  {author} {\bibinfo {author} {\bibfnamefont {D.~J.}\ \bibnamefont {Bosworth}}, \bibinfo {author} {\bibfnamefont {M.~T.}\ \bibnamefont {Eiles}},\ and\ \bibinfo {author} {\bibfnamefont {P.}~\bibnamefont {Schmelcher}},\ }\bibfield  {title} {\bibinfo {title} {Metastable doubly charged {Rydberg} trimers},\ }\href@noop {} {\bibfield  {journal} {\bibinfo  {journal} {Physical Review Research}\ }\textbf {\bibinfo {volume} {6}},\ \bibinfo {pages} {043164} (\bibinfo {year} {2024})}\BibitemShut {NoStop}%
\bibitem [{\citenamefont {Duspayev}\ and\ \citenamefont {Raithel}(2022)}]{Duspayev2022NonadiabaticRAIM}%
  \BibitemOpen
  \bibfield  {author} {\bibinfo {author} {\bibfnamefont {A.}~\bibnamefont {Duspayev}}\ and\ \bibinfo {author} {\bibfnamefont {G.}~\bibnamefont {Raithel}},\ }\bibfield  {title} {\bibinfo {title} {Nonadiabatic decay of {Rydberg}-atom--ion molecules},\ }\href {https://doi.org/10.1103/PhysRevA.105.012810} {\bibfield  {journal} {\bibinfo  {journal} {Physical Review A}\ }\textbf {\bibinfo {volume} {105}},\ \bibinfo {pages} {012810} (\bibinfo {year} {2022})}\BibitemShut {NoStop}%
\bibitem [{\citenamefont {Child}(1971)}]{Child1971CurveCrossing}%
  \BibitemOpen
  \bibfield  {author} {\bibinfo {author} {\bibfnamefont {M.~S.}\ \bibnamefont {Child}},\ }\bibfield  {title} {\bibinfo {title} {Curve-crossing and the {WKB} approximation},\ }\href {https://doi.org/10.1080/00268977100100181} {\bibfield  {journal} {\bibinfo  {journal} {Molecular Physics}\ }\textbf {\bibinfo {volume} {20}},\ \bibinfo {pages} {171} (\bibinfo {year} {1971})}\BibitemShut {NoStop}%
\bibitem [{\citenamefont {Child}(1974{\natexlab{a}})}]{Child1974Semiclassical}%
  \BibitemOpen
  \bibfield  {author} {\bibinfo {author} {\bibfnamefont {M.~S.}\ \bibnamefont {Child}},\ }\bibfield  {title} {\bibinfo {title} {Semiclassical theory of tunneling and curve-crossing problems: A diagrammatic approach},\ }\href {https://doi.org/10.1016/0022-2852(74)90132-5} {\bibfield  {journal} {\bibinfo  {journal} {Journal of Molecular Spectroscopy}\ }\textbf {\bibinfo {volume} {53}},\ \bibinfo {pages} {280} (\bibinfo {year} {1974}{\natexlab{a}})}\BibitemShut {NoStop}%
\bibitem [{\citenamefont {Child}(1976)}]{Child1976Predissociation}%
  \BibitemOpen
  \bibfield  {author} {\bibinfo {author} {\bibfnamefont {M.~S.}\ \bibnamefont {Child}},\ }\bibfield  {title} {\bibinfo {title} {Predissociation and photodissociation of {IBr}: A case of intermediate coupling strength},\ }\href {https://doi.org/10.1080/00268977600102651} {\bibfield  {journal} {\bibinfo  {journal} {Molecular Physics}\ }\textbf {\bibinfo {volume} {32}},\ \bibinfo {pages} {1495} (\bibinfo {year} {1976})}\BibitemShut {NoStop}%
\bibitem [{\citenamefont {Kn{\"o}ckel}\ \emph {et~al.}(1981)\citenamefont {Kn{\"o}ckel}, \citenamefont {Tiemann},\ and\ \citenamefont {Zoglowek}}]{Knockel1981}%
  \BibitemOpen
  \bibfield  {author} {\bibinfo {author} {\bibfnamefont {H.}~\bibnamefont {Kn{\"o}ckel}}, \bibinfo {author} {\bibfnamefont {E.}~\bibnamefont {Tiemann}},\ and\ \bibinfo {author} {\bibfnamefont {D.}~\bibnamefont {Zoglowek}},\ }\bibfield  {title} {\bibinfo {title} {Line broadening by predissociation in the spectrum of {IBr}},\ }\href {https://doi.org/10.1016/0022-2852(81)90321-0} {\bibfield  {journal} {\bibinfo  {journal} {Journal of Molecular Spectroscopy}\ }\textbf {\bibinfo {volume} {85}},\ \bibinfo {pages} {225} (\bibinfo {year} {1981})}\BibitemShut {NoStop}%
\bibitem [{\citenamefont {Sadeghpour}(2000)}]{Sadeghpour2000ChannelInteraction}%
  \BibitemOpen
  \bibfield  {author} {\bibinfo {author} {\bibfnamefont {H.~R.}\ \bibnamefont {Sadeghpour}},\ }\bibfield  {title} {\bibinfo {title} {Channel interaction in alkali-halide photodissociation: Interferometric lineshapes and dynamic switching},\ }\href@noop {} {\bibfield  {journal} {\bibinfo  {journal} {Int. J. Quantum Chem.}\ }\textbf {\bibinfo {volume} {80}},\ \bibinfo {pages} {958} (\bibinfo {year} {2000})}\BibitemShut {NoStop}%
\bibitem [{\citenamefont {Balakrishnan}\ \emph {et~al.}(1999)\citenamefont {Balakrishnan}, \citenamefont {Esry}, \citenamefont {Sadeghpour}, \citenamefont {Cornett},\ and\ \citenamefont {Cavagnero}}]{Balakrishnan1999QuantumWavePacket}%
  \BibitemOpen
  \bibfield  {author} {\bibinfo {author} {\bibfnamefont {N.}~\bibnamefont {Balakrishnan}}, \bibinfo {author} {\bibfnamefont {B.~D.}\ \bibnamefont {Esry}}, \bibinfo {author} {\bibfnamefont {H.~R.}\ \bibnamefont {Sadeghpour}}, \bibinfo {author} {\bibfnamefont {S.~T.}\ \bibnamefont {Cornett}},\ and\ \bibinfo {author} {\bibfnamefont {M.~J.}\ \bibnamefont {Cavagnero}},\ }\bibfield  {title} {\bibinfo {title} {Quantum wave-packet dynamics of the photodissociation of {LiF}},\ }\href@noop {} {\bibfield  {journal} {\bibinfo  {journal} {Phys. Rev. A}\ }\textbf {\bibinfo {volume} {60}} (\bibinfo {year} {1999})}\BibitemShut {NoStop}%
\bibitem [{\citenamefont {Cornett}\ \emph {et~al.}(1999)\citenamefont {Cornett}, \citenamefont {Sadeghpour},\ and\ \citenamefont {Cavagnero}}]{Cornett1999Interferometric}%
  \BibitemOpen
  \bibfield  {author} {\bibinfo {author} {\bibfnamefont {S.~T.}\ \bibnamefont {Cornett}}, \bibinfo {author} {\bibfnamefont {H.~R.}\ \bibnamefont {Sadeghpour}},\ and\ \bibinfo {author} {\bibfnamefont {M.~J.}\ \bibnamefont {Cavagnero}},\ }\bibfield  {title} {\bibinfo {title} {Interferometric line shape modulation in alkali-halide photoabsorption},\ }\href@noop {} {\bibfield  {journal} {\bibinfo  {journal} {Phys. Rev. Lett.}\ }\textbf {\bibinfo {volume} {82}} (\bibinfo {year} {1999})}\BibitemShut {NoStop}%
\bibitem [{\citenamefont {Engel}\ \emph {et~al.}(1988)\citenamefont {Engel}, \citenamefont {Metiu}, \citenamefont {Almeida}, \citenamefont {Marcus},\ and\ \citenamefont {Zewail}}]{Engel1988NaIDissociation}%
  \BibitemOpen
  \bibfield  {author} {\bibinfo {author} {\bibfnamefont {V.}~\bibnamefont {Engel}}, \bibinfo {author} {\bibfnamefont {H.}~\bibnamefont {Metiu}}, \bibinfo {author} {\bibfnamefont {R.}~\bibnamefont {Almeida}}, \bibinfo {author} {\bibfnamefont {R.~A.}\ \bibnamefont {Marcus}},\ and\ \bibinfo {author} {\bibfnamefont {A.~H.}\ \bibnamefont {Zewail}},\ }\bibfield  {title} {\bibinfo {title} {Molecular state evolution after excitation with an ultra-short laser pulse: A quantum analysis of {NaI} and {NaBr} dissociation},\ }\href {https://doi.org/10.1016/0009-2614(88)87165-3} {\bibfield  {journal} {\bibinfo  {journal} {Chem. Phys. Lett.}\ }\textbf {\bibinfo {volume} {152}},\ \bibinfo {pages} {1} (\bibinfo {year} {1988})}\BibitemShut {NoStop}%
\bibitem [{\citenamefont {Choi}\ and\ \citenamefont {Light}(1989)}]{Choi1989NaIPredissociation}%
  \BibitemOpen
  \bibfield  {author} {\bibinfo {author} {\bibfnamefont {S.~E.}\ \bibnamefont {Choi}}\ and\ \bibinfo {author} {\bibfnamefont {J.~C.}\ \bibnamefont {Light}},\ }\bibfield  {title} {\bibinfo {title} {Use of the discrete variable representation in the quantum dynamics by a wave packet propagation: Predissociation of {NaI}($^{1}\sigma^{+}$)$\rightarrow${NaI}($0^{+}$)$\rightarrow${Na}($^{2}s$)+{I}($^{2}p$)},\ }\href {https://doi.org/10.1063/1.455957} {\bibfield  {journal} {\bibinfo  {journal} {J. Chem. Phys.}\ }\textbf {\bibinfo {volume} {90}},\ \bibinfo {pages} {2593} (\bibinfo {year} {1989})}\BibitemShut {NoStop}%
\bibitem [{\citenamefont {Mizuno}\ and\ \citenamefont {Fukushima}(2018)}]{Mizuno2018}%
  \BibitemOpen
  \bibfield  {author} {\bibinfo {author} {\bibfnamefont {Y.}~\bibnamefont {Mizuno}}\ and\ \bibinfo {author} {\bibfnamefont {K.}~\bibnamefont {Fukushima}},\ }\bibfield  {title} {\bibinfo {title} {Power-law decay in the nonadiabatic photodissociation dynamics of alkali halides due to quantum wavepacket interference},\ }\href {https://doi.org/10.1063/1.5048957} {\bibfield  {journal} {\bibinfo  {journal} {The Journal of Chemical Physics}\ }\textbf {\bibinfo {volume} {149}},\ \bibinfo {pages} {174313} (\bibinfo {year} {2018})}\BibitemShut {NoStop}%
\bibitem [{\citenamefont {Liane}\ \emph {et~al.}(2024)\citenamefont {Liane}, \citenamefont {Simmermacher},\ and\ \citenamefont {Kirrander}}]{Liane2024}%
  \BibitemOpen
  \bibfield  {author} {\bibinfo {author} {\bibfnamefont {E.~M.}\ \bibnamefont {Liane}}, \bibinfo {author} {\bibfnamefont {M.}~\bibnamefont {Simmermacher}},\ and\ \bibinfo {author} {\bibfnamefont {A.}~\bibnamefont {Kirrander}},\ }\bibfield  {title} {\bibinfo {title} {Ultrafast x-ray scattering and electronic coherence at avoided crossings: Complete isotropic signals},\ }\href {https://doi.org/10.1088/1361-6455/ad53af} {\bibfield  {journal} {\bibinfo  {journal} {Journal of Physics B: Atomic, Molecular and Optical Physics}\ }\textbf {\bibinfo {volume} {57}},\ \bibinfo {pages} {145602} (\bibinfo {year} {2024})}\BibitemShut {NoStop}%
\bibitem [{\citenamefont {Cong}\ \emph {et~al.}(1996)\citenamefont {Cong}, \citenamefont {Roberts}, \citenamefont {Herek}, \citenamefont {Mokhtari},\ and\ \citenamefont {Zewail}}]{Cong1996NaIPredissociation}%
  \BibitemOpen
  \bibfield  {author} {\bibinfo {author} {\bibfnamefont {P.}~\bibnamefont {Cong}}, \bibinfo {author} {\bibfnamefont {G.}~\bibnamefont {Roberts}}, \bibinfo {author} {\bibfnamefont {J.~L.}\ \bibnamefont {Herek}}, \bibinfo {author} {\bibfnamefont {A.}~\bibnamefont {Mokhtari}},\ and\ \bibinfo {author} {\bibfnamefont {A.~H.}\ \bibnamefont {Zewail}},\ }\bibfield  {title} {\bibinfo {title} {Femtosecond real-time probing of reactions. 18. experimental and theoretical mapping of trajectories and potentials in the {NaI} dissociation reaction},\ }\href {https://doi.org/10.1021/jp9534034} {\bibfield  {journal} {\bibinfo  {journal} {J. Phys. Chem.}\ }\textbf {\bibinfo {volume} {100}},\ \bibinfo {pages} {7832} (\bibinfo {year} {1996})}\BibitemShut {NoStop}%
\bibitem [{\citenamefont {de~Vries}\ \emph {et~al.}(1980)\citenamefont {de~Vries}, \citenamefont {van Veen}, \citenamefont {Hutchinson},\ and\ \citenamefont {de~Vries}}]{deVries1980}%
  \BibitemOpen
  \bibfield  {author} {\bibinfo {author} {\bibfnamefont {M.~S.}\ \bibnamefont {de~Vries}}, \bibinfo {author} {\bibfnamefont {N.~J.~A.}\ \bibnamefont {van Veen}}, \bibinfo {author} {\bibfnamefont {M.}~\bibnamefont {Hutchinson}},\ and\ \bibinfo {author} {\bibfnamefont {A.~E.}\ \bibnamefont {de~Vries}},\ }\bibfield  {title} {\bibinfo {title} {Curve crossing and repulsive potentials probed by laser induced half collisions: {ICl} and {IBr}},\ }\href {https://doi.org/10.1016/0301-0104(80)80091-7} {\bibfield  {journal} {\bibinfo  {journal} {Chemical Physics}\ }\textbf {\bibinfo {volume} {51}},\ \bibinfo {pages} {159} (\bibinfo {year} {1980})}\BibitemShut {NoStop}%
\bibitem [{\citenamefont {Zou}\ \emph {et~al.}(2023)\citenamefont {Zou}, \citenamefont {Berngruber}, \citenamefont {Anasuri}, \citenamefont {Zuber}, \citenamefont {Meinert}, \citenamefont {L{\"o}w},\ and\ \citenamefont {Pfau}}]{Zou2023VibrationalRAIM}%
  \BibitemOpen
  \bibfield  {author} {\bibinfo {author} {\bibfnamefont {Y.-Q.}\ \bibnamefont {Zou}}, \bibinfo {author} {\bibfnamefont {M.}~\bibnamefont {Berngruber}}, \bibinfo {author} {\bibfnamefont {V.~S.~V.}\ \bibnamefont {Anasuri}}, \bibinfo {author} {\bibfnamefont {N.}~\bibnamefont {Zuber}}, \bibinfo {author} {\bibfnamefont {F.}~\bibnamefont {Meinert}}, \bibinfo {author} {\bibfnamefont {R.}~\bibnamefont {L{\"o}w}},\ and\ \bibinfo {author} {\bibfnamefont {T.}~\bibnamefont {Pfau}},\ }\bibfield  {title} {\bibinfo {title} {Observation of vibrational dynamics of oriented {Rydberg}-atom-ion molecules},\ }\href {https://doi.org/10.1103/PhysRevLett.130.023002} {\bibfield  {journal} {\bibinfo  {journal} {Physical Review Letters}\ }\textbf {\bibinfo {volume} {130}},\ \bibinfo {pages} {023002} (\bibinfo {year} {2023})}\BibitemShut {NoStop}%
\bibitem [{\citenamefont {Berngruber}\ \emph {et~al.}(2024{\natexlab{a}})\citenamefont {Berngruber}, \citenamefont {Bosworth}, \citenamefont {Herrera-Sancho}, \citenamefont {Anasuri}, \citenamefont {Zuber}, \citenamefont {Hummel}, \citenamefont {Krauter}, \citenamefont {Meinert}, \citenamefont {L{\"o}w}, \citenamefont {Schmelcher},\ and\ \citenamefont {Pfau}}]{Berngruber2024InSituCollisionDynamics}%
  \BibitemOpen
  \bibfield  {author} {\bibinfo {author} {\bibfnamefont {M.}~\bibnamefont {Berngruber}}, \bibinfo {author} {\bibfnamefont {D.~J.}\ \bibnamefont {Bosworth}}, \bibinfo {author} {\bibfnamefont {O.~A.}\ \bibnamefont {Herrera-Sancho}}, \bibinfo {author} {\bibfnamefont {V.~S.~V.}\ \bibnamefont {Anasuri}}, \bibinfo {author} {\bibfnamefont {N.}~\bibnamefont {Zuber}}, \bibinfo {author} {\bibfnamefont {F.}~\bibnamefont {Hummel}}, \bibinfo {author} {\bibfnamefont {J.}~\bibnamefont {Krauter}}, \bibinfo {author} {\bibfnamefont {F.}~\bibnamefont {Meinert}}, \bibinfo {author} {\bibfnamefont {R.}~\bibnamefont {L{\"o}w}}, \bibinfo {author} {\bibfnamefont {P.}~\bibnamefont {Schmelcher}},\ and\ \bibinfo {author} {\bibfnamefont {T.}~\bibnamefont {Pfau}},\ }\bibfield  {title} {\bibinfo {title} {In situ observation of nonpolar to strongly polar atom-ion collision dynamics},\ }\href {https://doi.org/10.1103/PhysRevLett.133.083001} {\bibfield  {journal} {\bibinfo  {journal} {Physical Review Letters}\ }\textbf {\bibinfo {volume}
  {133}},\ \bibinfo {pages} {083001} (\bibinfo {year} {2024}{\natexlab{a}})}\BibitemShut {NoStop}%
\bibitem [{\citenamefont {Zuber}\ \emph {et~al.}(2022)\citenamefont {Zuber}, \citenamefont {Anasuri}, \citenamefont {Berngruber}, \citenamefont {Zou}, \citenamefont {Meinert}, \citenamefont {L{\"o}w},\ and\ \citenamefont {Pfau}}]{Zuber2022}%
  \BibitemOpen
  \bibfield  {author} {\bibinfo {author} {\bibfnamefont {N.}~\bibnamefont {Zuber}}, \bibinfo {author} {\bibfnamefont {V.~S.~V.}\ \bibnamefont {Anasuri}}, \bibinfo {author} {\bibfnamefont {M.}~\bibnamefont {Berngruber}}, \bibinfo {author} {\bibfnamefont {Y.-Q.}\ \bibnamefont {Zou}}, \bibinfo {author} {\bibfnamefont {F.}~\bibnamefont {Meinert}}, \bibinfo {author} {\bibfnamefont {R.}~\bibnamefont {L{\"o}w}},\ and\ \bibinfo {author} {\bibfnamefont {T.}~\bibnamefont {Pfau}},\ }\bibfield  {title} {\bibinfo {title} {Observation of a molecular bond between ions and {Rydberg} atoms},\ }\href@noop {} {\bibfield  {journal} {\bibinfo  {journal} {Nature}\ }\textbf {\bibinfo {volume} {605}},\ \bibinfo {pages} {453} (\bibinfo {year} {2022})}\BibitemShut {NoStop}%
\bibitem [{\citenamefont {Le~Roy}(1974)}]{LeRoy1974RKR}%
  \BibitemOpen
  \bibfield  {author} {\bibinfo {author} {\bibfnamefont {R.~J.}\ \bibnamefont {Le~Roy}},\ }\bibfield  {title} {\bibinfo {title} {Long-range potential coefficients from rkr turning points: {$C_6$} and {$C_8$} for ${B}(^3\pi_{0_u}^+)$-state {Cl$_2$}, {Br$_2$}, and {I$_2$}},\ }\href@noop {} {\bibfield  {journal} {\bibinfo  {journal} {Canadian Journal of Physics}\ }\textbf {\bibinfo {volume} {52}},\ \bibinfo {pages} {246} (\bibinfo {year} {1974})}\BibitemShut {NoStop}%
\bibitem [{\citenamefont {Jackson}(1998)}]{Jackson1998}%
  \BibitemOpen
  \bibfield  {author} {\bibinfo {author} {\bibfnamefont {J.~D.}\ \bibnamefont {Jackson}},\ }\href@noop {} {\emph {\bibinfo {title} {Classical Electrodynamics}}},\ \bibinfo {edition} {3rd}\ ed.\ (\bibinfo  {publisher} {John Wiley \& Sons},\ \bibinfo {year} {1998})\BibitemShut {NoStop}%
\bibitem [{\citenamefont {Born}\ and\ \citenamefont {Oppenheimer}(1927)}]{Born1927}%
  \BibitemOpen
  \bibfield  {author} {\bibinfo {author} {\bibfnamefont {M.}~\bibnamefont {Born}}\ and\ \bibinfo {author} {\bibfnamefont {R.}~\bibnamefont {Oppenheimer}},\ }\bibfield  {title} {\bibinfo {title} {Zur quantentheorie der molekeln},\ }\href@noop {} {\bibfield  {journal} {\bibinfo  {journal} {Annalen der Physik}\ }\textbf {\bibinfo {volume} {389}},\ \bibinfo {pages} {457} (\bibinfo {year} {1927})}\BibitemShut {NoStop}%
\bibitem [{\citenamefont {Born}\ and\ \citenamefont {Huang}(1954)}]{Born1954DynamicalTheory}%
  \BibitemOpen
  \bibfield  {author} {\bibinfo {author} {\bibfnamefont {M.}~\bibnamefont {Born}}\ and\ \bibinfo {author} {\bibfnamefont {K.}~\bibnamefont {Huang}},\ }\href@noop {} {\emph {\bibinfo {title} {Dynamical Theory of Crystal Lattices}}}\ (\bibinfo  {publisher} {Oxford University Press},\ \bibinfo {address} {Oxford},\ \bibinfo {year} {1954})\BibitemShut {NoStop}%
\bibitem [{\citenamefont {Clark}(1979)}]{Clark1979}%
  \BibitemOpen
  \bibfield  {author} {\bibinfo {author} {\bibfnamefont {C.~W.}\ \bibnamefont {Clark}},\ }\bibfield  {title} {\bibinfo {title} {The calculation of non-adiabatic transition probabilities},\ }\href {https://doi.org/10.1016/0375-9601(79)90070-3} {\bibfield  {journal} {\bibinfo  {journal} {Physics Letters A}\ }\textbf {\bibinfo {volume} {70}},\ \bibinfo {pages} {295} (\bibinfo {year} {1979})}\BibitemShut {NoStop}%
\bibitem [{\citenamefont {Srikumar}\ \emph {et~al.}(2025)\citenamefont {Srikumar}, \citenamefont {Exner}, \citenamefont {Bl\"{a}ttner}, \citenamefont {Schmelcher}, \citenamefont {Eiles},\ and\ \citenamefont {Ott}}]{Srikumar2025Trilobite}%
  \BibitemOpen
  \bibfield  {author} {\bibinfo {author} {\bibfnamefont {R.}~\bibnamefont {Srikumar}}, \bibinfo {author} {\bibfnamefont {M.}~\bibnamefont {Exner}}, \bibinfo {author} {\bibfnamefont {R.}~\bibnamefont {Bl\"{a}ttner}}, \bibinfo {author} {\bibfnamefont {P.}~\bibnamefont {Schmelcher}}, \bibinfo {author} {\bibfnamefont {M.~T.}\ \bibnamefont {Eiles}},\ and\ \bibinfo {author} {\bibfnamefont {H.}~\bibnamefont {Ott}},\ }\bibfield  {title} {\bibinfo {title} {Vibrationally highly excited trilobite molecules stabilized by nonadiabatic coupling},\ }\href {https://doi.org/10.1103/PhysRevResearch.7.043103} {\bibfield  {journal} {\bibinfo  {journal} {Phys. Rev. Res.}\ }\textbf {\bibinfo {volume} {7}},\ \bibinfo {pages} {043103} (\bibinfo {year} {2025})}\BibitemShut {NoStop}%
\bibitem [{\citenamefont {Exner}\ \emph {et~al.}(2026)\citenamefont {Exner}, \citenamefont {Srikumar}, \citenamefont {Bl{\"a}ttner}, \citenamefont {Schmelcher}, \citenamefont {Sadeghpour}, \citenamefont {Eiles},\ and\ \citenamefont {Ott}}]{Exner2026SingletButterfly}%
  \BibitemOpen
  \bibfield  {author} {\bibinfo {author} {\bibfnamefont {M.}~\bibnamefont {Exner}}, \bibinfo {author} {\bibfnamefont {R.}~\bibnamefont {Srikumar}}, \bibinfo {author} {\bibfnamefont {R.}~\bibnamefont {Bl{\"a}ttner}}, \bibinfo {author} {\bibfnamefont {P.}~\bibnamefont {Schmelcher}}, \bibinfo {author} {\bibfnamefont {H.~R.}\ \bibnamefont {Sadeghpour}}, \bibinfo {author} {\bibfnamefont {M.~T.}\ \bibnamefont {Eiles}},\ and\ \bibinfo {author} {\bibfnamefont {H.}~\bibnamefont {Ott}},\ }\bibfield  {title} {\bibinfo {title} {Observation of spin-singlet butterfly {Rydberg} molecules in an ultracold atomic {Rb} gas},\ }\href {https://doi.org/10.1103/PhysRevLett.136.243002} {\bibfield  {journal} {\bibinfo  {journal} {Phys. Rev. Lett.}\ }\textbf {\bibinfo {volume} {136}},\ \bibinfo {pages} {243002} (\bibinfo {year} {2026})}\BibitemShut {NoStop}%
\bibitem [{\citenamefont {Guttridge}\ \emph {et~al.}(2025)\citenamefont {Guttridge}, \citenamefont {Hepworth}, \citenamefont {Ruttley}, \citenamefont {Durst}, \citenamefont {Eiles},\ and\ \citenamefont {Cornish}}]{Guttridge2025Tweezers}%
  \BibitemOpen
  \bibfield  {author} {\bibinfo {author} {\bibfnamefont {A.}~\bibnamefont {Guttridge}}, \bibinfo {author} {\bibfnamefont {T.~R.}\ \bibnamefont {Hepworth}}, \bibinfo {author} {\bibfnamefont {D.~K.}\ \bibnamefont {Ruttley}}, \bibinfo {author} {\bibfnamefont {A.~A.~T.}\ \bibnamefont {Durst}}, \bibinfo {author} {\bibfnamefont {M.~T.}\ \bibnamefont {Eiles}},\ and\ \bibinfo {author} {\bibfnamefont {S.~L.}\ \bibnamefont {Cornish}},\ }\bibfield  {title} {\bibinfo {title} {Individual assembly of two-species {Rydberg} molecules using optical tweezers},\ }\href {https://doi.org/10.1103/PhysRevLett.134.133401} {\bibfield  {journal} {\bibinfo  {journal} {Phys. Rev. Lett.}\ }\textbf {\bibinfo {volume} {134}},\ \bibinfo {pages} {133401} (\bibinfo {year} {2025})}\BibitemShut {NoStop}%
\bibitem [{\citenamefont {Aymar}\ \emph {et~al.}(1996)\citenamefont {Aymar}, \citenamefont {Greene},\ and\ \citenamefont {Luc-Koenig}}]{Aymar1996}%
  \BibitemOpen
  \bibfield  {author} {\bibinfo {author} {\bibfnamefont {M.}~\bibnamefont {Aymar}}, \bibinfo {author} {\bibfnamefont {C.~H.}\ \bibnamefont {Greene}},\ and\ \bibinfo {author} {\bibfnamefont {E.}~\bibnamefont {Luc-Koenig}},\ }\bibfield  {title} {\bibinfo {title} {Multichannel {Rydberg} spectroscopy of complex atoms},\ }\href {https://doi.org/10.1103/RevModPhys.68.1015} {\bibfield  {journal} {\bibinfo  {journal} {Reviews of Modern Physics}\ }\textbf {\bibinfo {volume} {68}},\ \bibinfo {pages} {1015} (\bibinfo {year} {1996})}\BibitemShut {NoStop}%
\bibitem [{\citenamefont {Lane}\ and\ \citenamefont {Thomas}(1958)}]{Lane1958}%
  \BibitemOpen
  \bibfield  {author} {\bibinfo {author} {\bibfnamefont {A.~M.}\ \bibnamefont {Lane}}\ and\ \bibinfo {author} {\bibfnamefont {R.~G.}\ \bibnamefont {Thomas}},\ }\bibfield  {title} {\bibinfo {title} {R-matrix theory of nuclear reactions},\ }\href {https://doi.org/10.1103/RevModPhys.30.257} {\bibfield  {journal} {\bibinfo  {journal} {Reviews of Modern Physics}\ }\textbf {\bibinfo {volume} {30}},\ \bibinfo {pages} {257} (\bibinfo {year} {1958})}\BibitemShut {NoStop}%
\bibitem [{\citenamefont {Burke}(2011)}]{Burke2011}%
  \BibitemOpen
  \bibfield  {author} {\bibinfo {author} {\bibfnamefont {P.~G.}\ \bibnamefont {Burke}},\ }\href {https://doi.org/10.1007/978-3-642-15931-2} {\emph {\bibinfo {title} {R-Matrix Theory of Atomic Collisions: Application to Atomic, Molecular and Optical Processes}}},\ \bibinfo {series} {Springer Series on Atomic, Optical, and Plasma Physics}, Vol.~\bibinfo {volume} {61}\ (\bibinfo  {publisher} {Springer-Verlag},\ \bibinfo {address} {Berlin, Heidelberg},\ \bibinfo {year} {2011})\BibitemShut {NoStop}%
\bibitem [{\citenamefont {Wang}(2012)}]{WangND}%
  \BibitemOpen
  \bibfield  {author} {\bibinfo {author} {\bibfnamefont {J.}~\bibnamefont {Wang}},\ }\emph {\bibinfo {title} {Hyperspherical Approach to Quantal Three-body Theory}},\ \href@noop {} {Ph.D. thesis},\ \bibinfo  {school} {University of Colorado} (\bibinfo {year} {2012})\BibitemShut {NoStop}%
\bibitem [{\citenamefont {Jr.}(1999)}]{Burke1999}%
  \BibitemOpen
  \bibfield  {author} {\bibinfo {author} {\bibfnamefont {J.~P.~B.}\ \bibnamefont {Jr.}},\ }\emph {\bibinfo {title} {Theoretical Investigation of Cold Alkali Atom Collisions}},\ \href@noop {} {Ph.D. thesis},\ \bibinfo  {school} {University of Colorado} (\bibinfo {year} {1999})\BibitemShut {NoStop}%
\bibitem [{\citenamefont {Watson}(1922)}]{Watson1922}%
  \BibitemOpen
  \bibfield  {author} {\bibinfo {author} {\bibfnamefont {G.~N.}\ \bibnamefont {Watson}},\ }\href@noop {} {\emph {\bibinfo {title} {A Treatise on the Theory of Bessel Functions}}}\ (\bibinfo  {publisher} {Cambridge University Press},\ \bibinfo {address} {Cambridge},\ \bibinfo {year} {1922})\BibitemShut {NoStop}%
\bibitem [{\citenamefont {Wigner}(1955)}]{Wigner1955}%
  \BibitemOpen
  \bibfield  {author} {\bibinfo {author} {\bibfnamefont {E.~P.}\ \bibnamefont {Wigner}},\ }\bibfield  {title} {\bibinfo {title} {Lower limit for the energy derivative of the scattering phase shift},\ }\href {https://doi.org/10.1103/PhysRev.98.145} {\bibfield  {journal} {\bibinfo  {journal} {Physical Review}\ }\textbf {\bibinfo {volume} {98}},\ \bibinfo {pages} {145} (\bibinfo {year} {1955})}\BibitemShut {NoStop}%
\bibitem [{\citenamefont {Smith}(1960)}]{Smith1960}%
  \BibitemOpen
  \bibfield  {author} {\bibinfo {author} {\bibfnamefont {F.~T.}\ \bibnamefont {Smith}},\ }\bibfield  {title} {\bibinfo {title} {Lifetime matrix in collision theory},\ }\href@noop {} {\bibfield  {journal} {\bibinfo  {journal} {Physical Review}\ }\textbf {\bibinfo {volume} {118}},\ \bibinfo {pages} {349} (\bibinfo {year} {1960})}\BibitemShut {NoStop}%
\bibitem [{\citenamefont {Landau}(1932)}]{Landau1932}%
  \BibitemOpen
  \bibfield  {author} {\bibinfo {author} {\bibfnamefont {L.~D.}\ \bibnamefont {Landau}},\ }\bibfield  {title} {\bibinfo {title} {On the theory of transfer of energy at collisions ii},\ }\href@noop {} {\bibfield  {journal} {\bibinfo  {journal} {Physikalische Zeitschrift der Sowjetunion}\ }\textbf {\bibinfo {volume} {2}},\ \bibinfo {pages} {46} (\bibinfo {year} {1932})}\BibitemShut {NoStop}%
\bibitem [{\citenamefont {Zener}(1932)}]{Zener1932}%
  \BibitemOpen
  \bibfield  {author} {\bibinfo {author} {\bibfnamefont {C.}~\bibnamefont {Zener}},\ }\bibfield  {title} {\bibinfo {title} {Non-adiabatic crossing of energy levels},\ }\href {https://doi.org/10.1098/rspa.1932.0165} {\bibfield  {journal} {\bibinfo  {journal} {Proceedings of the Royal Society of London A}\ }\textbf {\bibinfo {volume} {137}},\ \bibinfo {pages} {696} (\bibinfo {year} {1932})}\BibitemShut {NoStop}%
\bibitem [{\citenamefont {Stückelberg}(1932)}]{Stueckelberg1932}%
  \BibitemOpen
  \bibfield  {author} {\bibinfo {author} {\bibfnamefont {E.~C.~G.}\ \bibnamefont {Stückelberg}},\ }\bibfield  {title} {\bibinfo {title} {Theorie der unelastischen {Stösse} zwischen {Atomen}},\ }\href@noop {} {\bibfield  {journal} {\bibinfo  {journal} {Helvetica Physica Acta}\ }\textbf {\bibinfo {volume} {5}},\ \bibinfo {pages} {369} (\bibinfo {year} {1932})}\BibitemShut {NoStop}%
\bibitem [{\citenamefont {Child}(1974{\natexlab{b}})}]{Child1974MolecularCollisionTheory}%
  \BibitemOpen
  \bibfield  {author} {\bibinfo {author} {\bibfnamefont {M.~S.}\ \bibnamefont {Child}},\ }\href@noop {} {\emph {\bibinfo {title} {Molecular Collision Theory}}},\ Theoretical Chemistry: A Series of Monographs\ (\bibinfo  {publisher} {Academic Press},\ \bibinfo {address} {London},\ \bibinfo {year} {1974})\BibitemShut {NoStop}%
\bibitem [{\citenamefont {Zhu}\ and\ \citenamefont {Nakamura}(1997{\natexlab{a}})}]{ZhuNakamura1997}%
  \BibitemOpen
  \bibfield  {author} {\bibinfo {author} {\bibfnamefont {C.}~\bibnamefont {Zhu}}\ and\ \bibinfo {author} {\bibfnamefont {H.}~\bibnamefont {Nakamura}},\ }\bibfield  {title} {\bibinfo {title} {Usefulness of the newly completed semiclassical theory of curve crossing: Multi-channel resonant scattering},\ }\href {https://doi.org/10.1016/S0009-2614(97)00593-9} {\bibfield  {journal} {\bibinfo  {journal} {Chemical Physics Letters}\ }\textbf {\bibinfo {volume} {274}},\ \bibinfo {pages} {205} (\bibinfo {year} {1997}{\natexlab{a}})}\BibitemShut {NoStop}%
\bibitem [{\citenamefont {Zhu}\ and\ \citenamefont {Nakamura}(1997{\natexlab{b}})}]{ZhuNakamura21997}%
  \BibitemOpen
  \bibfield  {author} {\bibinfo {author} {\bibfnamefont {C.}~\bibnamefont {Zhu}}\ and\ \bibinfo {author} {\bibfnamefont {H.}~\bibnamefont {Nakamura}},\ }\bibfield  {title} {\bibinfo {title} {Semiclassical theory of multi-channel curve crossing problems: {Landau--Zener} case},\ }\href {https://doi.org/10.1063/1.473115} {\bibfield  {journal} {\bibinfo  {journal} {The Journal of Chemical Physics}\ }\textbf {\bibinfo {volume} {106}},\ \bibinfo {pages} {2599} (\bibinfo {year} {1997}{\natexlab{b}})}\BibitemShut {NoStop}%
\bibitem [{\citenamefont {Zhu}\ \emph {et~al.}(1992)\citenamefont {Zhu}, \citenamefont {Nakamura}, \citenamefont {Re},\ and\ \citenamefont {Aquilanti}}]{ZhuNakamura1992_Stokes}%
  \BibitemOpen
  \bibfield  {author} {\bibinfo {author} {\bibfnamefont {C.}~\bibnamefont {Zhu}}, \bibinfo {author} {\bibfnamefont {H.}~\bibnamefont {Nakamura}}, \bibinfo {author} {\bibfnamefont {N.}~\bibnamefont {Re}},\ and\ \bibinfo {author} {\bibfnamefont {V.}~\bibnamefont {Aquilanti}},\ }\bibfield  {title} {\bibinfo {title} {{The two-state linear curve crossing problems revisited. I. Analysis of Stokes phenomenon and expressions for scattering matrices}},\ }\href {https://doi.org/10.1063/1.463178} {\bibfield  {journal} {\bibinfo  {journal} {The Journal of Chemical Physics}\ }\textbf {\bibinfo {volume} {97}},\ \bibinfo {pages} {1892} (\bibinfo {year} {1992})}\BibitemShut {NoStop}%
\bibitem [{\citenamefont {Zhu}\ and\ \citenamefont {Nakamura}(1992)}]{ZhuNakamura1992_LZ}%
  \BibitemOpen
  \bibfield  {author} {\bibinfo {author} {\bibfnamefont {C.}~\bibnamefont {Zhu}}\ and\ \bibinfo {author} {\bibfnamefont {H.}~\bibnamefont {Nakamura}},\ }\bibfield  {title} {\bibinfo {title} {The two-state linear curve crossing problems revisited. {II}. {Analytical} approximations for the {Stokes} constant and scattering matrix: The {Landau--Zener} case},\ }\href {https://doi.org/10.1063/1.463368} {\bibfield  {journal} {\bibinfo  {journal} {The Journal of Chemical Physics}\ }\textbf {\bibinfo {volume} {97}},\ \bibinfo {pages} {8497} (\bibinfo {year} {1992})}\BibitemShut {NoStop}%
\bibitem [{\citenamefont {Nakamura}(2002)}]{Nakamura2002Nonadiabatic}%
  \BibitemOpen
  \bibfield  {author} {\bibinfo {author} {\bibfnamefont {H.}~\bibnamefont {Nakamura}},\ }\href@noop {} {\emph {\bibinfo {title} {Nonadiabatic Transition: Concepts, Basic Theories and Applications}}}\ (\bibinfo  {publisher} {World Scientific},\ \bibinfo {address} {Singapore},\ \bibinfo {year} {2002})\BibitemShut {NoStop}%
\bibitem [{\citenamefont {Giannakeas}\ and\ \citenamefont {Greene}(2020)}]{Giannakeas2020EfimovLineshapes}%
  \BibitemOpen
  \bibfield  {author} {\bibinfo {author} {\bibfnamefont {P.}~\bibnamefont {Giannakeas}}\ and\ \bibinfo {author} {\bibfnamefont {C.~H.}\ \bibnamefont {Greene}},\ }\bibfield  {title} {\bibinfo {title} {Asymmetric lineshapes of efimov resonances in mass-imbalanced ultracold gases},\ }\href {https://doi.org/10.3390/atoms8040089} {\bibfield  {journal} {\bibinfo  {journal} {Atoms}\ }\textbf {\bibinfo {volume} {8}},\ \bibinfo {pages} {89} (\bibinfo {year} {2020})}\BibitemShut {NoStop}%
\bibitem [{\citenamefont {Child}(2014)}]{Child1991Semiclassical}%
  \BibitemOpen
  \bibfield  {author} {\bibinfo {author} {\bibfnamefont {M.~S.}\ \bibnamefont {Child}},\ }\href@noop {} {\emph {\bibinfo {title} {Semiclassical Mechanics with Molecular Applications}}},\ \bibinfo {edition} {2nd}\ ed.\ (\bibinfo  {publisher} {Oxford University Press},\ \bibinfo {address} {Oxford},\ \bibinfo {year} {2014})\BibitemShut {NoStop}%
\bibitem [{\citenamefont {Landau}\ and\ \citenamefont {Lifshitz}(1977)}]{LandauLifshitzQM}%
  \BibitemOpen
  \bibfield  {author} {\bibinfo {author} {\bibfnamefont {L.~D.}\ \bibnamefont {Landau}}\ and\ \bibinfo {author} {\bibfnamefont {E.~M.}\ \bibnamefont {Lifshitz}},\ }\href@noop {} {\emph {\bibinfo {title} {Quantum Mechanics: Non-Relativistic Theory}}},\ \bibinfo {edition} {3rd}\ ed.\ (\bibinfo  {publisher} {Pergamon Press},\ \bibinfo {year} {1977})\BibitemShut {NoStop}%
\bibitem [{\citenamefont {Sadeghpour}\ \emph {et~al.}(1992)\citenamefont {Sadeghpour}, \citenamefont {Greene},\ and\ \citenamefont {Cavagnero}}]{SadeghpourGreeneCavagnero1992}%
  \BibitemOpen
  \bibfield  {author} {\bibinfo {author} {\bibfnamefont {H.~R.}\ \bibnamefont {Sadeghpour}}, \bibinfo {author} {\bibfnamefont {C.~H.}\ \bibnamefont {Greene}},\ and\ \bibinfo {author} {\bibfnamefont {M.}~\bibnamefont {Cavagnero}},\ }\bibfield  {title} {\bibinfo {title} {Extensive eigenchannel {R}-matrix study of the {H}$^{-}$ photodetachment spectrum},\ }\href {https://doi.org/10.1103/PhysRevA.45.1587} {\bibfield  {journal} {\bibinfo  {journal} {Physical Review A}\ }\textbf {\bibinfo {volume} {45}},\ \bibinfo {pages} {1587} (\bibinfo {year} {1992})}\BibitemShut {NoStop}%
\bibitem [{\citenamefont {Giannakeas}\ and\ \citenamefont {Greene}(2018)}]{Giannakeas2018EfimovStueckelberg}%
  \BibitemOpen
  \bibfield  {author} {\bibinfo {author} {\bibfnamefont {P.}~\bibnamefont {Giannakeas}}\ and\ \bibinfo {author} {\bibfnamefont {C.~H.}\ \bibnamefont {Greene}},\ }\bibfield  {title} {\bibinfo {title} {Ultracold heteronuclear three-body systems: How diabaticity limits the universality of recombination into shallow dimers},\ }\href {https://doi.org/10.1103/PhysRevLett.120.023401} {\bibfield  {journal} {\bibinfo  {journal} {Physical Review Letters}\ }\textbf {\bibinfo {volume} {120}},\ \bibinfo {pages} {023401} (\bibinfo {year} {2018})}\BibitemShut {NoStop}%
\bibitem [{\citenamefont {Fano}(1961)}]{Fano1961}%
  \BibitemOpen
  \bibfield  {author} {\bibinfo {author} {\bibfnamefont {U.}~\bibnamefont {Fano}},\ }\bibfield  {title} {\bibinfo {title} {Effects of configuration interaction on intensities and phase shifts},\ }\href {https://doi.org/10.1103/PhysRev.124.1866} {\bibfield  {journal} {\bibinfo  {journal} {Physical Review}\ }\textbf {\bibinfo {volume} {124}},\ \bibinfo {pages} {1866} (\bibinfo {year} {1961})}\BibitemShut {NoStop}%
\bibitem [{\citenamefont {Feshbach}(1962)}]{Feshbach1962}%
  \BibitemOpen
  \bibfield  {author} {\bibinfo {author} {\bibfnamefont {H.}~\bibnamefont {Feshbach}},\ }\bibfield  {title} {\bibinfo {title} {A unified theory of nuclear reactions},\ }\href@noop {} {\bibfield  {journal} {\bibinfo  {journal} {Annals of Physics}\ }\textbf {\bibinfo {volume} {19}},\ \bibinfo {pages} {287} (\bibinfo {year} {1962})}\BibitemShut {NoStop}%
\bibitem [{Note1()}]{Note1}%
  \BibitemOpen
  \bibinfo {note} {This could be the electronic ground-state for a direct transition to an electronic $p$-state, or a transition from an intermediate state to the weak $s$- and $d$-state admixture in the molecular configuration as was done in Ref.~\cite {Zuber2022}. The initial state of the nuclei is a low-energy scattering wave function.}\BibitemShut {Stop}%
\bibitem [{\citenamefont {Connerade}\ and\ \citenamefont {Lane}(1988)}]{Connerade1988InteractingResonances}%
  \BibitemOpen
  \bibfield  {author} {\bibinfo {author} {\bibfnamefont {J.-P.}\ \bibnamefont {Connerade}}\ and\ \bibinfo {author} {\bibfnamefont {A.~M.}\ \bibnamefont {Lane}},\ }\bibfield  {title} {\bibinfo {title} {Interacting resonances in atomic spectroscopy},\ }\href {https://doi.org/10.1088/0034-4885/51/11/002} {\bibfield  {journal} {\bibinfo  {journal} {Reports on Progress in Physics}\ }\textbf {\bibinfo {volume} {51}},\ \bibinfo {pages} {1439} (\bibinfo {year} {1988})}\BibitemShut {NoStop}%
\bibitem [{\citenamefont {Connerade}\ and\ \citenamefont {Lane}(1987)}]{Connerade1987QReversals}%
  \BibitemOpen
  \bibfield  {author} {\bibinfo {author} {\bibfnamefont {J.~P.}\ \bibnamefont {Connerade}}\ and\ \bibinfo {author} {\bibfnamefont {A.~M.}\ \bibnamefont {Lane}},\ }\bibfield  {title} {\bibinfo {title} {$q$ reversals in a {Rydberg} series caused by an interloper level},\ }\href {https://doi.org/10.1088/0022-3700/20/8/015} {\bibfield  {journal} {\bibinfo  {journal} {Journal of Physics B: Atomic and Molecular Physics}\ }\textbf {\bibinfo {volume} {20}},\ \bibinfo {pages} {1757} (\bibinfo {year} {1987})}\BibitemShut {NoStop}%
\bibitem [{\citenamefont {Mizuno}\ and\ \citenamefont {Hukushima}(2018)}]{mizuno2018power}%
  \BibitemOpen
  \bibfield  {author} {\bibinfo {author} {\bibfnamefont {Y.}~\bibnamefont {Mizuno}}\ and\ \bibinfo {author} {\bibfnamefont {K.}~\bibnamefont {Hukushima}},\ }\bibfield  {title} {\bibinfo {title} {Power-law decay in the nonadiabatic photodissociation dynamics of alkali halides due to quantum wavepacket interference},\ }\href@noop {} {\bibfield  {journal} {\bibinfo  {journal} {The Journal of Chemical Physics}\ }\textbf {\bibinfo {volume} {149}} (\bibinfo {year} {2018})}\BibitemShut {NoStop}%
\bibitem [{Note2()}]{Note2}%
  \BibitemOpen
  \bibinfo {note} {A full dataset to compare with the results of Ref.~\cite {Duspayev2022NonadiabaticRAIM} can be found in \cite {Abraham2026Thesis}.}\BibitemShut {Stop}%
\bibitem [{\citenamefont {Berngruber}\ \emph {et~al.}(2024{\natexlab{b}})\citenamefont {Berngruber}, \citenamefont {Bosworth}, \citenamefont {Herrera-Sancho}, \citenamefont {Anasuri}, \citenamefont {Zuber}, \citenamefont {Hummel}, \citenamefont {Krauter}, \citenamefont {Meinert}, \citenamefont {L{\"o}w}, \citenamefont {Schmelcher},\ and\ \citenamefont {Pfau}}]{Berngruber2024}%
  \BibitemOpen
  \bibfield  {author} {\bibinfo {author} {\bibfnamefont {M.}~\bibnamefont {Berngruber}}, \bibinfo {author} {\bibfnamefont {D.~J.}\ \bibnamefont {Bosworth}}, \bibinfo {author} {\bibfnamefont {O.~A.}\ \bibnamefont {Herrera-Sancho}}, \bibinfo {author} {\bibfnamefont {V.~S.~V.}\ \bibnamefont {Anasuri}}, \bibinfo {author} {\bibfnamefont {N.}~\bibnamefont {Zuber}}, \bibinfo {author} {\bibfnamefont {F.}~\bibnamefont {Hummel}}, \bibinfo {author} {\bibfnamefont {J.}~\bibnamefont {Krauter}}, \bibinfo {author} {\bibfnamefont {F.}~\bibnamefont {Meinert}}, \bibinfo {author} {\bibfnamefont {R.}~\bibnamefont {L{\"o}w}}, \bibinfo {author} {\bibfnamefont {P.}~\bibnamefont {Schmelcher}},\ and\ \bibinfo {author} {\bibfnamefont {T.}~\bibnamefont {Pfau}},\ }\bibfield  {title} {\bibinfo {title} {In situ observation of nonpolar to strongly polar atom-ion collision dynamics},\ }\href {https://doi.org/10.1103/PhysRevLett.133.083001} {\bibfield  {journal} {\bibinfo  {journal} {Phys. Rev. Lett.}\ }\textbf {\bibinfo {volume} {133}},\
  \bibinfo {pages} {083001} (\bibinfo {year} {2024}{\natexlab{b}})}\BibitemShut {NoStop}%
\bibitem [{\citenamefont {G\'en\'evriez}\ \emph {et~al.}(2025)\citenamefont {G\'en\'evriez}, \citenamefont {Jungers}, \citenamefont {Rosen},\ and\ \citenamefont {Eichmann}}]{Genevriez2025}%
  \BibitemOpen
  \bibfield  {author} {\bibinfo {author} {\bibfnamefont {M.}~\bibnamefont {G\'en\'evriez}}, \bibinfo {author} {\bibfnamefont {M.}~\bibnamefont {Jungers}}, \bibinfo {author} {\bibfnamefont {C.}~\bibnamefont {Rosen}},\ and\ \bibinfo {author} {\bibfnamefont {U.}~\bibnamefont {Eichmann}},\ }\bibfield  {title} {\bibinfo {title} {Series of molecularlike doubly excited states of a quasi-three-body coulomb system},\ }\href {https://doi.org/10.1103/svsd-9mj3} {\bibfield  {journal} {\bibinfo  {journal} {Phys. Rev. Lett.}\ }\textbf {\bibinfo {volume} {135}},\ \bibinfo {pages} {153002} (\bibinfo {year} {2025})}\BibitemShut {NoStop}%
\bibitem [{\citenamefont {Tiwari}\ \emph {et~al.}(2026)\citenamefont {Tiwari}, \citenamefont {Karman},\ and\ \citenamefont {Walther}}]{tiwari2026low}%
  \BibitemOpen
  \bibfield  {author} {\bibinfo {author} {\bibfnamefont {S.~K.}\ \bibnamefont {Tiwari}}, \bibinfo {author} {\bibfnamefont {T.}~\bibnamefont {Karman}},\ and\ \bibinfo {author} {\bibfnamefont {V.}~\bibnamefont {Walther}},\ }\bibfield  {title} {\bibinfo {title} {Low-energy purification of crystal defects by {Rydberg} excitons},\ }\href@noop {} {\bibfield  {journal} {\bibinfo  {journal} {arXiv preprint arXiv:2604.27331}\ } (\bibinfo {year} {2026})}\BibitemShut {NoStop}%
\bibitem [{\citenamefont {Berry}(1966)}]{Berry1966PhaseShifts}%
  \BibitemOpen
  \bibfield  {author} {\bibinfo {author} {\bibfnamefont {M.~V.}\ \bibnamefont {Berry}},\ }\bibfield  {title} {\bibinfo {title} {Semi-classical scattering phase shifts in the presence of metastable states},\ }\href {https://doi.org/10.1088/0370-1328/88/2/304} {\bibfield  {journal} {\bibinfo  {journal} {Proceedings of the Physical Society}\ }\textbf {\bibinfo {volume} {88}},\ \bibinfo {pages} {285} (\bibinfo {year} {1966})}\BibitemShut {NoStop}%
\bibitem [{\citenamefont {Zhu}\ and\ \citenamefont {Nakamura}(1995)}]{Zhu1995}%
  \BibitemOpen
  \bibfield  {author} {\bibinfo {author} {\bibfnamefont {C.}~\bibnamefont {Zhu}}\ and\ \bibinfo {author} {\bibfnamefont {H.}~\bibnamefont {Nakamura}},\ }\bibfield  {title} {\bibinfo {title} {{Theory of Nonadiabatic Transition for General Two-State Curve Crossing Problems. II. Landau--Zener Case}},\ }\href {https://doi.org/10.1063/1.469057} {\bibfield  {journal} {\bibinfo  {journal} {The Journal of Chemical Physics}\ }\textbf {\bibinfo {volume} {102}},\ \bibinfo {pages} {7448} (\bibinfo {year} {1995})}\BibitemShut {NoStop}%
\bibitem [{\citenamefont {Sakurai}\ and\ \citenamefont {Napolitano}(2011)}]{Sakurai2011ModernQM}%
  \BibitemOpen
  \bibfield  {author} {\bibinfo {author} {\bibfnamefont {J.~J.}\ \bibnamefont {Sakurai}}\ and\ \bibinfo {author} {\bibfnamefont {J.}~\bibnamefont {Napolitano}},\ }\href@noop {} {\emph {\bibinfo {title} {Modern Quantum Mechanics}}},\ \bibinfo {edition} {2nd}\ ed.\ (\bibinfo  {publisher} {Addison-Wesley},\ \bibinfo {address} {Boston},\ \bibinfo {year} {2011})\BibitemShut {NoStop}%
\bibitem [{\citenamefont {Abraham}(2026)}]{Abraham2026Thesis}%
  \BibitemOpen
  \bibfield  {author} {\bibinfo {author} {\bibfnamefont {N.}~\bibnamefont {Abraham}},\ }\emph {\bibinfo {title} {{Non-adiabatic Predissociation of Ultralong-Range Rydberg Atom-Ion Molecules}}},\ \href@noop {} {Ph.D. thesis},\ \bibinfo  {school} {Max Planck Institute for the Physics of Complex Systems} (\bibinfo {year} {2026}),\ \bibinfo {note} {{Ph.D.} thesis, submitted}\BibitemShut {NoStop}%
\end{thebibliography}
%

\end{document}